\documentclass[11pt]{article}
\usepackage{authblk, amsmath, openbib, epsf, graphicx}

\ExplSyntaxOn
\keys_define:nn { manuscript }
  {
    blind .choice:,
    blind / yes .code:n = { \cs_set_eq:NN \nonblind \use_none:n },
    blind / no  .code:n = { \cs_set_eq:NN \nonblind \use:n },
    blind .initial:n = no
  }
\newcommand{\papersetup}[1]{\keys_set:nn { manuscript } {#1}}
\ExplSyntaxOff

\papersetup{blind=no} 

\usepackage[round,authoryear]{natbib}

\usepackage[colorlinks=true,
            citecolor=blue,
            linkcolor=blue,
            urlcolor=blue]{hyperref}
 \usepackage[normalem]{ulem}

\bibpunct[, ]{(}{)}{;}{a}{}{,}

\usepackage{doi}
\usepackage{amsmath,amssymb}
\usepackage{mathrsfs}
\usepackage{booktabs,dcolumn}
\usepackage{caption}
\usepackage{subcaption}
\usepackage{threeparttable}
\usepackage{booktabs}
\usepackage{rotating,booktabs}

\usepackage{indentfirst}
\usepackage{array}
\usepackage{times}
\usepackage[usenames]{color}
\usepackage{psfrag,graphicx}
\usepackage{eso-pic}
\usepackage{graphicx}
\usepackage{type1cm}
\usepackage{bm}
\usepackage{float}
\usepackage{multirow}
\usepackage{silence}
\usepackage{color}
\usepackage{appendix} 
\usepackage{orcidlink} 

\usepackage{cleveref}
\usepackage{comment}

\usepackage{newtxtext,newtxmath} 

\usepackage{amsmath, amssymb}
\usepackage{xcolor}  

\definecolor{forestgreen}{RGB}{34,139,34}

\usepackage{algorithm}
\usepackage{algpseudocode}
\usepackage{enumitem}

\newcommand{\utwi}[1]{\mbox{\boldmath $ #1$}}

\usepackage{geometry}
\usepackage{multibib}          
\newcites{supp}{References}   

\date{}
\renewcommand{\baselinestretch}{1.5}

\begin{document}
\begin{titlepage}
\title{Modeling Joint Tail Risk and Dependence for Cryptocurrencies via a Copula-Based Semiparametric Framework}
\author{
\nonblind{ 
Cathy W.S. Chen\thanks{Corresponding author: Cathy W.S. Chen, E-mail: chenws@mail.fcu.edu.tw}$\:\:^{1}$,
Takaaki Koike$^{2}$, and Aaron E. P. Lin$^{1}$
\\
$^{1}$Department of Statistics, Feng Chia University, Taiwan\\
$^{2}$Graduate School of Economics, Hitotsubashi University, Japan
}
}

\maketitle
\setcounter{page}{1}

\begin{abstract}
This study proposes a Bayesian copula-based bivariate semiparametric framework for jointly forecasting Value-at-Risk (VaR) and Expected Shortfall (ES) while modeling cross-asset tail dependence. The marginal dynamics are specified using four ES-CAViaR-type models, including extensions incorporating Rogers-Satchell and Garman-Klass volatility measures. To construct the copula inputs, we use the empirical ranks of standardized residuals, which depend on the marginal parameters, thereby linking marginal tail dynamics and dependence modeling without requiring fully specified marginal distributions. Clayton and BB1 copulas are considered, and adaptive Markov chain Monte Carlo methods are used for estimation and joint tail-risk forecasting. A simulation study under a DCC-GARCH-$t$ data-generating process evaluates the finite-sample forecasting performance of the proposed framework. 
Using Bitcoin and Ethereum data, we evaluate one-step-ahead
forecasts from 2020 to 2025 through a rolling-window
forecasting scheme and compare the proposed models with
Bayesian DCC-GARCH-N and DCC-GARCH-$t$ benchmarks.
Rolling copula estimates also reveal time-varying asymmetry in
Bitcoin-Ethereum tail dependence, with lower-tail dependence remaining higher than upper-tail dependence for most of the period. The proposed
framework delivers competitive forecasting performance for both
marginal and joint tail risk.

\vspace{1cm} 
\noindent{\textbf{Keywords:} Value-at-Risk, Expected Shortfall, ES-CAViaR, Copulas, Tail dependence, Semiparametric copula inference, 
Bayesian inference.
}
\end{abstract}
\end{titlepage}

\setcounter{page}{2}
\section{Introduction}

Tail risk plays a central role in financial analysis because extreme negative movements can arise abruptly and lead to substantial losses. Value-at-Risk (VaR) and Expected Shortfall (ES) are widely used measures for quantifying tail risk. VaR characterizes the loss threshold at a given confidence level, while ES measures the expected loss beyond this threshold in the lower tail. 
ES complements VaR by capturing the average loss conditional on losses exceeding the VaR level at the same confidence level $\alpha$. By focusing on the tail beyond the quantile, ES reflects the severity of extreme losses and provides a more comprehensive assessment of downside risk, particularly when return distributions exhibit heavy tails or asymmetry. Consistent with their importance, both VaR and ES remain integral components of the market risk capital framework under Basel~III \citep{Basel2016}.

Methods for VaR and ES modeling for single assets and portfolios can be broadly classified into parametric, semiparametric, and nonparametric approaches. Parametric approaches typically specify the conditional return distribution through GARCH-type models, from which VaR and ES are obtained via analytical or likelihood-based calculations under an assumed innovation distribution \citep{Engle1982EJES,Bollerslev1986JoE}.
Semiparametric approaches retain a parametric structure for volatility or quantile dynamics while allowing greater flexibility in modeling tail behavior, thereby avoiding full specification of the return distribution. A representative example is the conditional autoregressive VaR (CAViaR) framework of \citet{Engle2004}, which directly models conditional quantiles but is primarily designed for VaR forecasting.

To enable coherent inference on both VaR and ES, extensions such as ES-CAViaR introduce additional dynamic structures that link ES to the VaR process \citep{Taylor2019JBES}. This framework has been further extended to incorporate realized measures and other covariates to improve tail risk forecasting \citep{ChenHsu2023,ChenKoike2024JF,ChenChen2025}.
Despite these advances, most existing approaches focus on individual assets and do not explicitly account for cross-asset dependence. In practice, extreme losses often occur jointly across assets, making joint modeling of tail risk an important task.

Financial assets are exposed to common economic shocks and contagion effects, which induce tail dependence \citep{Longin2001JoF}. Joint modeling of VaR and ES across assets captures extreme co-movements and provides more reliable portfolio-level risk evaluation, particularly during periods of market stress. Such a framework requires an explicit specification of cross-asset tail dependence. 
Conventional multivariate models embed dependence in dynamic correlation structures, such as the dynamic conditional correlation (DCC) model of \citet{Engle2002JBES}, a multivariate GARCH model with time-varying correlations of \citet{TseTsui2002JBES}, or bivariate asymmetric GARCH-type models with heavy-tailed innovations (\citealt*{Choy2014QF}; \citealt{Chen2019ASMBI_bvi_garch}). In these specifications, conditional variances and correlations jointly determine the dependence structure, which may be inadequate for capturing nonlinear and extremal dependence.
Multivariate semiparametric risk modeling is also possible
through a DCC-based formulation \citep{StortiWang2025}.
More recently, \citet{YANG_IJF} propose a dynamic joint
estimation approach for CoVaR and CoES using quantile
regression and score-driven dynamics to assess systemic risk.

Copula-based models provide a flexible framework for separating marginal dynamics from the dependence structure, allowing for a more accurate characterization of joint tail risk, especially in the presence of asymmetric and heavy-tailed behavior.
Different copulas capture distinct features of dependence, particularly in the tails. 
We refer readers to~\citet{Nelsen2006Springer,Joe2014CRC} for comprehensive overviews of copulas.
The Clayton copula captures lower-tail dependence, whereas the Gumbel copula
emphasizes upper-tail dependence and the Student-$t$ copula imposes symmetric
dependence in both tails. Since our primary interest is downside risk through
VaR and ES, we focus on lower-tail dependence. The Clayton copula therefore
provides a parsimonious benchmark for joint extreme losses. To allow greater
flexibility, we also consider the BB1 (Clayton-Gumbel) copula
\citep{joe1996multivariate,joe1997multivariate}, which accommodates asymmetric
lower- and upper-tail dependence. We do not include the Student-$t$ copula
because its symmetric tail-dependence structure does not directly target the
asymmetric lower-tail dependence emphasized in this study.

We develop a Bayesian semiparametric bivariate framework that jointly models
marginal tail dynamics, cross-asset tail dependence, and VaR-ES risk
forecasts. 
The Bayesian approach allows the incorporation of prior information and produces posterior simulations that facilitate interval estimation and predictive inference. 
The framework incorporates four ES-CAViaR-type marginal
specifications and links them through copulas that capture nonlinear and
tail dependence across assets. Copula parameters are estimated using a
rank-based quasi-likelihood constructed from empirical distributions, which
reduces sensitivity to marginal misspecification.
The resulting composite pseudo-posterior provides a unified framework for
estimating marginal tail risk and dependence parameters while quantifying
parameter uncertainty.

Semiparametric copula inference commonly replaces unknown marginal
transforms by empirical distribution functions
\citep{GenestGhoudiRivest1995}; for multivariate time series, the use of
estimated standardized residuals is studied by
\citet{ChenFan2006SCOMDY}. Related Bayesian approaches include Gaussian
copula regression \citep{Pitt2006}, the extended rank likelihood
\citep{Hoff2007}, and inference on copula functionals using Bayesian
exponentially tilted empirical likelihood \citep{GrazianLiseo2017BA}.
These approaches provide useful background, but they do not directly justify our joint pseudo-posterior, in which the empirical residual ranks depend on the marginal VaR--ES parameters. We distinguish computation under the specified pseudo-posterior from claims of exact Bayesian inference or asymptotically calibrated inference.

\citet{Ausin2010CSDA} develop full Bayesian inference
for copula-GARCH models by jointly estimating the
marginal and copula parameters.
Our framework combines semiparametric ES-CAViaR
marginal modeling with a copula-based dependence structure through a composite
pseudo-posterior that retains distinct marginal and dependence components.
This construction allows the framework to characterize both marginal tail risk
and cross-asset tail dependence while accounting for parameter uncertainty.
Moreover, the ES-CAViaR specifications directly target VaR and ES dynamics,
making them particularly suitable for joint tail-risk forecasting.

The aim is to develop a flexible framework for jointly modeling and forecasting
tail risk across multiple assets by integrating semiparametric ES-CAViaR
models with copula-based dependence structures. The main contributions are
threefold.

First, we extend the ES-CAViaR framework to a bivariate setting and compare
four marginal specifications. These include the benchmark ES-CAViaR model of
\citet{Taylor2019JBES}, a Volatility-standardized exceedance specification, and two
extensions incorporating the Rogers-Satchell and Garman-Klass volatility
measures \citep{RS1991,Garman1980} to capture time-varying volatility
and extreme market movements. We also include Bayesian DCC-GARCH models
with normal and Student-$t$ innovations as external parametric benchmarks.

Second, we employ the Clayton and BB1 copulas to characterize cross-asset
tail dependence without requiring fully specified conditional marginal
distributions. By re-estimating the copula parameters over rolling windows,
we examine the evolution of lower- and upper-tail dependence and their
asymmetry. In particular, the BB1 specification reveals a consistent dominance of lower-tail dependence over upper-tail dependence.
Beyond marginal VaR and ES forecasts, the copula
component enables us to predict the probability of simultaneous VaR
violations, providing an additional assessment of joint downside risk.

Third, we develop a sequential Bayesian estimation approach using MCMC
that integrates marginal VaR-ES dynamics with copula-based dependence.
The marginal parameters are estimated through AL
working likelihoods without feedback from the copula pseudo-likelihood,
thereby preserving the marginal VaR-ES estimating criterion.
Marginal parameter uncertainty enters copula estimation through
updated ranks of standardized residuals, allowing dependence
parameters to be re-estimated at each forecast origin.

The remainder of this paper is organized as follows.
Section~\ref{sec:model:specifications} describes the model specification,
including the marginal ES-CAViaR-type models and the copula-based dependence
structure. Section~\ref{sec:Bayesian:inference} develops the Bayesian
estimation framework and introduces the forecast evaluation criteria.
Section~\ref{sec:simulation} reports a simulation study based on data generated from a
DCC-GARCH-$t$ model, with one-step-ahead VaR and ES forecasts obtained from
the proposed copula-based ES-CAViaR-st model.
Section~\ref{sec:empirical:study} examines the cryptocurrency market,
including the data description, in-sample estimation and out-of-sample
forecasting performance evaluation. Section~\ref{sec:conclusion} concludes the paper.

\section{Model Specifications}\label{sec:model:specifications}

We consider four ES-CAViaR-type specifications
to characterize marginal tail-risk dynamics
and jointly forecast VaR and ES, including the
benchmark ES-CAViaR model and three extensions.
 
\subsection{Marginal Specifications}\label{sec:marginal:specificification}

This study extends four ES-CAViaR-type models to characterize marginal tail dynamics and jointly forecast VaR and ES. Specifically, we redefine the exceedance component in the ES dynamics as a scaled exceedance ratio to improve the modeling of tail behavior. 
Below, $r_{i,t}$ denotes the log return of Asset $i$ at time $t$, that is, $r_{i,t}=\{\log (P_{i,t}^{\mathrm{C}})-\log (P_{i,t-1}^{\mathrm{C}})\}\times 100$, where $P_{i,t}^{\mathrm{C}}$ is the closing price of Asset $i$ at time $t$.
In addition, $\mathcal F_{t-1}$ denotes the past information up to time $t-1$.
For each $i$ and $t$, the VaR at level $\alpha\in (0,1)$ conditional on $\mathcal F_{t-1}$, denoted by $Q_{i,t}$, is the $\alpha$-quantile of $r_{i,t}$ conditional on $\mathcal F_{t-1}$, which satisfies $\mathbb{P}\!\left( r_{i,t} < Q_{i,t} \mid \mathcal{F}_{t-1} \right) = \alpha$. 
 Analogously, the ES at level $\alpha\in (0,1)$ conditional on $\mathcal F_{t-1}$, denoted by $e_{i,t}$, is given by $e_{i,t}
 =
 \mathbb{E}\!\left( r_{i,t} \mid r_{i,t} < Q_{i,t},\, \mathcal{F}_{t-1} \right)$.

\paragraph{ES-CAViaR-type Models}

We use the ES-CAViaR model of \citet{Taylor2019JBES} as the benchmark marginal specification for joint VaR and ES dynamics. For Asset $i$, the model is given by
\begin{equation*}
	\begin{aligned}
    \hspace{-3.5cm} \mbox{ES-CAViaR} &\\
\quad 
Q_{i,t}&= \beta_{i,1}+ \beta_{i,2} Q_{i,t-1}+\beta_{i,3} \left| r_{i,t-1}\right| ,\\
		e_{i,t}&= Q_{i,t}-w_{i,t},\\
		w_{i,t}&= 
		\begin{cases}
			\gamma_{i,1} + \gamma_{i,2} \left ({Q_{i,t-1} - r_{i,t-1}}\right)+ \gamma_{i,3} w_{i,t-1}, & \: \text{if } r_{i,t-1} \leq Q_{i,t-1}, \\
			w_{i,t-1}, &\:\text{otherwise}.
		\end{cases}
	\end{aligned}
\end{equation*}
The parameter restrictions of this model and the subsequent three models are specified in the paragraph \textit{Parameter restrictions and initialization} below.

To account for differences in market volatility, we consider an alternative specification that standardizes VaR exceedances by recent conditional volatility. A VaR exceedance of the same magnitude may have different implications during calm and highly volatile market periods. The resulting ES-CAViaR-st model is
\begin{equation*}
	\begin{aligned}
\hspace{-3.5cm} \mbox{ES-CAViaR-st} &\\
\quad Q_{i,t} &= \beta_{i,1}+ \beta_{i,2} Q_{i,t-1}+\beta_{i,3} \left| r_{i,t-1}\right| ,\\
		e_{i,t} &= Q_{i,t}-w_{i,t},\\
		w_{i,t} &=
		\begin{cases}
			\gamma_{i,1} + \gamma_{i,2} \dfrac{Q_{i,t-1}-r_{i,t-1}}{\sigma_{i,t-1}}
			+ \gamma_{i,3} w_{i,t-1}, & \text{if } r_{i,t-1} \leq Q_{i,t-1}, \\
			w_{i,t-1}, & \text{otherwise}.
		\end{cases}
	\end{aligned}
\end{equation*}

The term
$(Q_{i,t-1}-r_{i,t-1})/\sigma_{i,t-1}$
measures the magnitude of a VaR violation
relative to recent market volatility.
We estimate $\sigma_{i,t}$ using the
exponentially weighted moving average recursion
\begin{equation*}
\sigma_{i,t}^{2}
=
\lambda \sigma_{i,t-1}^{2}
+
(1-\lambda)r_{i,t}^{2},
\qquad
0<\lambda<1,
\end{equation*}
where $\lambda$ is a smoothing parameter. Following the standard RiskMetrics specification, we fix $\lambda=0.94$ throughout the simulation and empirical analyses.

In the third model, we extend the VaR dynamics by
replacing $|r_{i,t-1}|$ with a daily volatility
proxy constructed from open, high, low, and close
prices. Unlike close-to-close returns, which may
overlook substantial intraday price fluctuations,
this range-based measure incorporates additional
information about daily price movements.
We construct $V_{i,t}$ using the Rogers-Satchell
measure \citep{RS1991}.
The resulting ES-CAViaR-v specification is
\begin{align}
\hspace{-2.5cm} \mbox{ES-CAViaR-v}
& \nonumber\\
Q_{i,t}
&=
\beta_{i,1}
+\beta_{i,2}Q_{i,t-1}
+\beta_{i,3}\log(V_{i,t-1}+1),
\label{eq:ES-CAViaR-v}\\
e_{i,t}
&=
Q_{i,t}-w_{i,t},
\nonumber\\
w_{i,t}
&=
\begin{cases}
\gamma_{i,1}
+\gamma_{i,2}\dfrac{Q_{i,t-1}-r_{i,t-1}}{\sigma_{i,t-1}}
+\gamma_{i,3}w_{i,t-1},
& \text{if } r_{i,t-1}\leq Q_{i,t-1},\\
w_{i,t-1},
& \text{otherwise},
\end{cases}
\nonumber\\
V_{i,t}
&=
\log\!\left(\frac{P_{i,t}^{\mathrm{H}}}{P_{i,t}^{\mathrm{O}}}\right)
\log\!\left(\frac{P_{i,t}^{\mathrm{H}}}{P_{i,t}^{\mathrm{C}}}\right)
+
\log\!\left(\frac{P_{i,t}^{\mathrm{L}}}{P_{i,t}^{\mathrm{O}}}\right)
\log\!\left(\frac{P_{i,t}^{\mathrm{L}}}{P_{i,t}^{\mathrm{C}}}\right),
\nonumber
\end{align}
where \(P_{i,t}^{\mathrm{O}}\), \(P_{i,t}^{\mathrm{H}}\), \(P_{i,t}^{\mathrm{L}}\), and \(P_{i,t}^{\mathrm{C}}\) denote the daily open, high, low, and close prices, respectively. 

For the fourth model, we incorporate the Garman-Klass volatility measure \citep{Garman1980} as a VaR covariate, following \citet*{ChenLin2026CE}. By combining information from the intraday high-low range and open-close price movements, the Garman-Klass measure captures multiple dimensions of daily price variation and provides an informative volatility input for tail-risk forecasting. The corresponding ES-CAViaR-R specification is
\begin{align}
\hspace{-1.cm} \mbox{ES-CAViaR-R}
& \nonumber\\
Q_{i,t}
&=
\beta_{i,1}
+\beta_{i,2}Q_{i,t-1}
+\beta_{i,3}\tilde{R}_{i,t-1},
\label{eq:ES-CAViaR-R}\\
e_{i,t}
&=
Q_{i,t}-w_{i,t},
\nonumber\\
w_{i,t}
&=
\begin{cases}
\gamma_{i,1}
+\gamma_{i,2}\dfrac{Q_{i,t-1}-r_{i,t-1}}{\sigma_{i,t-1}}
+\gamma_{i,3}w_{i,t-1},
& \text{if } r_{i,t-1}\leq Q_{i,t-1},\\
w_{i,t-1},
& \text{otherwise},
\end{cases}
\nonumber\\
\tilde{R}_{i,t}
&=
0.511
\left\{
\log\left(\frac{P_{i,t}^{\mathrm H}}
{P_{i,t}^{\mathrm L}}\right)
\right\}^{2}-0.383
\left\{
\log\left(\frac{P_{i,t}^{\mathrm C}}
{P_{i,t}^{\mathrm O}}\right)
\right\}^{2}
\nonumber\\
&-0.019
\Biggl[
\log\left(\frac{P_{i,t}^{\mathrm C}}
{P_{i,t}^{\mathrm O}}\right)
\log\left(
\frac{P_{i,t}^{\mathrm H}P_{i,t}^{\mathrm L}}
{(P_{i,t}^{\mathrm O})^{2}}
\right)
-2
\log\left(\frac{P_{i,t}^{\mathrm H}}
{P_{i,t}^{\mathrm O}}\right)
\log\left(\frac{P_{i,t}^{\mathrm L}}
{P_{i,t}^{\mathrm O}}\right)
\Biggr].
\nonumber
\end{align}

\paragraph{Parameter restrictions and initialization}

In our downside-risk analysis, $Q_{i,t}$ represents
a negative lower-tail VaR. We impose the following
parameter restrictions across all four ES-CAViaR
specifications:
\begin{equation*}
\beta_{i,1}<0,
\quad
0\leq\beta_{i,2}<1,
\quad
\beta_{i,3}<0,
\quad
\gamma_{i,1}>0,
\quad
\gamma_{i,2}>0,
\quad
0<\gamma_{i,3}<1.
\end{equation*}
The recursions start at $t=2$, as all ES-CAViaR
specifications depend on the lagged return $r_{i,t-1}$.
Given $Q_{i,1}<0$, the sign restrictions preserve
$Q_{i,t}<0$ recursively. Analogous restrictions apply
to the volatility extensions, whose covariates are
nonnegative. We initialize $w_{i,1}>0$, which,
together with the restrictions on $\boldsymbol{\gamma}_i$
and a strictly positive scale denominator, ensures
$w_{i,t}>0$ and hence $e_{i,t}<Q_{i,t}$ for $t\geq2$.
The conditions $0\leq\beta_{i,2}<1$ and
$0<\gamma_{i,3}<1$ control persistence in the VaR
and ES recursions, respectively.

\subsection{Copula Representation and Conditional Rank-Based Pseudo-Observations}
\label{sec:conditional:rank:pseudo}

Let \(H_t\) denote the conditional joint distribution of
\((r_{1,t},r_{2,t})^{\prime}\) given \(\mathcal F_{t-1}\), with continuous
marginal distributions \(F_{1,t}\) and \(F_{2,t}\). By Sklar's theorem,
\[
H_t(x_1,x_2)
=
C_t\!\left(
F_{1,t}(x_1),
F_{2,t}(x_2)
\right),
\]
where \(C_t\) is a bivariate copula. Our marginal models specify the conditional VaR \(Q_{i,t}\) and
ES \(e_{i,t}\), rather than the full conditional marginal
distribution. We therefore construct the copula pseudo-observations from
within-window empirical ranks of transformed residuals.
Conditional on the current draw of the marginal parameters, define
\begin{equation}
z_{i,t}
=
\frac{r_{i,t}-Q_{i,t}}{w_{i,t}},
\qquad
w_{i,t}=Q_{i,t}-e_{i,t}>0.
\label{eq:rank:standardization}
\end{equation}
Since \(w_{i,t}>0\), the transformation in
\eqref{eq:rank:standardization} is strictly increasing conditional on
\(\mathcal F_{t-1}\). Hence, the conditional copula of
\((r_{1,t},r_{2,t})^{\prime}\) is identical to that of
\((z_{1,t},z_{2,t})^{\prime}\).

For a given rolling estimation window, we suppress the window index and
let \(t\) denote observation dates within the window. We assume the working
representations:
\begin{equation}
F_{i,t}(x)
=
G_i\!\left(
\frac{x-Q_{i,t}}{w_{i,t}}
\right),
\quad i=1,2,
\qquad
C_t(u_1,u_2)=C(u_1,u_2),
\label{eq:rank_working_model}
\end{equation}
where $G_1$ and $G_2$ are continuous distribution
functions that remain parametrically unspecified and $C_t$ is the copula of $(r_{1,t},r_
{2,t})\mid \mathcal F_{t-1}$.
Under the working representations in~\eqref{eq:rank_working_model}, the transformed
residuals have common conditional marginal
distributions within each window, while their
conditional copula remains constant over that window.

Since $G_1$ and $G_2$ are unspecified, we construct the copula
pseudo-observations using the empirical ranks of
the standardized residuals:
\begin{equation}
z_{i,t}^{[j]}
=
\frac{r_{i,t}-Q_{i,t}^{[j]}}
     {Q_{i,t}^{[j]}-e_{i,t}^{[j]}},
\quad i=1,2,\quad t=2,\ldots,n,
\label{eq:MCMC:standardized:residuals}
\end{equation}
where $Q_{i,t}^{[j]}$ and $e_{i,t}^{[j]}$
denote the conditional VaR and ES evaluated
at the parameter draws from MCMC iteration $j$,
and $n$ denotes the rolling-window sample size.
The corresponding rank-based pseudo-observations
are computed as
\begin{equation}
u_{i,t}^{[j]}
=
\frac{1}{n}
\sum_{s=2}^{n}
\mathbb{I}\left(z_{i,s}^{[j]}\leq z_{i,t}^{[j]}\right),
\quad i=1,2,\quad t=2,\ldots,n,
\label{eq:MCMC:rank:pseudo}
\end{equation}
where $\mathbb{I}$ is the indicator.
Note that $u_{i,t}^{[j]}\in(0,1)$.

Unlike the conventional conditional probability integral
transform (PIT) $F_{i,t}(r_{i,t}\mid\mathcal F_{t-1})$, which require fully specified conditional marginal
distribution functions, our rank-based
pseudo-observations rely on the empirical
distributions of standardized residuals.
These pseudo-observations are not exact conditional
PITs, but provide approximately uniform marginal
inputs for copula estimation under the working
assumption that the standardized residual
distributions remain approximately stable within
each rolling estimation window. This construction
avoids imposing parametric assumptions on the
complete conditional return distributions.
Since the standardized residuals depend on the
marginal parameters, we recompute their empirical
ranks whenever these parameters are updated
during Bayesian estimation.

\subsection{Copula Families}\label{sec:copula:specification}

We consider the Clayton and BB1 copulas to model
cross-asset tail dependence, with particular emphasis
on the joint occurrence of extreme negative returns.
These copulas provide alternative representations of
tail dependence, allowing us to examine how
dependence asymmetry affects forecasts of simultaneous
VaR violations.

\paragraph{Clayton copula}
For $u_1,u_2\in(0,1)$, the Clayton copula is
\[
C^{\mathrm{Clayton}}(u_1,u_2\mid\theta)
=
(u_1^{-\theta}+u_2^{-\theta}-1)^{-1/\theta},
\quad \theta>0.
\]
It exhibits lower-tail dependence, characterized by the lower-tail dependence coefficient (TDC)
\[
\lambda_L^{\mathrm{Clayton}}=2^{-1/\theta},
\]
while its upper-TDC is zero.
The copula parameter is related to Kendall's tau through
\begin{equation}
\tau=\frac{\theta}{\theta+2},
\qquad
\theta=\frac{2\tau}{1-\tau}.
\label{eq:kendall_clayton}
\end{equation}
The corresponding copula density is
\[
c^{\mathrm{Clayton}}(u_1,u_2\mid\theta)
=
(1+\theta)(u_1u_2)^{-1-\theta}
(u_1^{-\theta}+u_2^{-\theta}-1)^{-2-1/\theta}.
\]

\paragraph{BB1 copula}
The BB1 copula extends the Clayton specification
by allowing both lower- and upper-tail dependence:
\[
C^{\mathrm{BB1}}(u_1,u_2\mid\theta,\delta)
=
\left[
1+
\left\{
(u_1^{-\theta}-1)^\delta
+
(u_2^{-\theta}-1)^\delta
\right\}^{1/\delta}
\right]^{-1/\theta},
\]
where $\theta>0$ and $\delta\geq 1$.
Its TDCs are
\[
\lambda_L^{\mathrm{BB1}}=2^{-1/(\delta\theta)},
\qquad
\lambda_U^{\mathrm{BB1}}=2-2^{1/\delta}.
\]
When $\delta=1$, the BB1 copula reduces to the
Clayton copula. The BB1 specification
provides a more flexible representation of
asymmetric tail dependence.

For notational convenience, define
\[
A=u_1^{-\theta}-1,\quad
B=u_2^{-\theta}-1,\quad
S=(A^\delta+B^\delta)^{1/\delta}.
\]
Then the density of the BB1 copula~\citep[p.~191]{Joe2014CRC} is
\[
\begin{aligned}
c^{\mathrm{BB1}}(u_1,u_2\mid\theta,\delta)
={}&u_1^{-(\theta+1)}u_2^{-(\theta+1)}
A^{\delta-1}B^{\delta-1}
(A^\delta+B^\delta)^{1/\delta-2}\\
&\times(1+S)^{-1/\theta-2}
\left[S(\delta\theta+1)+\theta(\delta-1)\right].
\end{aligned}
\]

\paragraph{Implications for joint tail-risk forecasting}
The distinction between these copulas is relevant
to forecasting simultaneous VaR violations. For
a given risk level $\alpha$, the conditional
probability of joint violations is
\begin{align}\label{eq:true:joint:prob}
p_{t}
:=\mathbb{P}(r_{1,t}\le Q_{1,t},r_{2,t}\le Q_{2,t}\mid \mathcal F_{t-1})
=
C_t(\alpha,\alpha;\boldsymbol\psi),
\end{align}
where $\boldsymbol\psi$ denotes the copula
parameter. The Clayton copula focuses on
lower-tail dependence, whereas the BB1 copula
allows additional flexibility in the overall
dependence structure. Comparing these
specifications enables us to assess whether
greater flexibility in modeling tail dependence
improves out-of-sample joint tail-event forecasts.

\section{Bayesian Inference for the Copula-Based ES-CAViaR models}\label{sec:Bayesian:inference}

We combine an asymmetric Laplace (AL) marginal
working likelihood with a rank-based copula
criterion. We first specify the priors, then
introduce the marginal working posterior and
conditional copula pseudo-posterior, and finally
describe the MCMC procedure, rolling forecasts,
and out-of-sample evaluation.

\subsection{Prior Specifications}\label{sec:prior}

Let
$\utwi{\vartheta}
=
\left(
\utwi{\beta}_{1}^{\prime},
\utwi{\gamma}_{1}^{\prime},
\utwi{\beta}_{2}^{\prime},
\utwi{\gamma}_{2}^{\prime},
\boldsymbol{\psi}^{\prime}
\right)^{\prime}
=
\left(
\utwi{\eta}^{\prime},
\boldsymbol{\psi}^{\prime}
\right)^{\prime}$
denote the full vector of unknown parameters in the
copula-based bivariate ES-CAViaR framework, where
$\utwi{\eta}
=
\left(
\utwi{\beta}_{1}^{\prime},
\utwi{\gamma}_{1}^{\prime},
\utwi{\beta}_{2}^{\prime},
\utwi{\gamma}_{2}^{\prime}
\right)^{\prime}$
collects the marginal VaR-ES parameters, and the
dependence parameter $\boldsymbol{\psi}$ is
$\theta$ for the Clayton copula and $(\theta,\delta)^{\prime}$
for the BB1 copula.
For $i=1,2$, let
$\utwi{\beta}_{i}
=(\beta_{i,1},\beta_{i,2},\beta_{i,3})'$
and
$\utwi{\gamma}_{i}
=(\gamma_{i,1},\gamma_{i,2},\gamma_{i,3})'$
denote the parameter vectors in the ES-CAViaR-type
models for Asset $i$.
The prior factorizes as
\begin{align*}
\pi(\boldsymbol\vartheta)
=
\pi_C(\boldsymbol\psi)
\prod_{i=1}^{2}
\pi_{\beta,i}(\boldsymbol\beta_i)
\pi_{\gamma,i}(\boldsymbol\gamma_i).
\end{align*}

\paragraph{Marginal dynamics}

The priors for the marginal parameters are specified
as truncated multivariate normal distributions.
For each $i$,
\[
\boldsymbol{\beta}_i
\sim
\operatorname{TN}_3\!\left(
\boldsymbol{\mu}_{\beta,i},
\Sigma_{\beta,i},
\mathcal{B}
\right),
\qquad
\boldsymbol{\gamma}_i
\sim
\operatorname{TN}_3\!\left(
\boldsymbol{\mu}_{\gamma,i},
\Sigma_{\gamma,i},
\mathcal G
\right),
\]
where $\operatorname{TN}(\cdot)$ denotes a truncated
normal distribution, and the parameters are
\[
\boldsymbol{\mu}_{\beta,i}
=
(\mu_{i,1},\mu_{i,2},\mu_{i,3})^{\prime},
\qquad
\Sigma_{\beta,i}
=
\mathrm{diag}(
\sigma_{i,1}^2,
\sigma_{i,2}^2,
\sigma_{i,3}^2
),
\]
\[
\boldsymbol{\mu}_{\gamma,i}
=
(\eta_{i,1},\eta_{i,2},\eta_{i,3})^{\prime},
\qquad
\Sigma_{\gamma,i}
=
\mathrm{diag}(
\zeta_{i,1}^2,
\zeta_{i,2}^2,
\zeta_{i,3}^2
).
\]
In addition, the truncation regions are defined as
\begin{align*}
\mathcal B
&=
\{\boldsymbol\beta_i:
\beta_{i,1}<0,\
0\leq\beta_{i,2}<1,\
\beta_{i,3}<0\},\\
\mathcal G
&=
\{\boldsymbol\gamma_i:
\gamma_{i,1}>0,\
\gamma_{i,2}>0,\
0<\gamma_{i,3}<1\}.
\end{align*}

\paragraph{Clayton copula}

For Bayesian inference, we parameterize the Clayton
copula in terms of Kendall's tau, as defined in
equation~(\ref{eq:kendall_clayton}). We assign the prior
\[
\tau\sim\operatorname{Beta}(a_\tau,b_\tau),
\qquad a_\tau,b_\tau>0.
\]
This prior ensures that $\tau$ lies in $(0,1)$
and provides flexibility in specifying prior
beliefs about the strength of dependence.

\paragraph{BB1 copula}

For the BB1 copula, we assign independent log-normal
priors to the dependence parameters:
\[
\theta \sim \operatorname{Lognormal}
(\mu_\theta,\sigma_\theta^2),
\qquad
\delta-1 \sim \operatorname{Lognormal}
(\mu_\delta,\sigma_\delta^2).
\]
These priors ensure that $\theta>0$ and $\delta>1$
while allowing flexibility in specifying prior beliefs
about the dependence structure.

Building on the adaptive MCMC approach of
\citet{ChenSo2006IJoF}, we develop a modified
Metropolis-Hastings (MH) sampler that jointly updates
the transformed parameter vector
$\boldsymbol{\phi}
=(\log\theta,\log(\delta-1))^{\prime}$.
After burn-in, the sampler combines an independence
Gaussian proposal and a
Gaussian random-walk proposal.
The former accounts for posterior dependence through
a covariance matrix estimated from the burn-in draws,
while the latter facilitates local exploration.
This modification aims to improve posterior mixing.

\subsection{Pseudo-Posterior Specification}\label{sec:posterior}

We extend the univariate ES-CAViaR framework to a bivariate
semiparametric setting by modeling cross-asset dependence through a copula.
For each margin, the VaR and ES dynamics are estimated using the
ES-CAViaR quasi-likelihood based on the AL representation
of quantile regression \citep{Koenker1978EJES,Koenker1999JASA}.

Define the check loss \begin{align*}
\rho_\alpha(v)=v\{\alpha-\mathbb I(v\leq0)\},\quad v\in \mathbb{R}.
\end{align*}
We retain the AL working kernel of the ES-CAViaR formulation
\citep{Taylor2019JBES}:
\begin{align*}
 f_{i,t}(r_{i,t};\boldsymbol\eta_i)
 &=\frac{1-\alpha}{-e_{i,t}(\boldsymbol\eta_i)}
 \exp\left\{
 \frac{\rho_\alpha(r_{i,t}-Q_{i,t}(\boldsymbol\eta_i))}
      {\alpha e_{i,t}(\boldsymbol\eta_i)}
 \right\},\\
 L_{M,i}(\boldsymbol\eta_i)&=\prod_{t=2}^{n}
                         f_{i,t}(r_{i,t};\boldsymbol\eta_i),
\end{align*}
where $(Q_{i,t}(\boldsymbol{\eta}_i),e_{i,t}(\boldsymbol{\eta}_i))^{\prime}$ follows one of the dynamics presented in Section~\ref{sec:marginal:specificification}.

Although $f_{i,t}$ is an AL density as a function of its observation,
we use it as an estimation criterion, not as an assumption about the
true conditional return distribution. Its interpretation as a joint
VaR--ES criterion uses the zero-conditional-mean working approximation;
the exact interpretation requires $\mathbb E[r_{i,t}\,|\,\mathcal F_{t-1}]=0$.

For the dependence component, we use the rank-based
pseudo-observations defined in Section~\ref{sec:conditional:rank:pseudo}.
These pseudo-observations depend on the marginal
ES-CAViaR parameters and enter the copula
pseudo-likelihood specified below.

\subsection{Bayesian Estimation}

Let $\boldsymbol{\eta}=(\boldsymbol{\eta}_1^{\prime},
\boldsymbol{\eta}_2^{\prime})^{\prime}$ collect the marginal VaR-ES
parameters, and let $\boldsymbol{\psi}$ denote
the copula parameters. Define $\boldsymbol{u}(\boldsymbol{\eta})
=
\{u_{i,t}(\boldsymbol{\eta}_i):
i=1,2,\;t=2,\ldots,n\}$.
The marginal working likelihood and copula
pseudo-likelihood are given by
\begin{align*}
L_M(\boldsymbol{\eta})
&=
\prod_{i=1}^{2}L_{M,i}(\boldsymbol{\eta}_i),
\\
L_C(\boldsymbol{\psi};
\boldsymbol{u}(\boldsymbol{\eta}))
&=
\prod_{t=2}^{n}
c\left(
u_{1,t}(\boldsymbol{\eta}_1),
u_{2,t}(\boldsymbol{\eta}_2);
\boldsymbol{\psi}
\right),
\end{align*}
where $c$ denotes the selected copula density.
We conduct Bayesian estimation by first updating
the marginal parameters using their respective
working likelihoods and prior distributions:
\begin{equation}
\pi_M(\boldsymbol{\eta}\mid\boldsymbol{r})
\propto
L_M(\boldsymbol{\eta})\pi_M(\boldsymbol{\eta}).
\label{eq:marginal-posterior}
\end{equation}
Conditional on the current marginal parameter draws,
we update the copula parameters using
\begin{equation}
\pi_C(\boldsymbol{\psi}\mid
\boldsymbol{u}(\boldsymbol{\eta}))
\propto
L_C(\boldsymbol{\psi};
\boldsymbol{u}(\boldsymbol{\eta}))
\pi_C(\boldsymbol{\psi}).
\label{eq:copula-posterior}
\end{equation}
The two posterior components define the following
joint distribution:
\begin{equation*}
\pi_{\mathrm{cut}}(
\boldsymbol{\eta},\boldsymbol{\psi}
\mid\boldsymbol{r})
=
\pi_M(\boldsymbol{\eta}\mid\boldsymbol{r})
\pi_C\left(
\boldsymbol{\psi}\mid
\boldsymbol{u}(\boldsymbol{\eta})
\right).
\end{equation*}

We adopt this sequential Bayesian approach to
avoid feedback from the copula pseudo-likelihood
to the marginal parameter updates. The ES-CAViaR
models estimate conditional VaR and ES through
an AL working likelihood without
specifying complete conditional return distributions.
The associated working criterion provides a basis
for joint VaR-ES estimation under the conditions
discussed by \citet{FZ2016} and
\citet{Taylor2019JBES}. Allowing the rank-based copula
criterion to influence marginal estimation would
change the VaR-ES estimating criterion and could
introduce sensitivity to misspecification of the
dependence structure.

This concern is particularly relevant at extremely low risk
levels, where few VaR violations
inform the ES recursion, whereas the copula
pseudo-likelihood uses ranks from the entire
estimation window. We therefore retain the
marginal working posterior while allowing
marginal parameter uncertainty to enter copula
estimation through updated residual ranks.
Under this no-feedback specification, the
marginal posterior does not depend on the
selected copula family, while the copula
component determines the associated joint
tail-event probability forecasts.

\subsection{Forecasting Procedure}
\label{sec:forecasting:procedure}

We implement Bayesian inference using MCMC based on
the marginal working posterior and conditional copula
pseudo-posterior in equations~\eqref{eq:marginal-posterior}
and \eqref{eq:copula-posterior}, respectively.

\paragraph{MCMC estimation and forecasting}
At each MCMC iteration, we recompute the
rank-based pseudo-observations defined in
Section~\ref{sec:conditional:rank:pseudo} using the current marginal parameter
draws and use them to update the copula parameters.
We then obtain one-step-ahead VaR and ES forecasts.
Let $N$ denote the total number of MCMC iterations.
At each iteration $j=1,\ldots,N$, we perform the following five
steps for the ES-CAViaR-st specification.

\begin{enumerate}[label=\textbf{Step \arabic*.},
leftmargin=*]

\item \textbf{Update the VaR parameters.}
For each asset $i=1,2$, update
$\boldsymbol{\beta}_i$ using an
MH step targeting
\[
\pi_M\!\left(
\boldsymbol{\beta}_i
\mid
\boldsymbol{\gamma}_i^{[j-1]},
\boldsymbol r
\right).
\]

\item \textbf{Update the ES parameters.}
For each asset $i=1,2$, update
$\boldsymbol{\gamma}_i$ using an
MH step targeting
\[
\pi_M\!\left(
\boldsymbol{\gamma}_i
\mid
\boldsymbol{\beta}_i^{[j]},
\boldsymbol r
\right).
\]

\item \textbf{Recompute the rank-based copula inputs.}
Using the current marginal parameter draws,
compute $z_{i,t}^{[j]}$ and $u_{i,t}^{[j]}$
from equations~
\eqref{eq:MCMC:standardized:residuals}
and \eqref{eq:MCMC:rank:pseudo}, respectively.
Let
\[
\boldsymbol u^{[j]}
=
\{(u_{1,t}^{[j]},u_{2,t}^{[j]})^{\prime}:
t=2,\ldots,n\}.
\]

\item \textbf{Update the copula parameters.}
Update $\boldsymbol{\psi}$ using an
MH step targeting
\[
\pi_C\!\left(
\boldsymbol{\psi}
\mid
\boldsymbol u^{[j]}
\right).
\]

\item \textbf{Compute the one-step-ahead
VaR and ES forecasts.}
Using the current marginal parameter draws,
compute
\[
\begin{aligned}
Q_{i,n+1}^{[j]}
&=
\beta_{i,1}^{[j]}
+\beta_{i,2}^{[j]}Q_{i,n}^{[j]}
+\beta_{i,3}^{[j]}|r_{i,n}|,\\
e_{i,n+1}^{[j]}
&=
Q_{i,n+1}^{[j]}-w_{i,n+1}^{[j]},\\
w_{i,n+1}^{[j]}
&=
\begin{cases}
\gamma_{i,1}^{[j]}
+\gamma_{i,2}^{[j]}
\dfrac{Q_{i,n}^{[j]}-r_{i,n}}{\sigma_{i,n}}
+\gamma_{i,3}^{[j]}w_{i,n}^{[j]},
& r_{i,n}\leq Q_{i,n}^{[j]},\\[4pt]
w_{i,n}^{[j]}, & \text{otherwise}.
\end{cases}
\end{aligned}
\]
\end{enumerate}

The forecasts are defined analogously for other marginal specifications in Section~\ref{sec:marginal:specificification}.
The marginal VaR and ES forecasts depend only on
the marginal parameter draws, which are obtained
from their respective working posteriors.
The copula parameters are estimated conditional
on the rank-based pseudo-observations constructed
from these draws. Consequently, the copula
likelihood does not affect marginal VaR and ES
estimation, while marginal parameter uncertainty
enters the estimation of cross-asset dependence.

Let $\mathcal J_{\mathrm{ret}}\subseteq\{1,\ldots,N\}$ denote the
set of iteration indices retained after burn-in and thinning, and
let $K=|\mathcal J_{\mathrm{ret}}|$. For each asset $i=1,2$, the
one-step-ahead VaR and ES forecasts are
\begin{align*}
\widehat Q_{i,n+1}
&=\frac{1}{K}\sum_{j\in\mathcal J_{\mathrm{ret}}}Q_{i,n+1}^{[j]},
\qquad
\widehat e_{i,n+1}
=\frac{1}{K}\sum_{j\in\mathcal J_{\mathrm{ret}}}e_{i,n+1}^{[j]}.
\end{align*}

\paragraph{Joint VaR violation probability forecasting}

In addition to marginal VaR and ES forecasts,
we construct one-step-ahead joint VaR violation
probability forecast for~\eqref{eq:true:joint:prob} as
 \begin{align}
 \widehat p_{n+1}
 &=
 C(\alpha,\alpha;\widehat{\boldsymbol\psi}),
 \qquad \alpha=0.01,
 \label{eq:joint_prob:forecast}
 \end{align}
where $\widehat{\boldsymbol\psi}$ denotes the
posterior mean of the copula parameters obtained
from the current rolling estimation window,
with $\widehat{\boldsymbol\psi}=\widehat\theta$
for the Clayton copula and
$\widehat{\boldsymbol\psi}
=(\widehat\theta,\widehat\delta)^{\prime}$
for the BB1 copula.
The resulting probability quantifies the predicted
likelihood of simultaneous VaR violations
across the two assets and provides a basis for
evaluating joint tail-event forecasting performance.

\subsection{Forecast Evaluation}\label{sec:forecast:evaluation}

For each out-of-sample date $t=n+1,\ldots,n+T$, the forecasts use observations from $t-n$ through $t-1$.

\paragraph{Marginal and joint violation rates}

For VaR forecast accuracy, the violation rate (VRate) is commonly adopted to evaluate a single asset. For each $i$, VRate is defined as
\begin{eqnarray*}
\operatorname{VRate}_i=\frac1T\sum_{t=n+1}^{n+T}I_{i,t},\quad\text{where}\quad
I_{i,t}=\mathbb I(r_{i,t}\leq\widehat Q_{i,t}).
\end{eqnarray*}
Correct conditional quantiles imply the marginal violation probability
$\alpha$ for $i=1,2$. 

While the VRate assesses marginal VaR coverage, it does not describe
simultaneous tail events across the two assets. We therefore report the joint
violation rate
\begin{align*}
V_{12}=\frac1T\sum_{t=n+1}^{n+T}J_t,\quad\text{where}\quad   J_t=I_{1,t}I_{2,t}.
\end{align*}
This quantity measures the observed frequency of
simultaneous 1\% VaR violations for Bitcoin and Ethereum.
We compare the observed joint violation rate $V_{12}$
with the mean predicted probability
\begin{align}\label{eq:mean:predicted:prob}
\frac{1}{T}\sum_{t=n+1}^{n+T}\widehat p_t,
\end{align}
where each $\widehat p_t$ denotes the
joint VaR violation probability computed
using the posterior mean of the copula
parameters, as defined in
Equation~\eqref{eq:joint_prob:forecast}.

We assess marginal VaR calibration by comparing
the observed violation rates with the nominal
level $\alpha=0.01$. For simultaneous VaR violations,
we compare the observed joint violation rate
$V_{12}$ with the mean predicted probability~\eqref{eq:mean:predicted:prob}.
Close agreement between these quantities
indicates satisfactory average calibration
of the joint probability forecasts.
We provide more formal calibration tests in the Supplementary Materials.

\paragraph{Comparative forecast evaluation}
For an observed return $y$ and forecasts $e<q<0$,
recall that $\rho_\alpha(u)=u\{\alpha-\mathbb I(u<0)\}$.
We consider the following scoring functions:
\begin{align*}
 S_{\mathrm{QS}}(q;y)
 &=\rho_\alpha(y-q),\\
 S_{\mathrm{ALS}}(q,e;y)
 &=-\log\left(\frac{1-\alpha}{-e}\right)
     -\frac{\rho_\alpha(y-q)}{\alpha e},\\
 S_{\mathrm{FZ}}(q,e;y)
 &=\frac{\mathbb I(y\leq q)(y-q)}{\alpha e}
     +\frac{q}{e}+\log(-e)-1,\\
 S_{\mathrm{FZG}}(q,e;y)
 &=q\{\mathbb I(y\leq q)-\alpha\}
     -y\mathbb I(y\leq q)\\
 &\quad+\frac{\exp(e)}{1+\exp(e)}
      \left\{e-q+
      \frac{\mathbb I(y\leq q)(q-y)}{\alpha}\right\}
     -\log\{1+\exp(e)\}+\log 2.
\end{align*}

The quantile score (QS) evaluates marginal VaR forecasts
\citep{Giacomini2005}. 
The asymmetric Laplace log score (ALS) corresponds
to the marginal working likelihood and requires
the zero-conditional-mean qualification for joint
VaR--ES consistency \citep{Taylor2019JBES}.
The FZ and FZG scores
belong to the class of consistent joint VaR--ES
scoring functions \citep{FZ2016,FZG2016,PattonZiegel2019JE}.

To evaluate the predicted probabilities of simultaneous
VaR violations, we use the Bernoulli logarithmic score~\citep{gneiting2007strictly}:
\begin{align*}
S_{\mathrm{Bern}}(p,q_1,q_2;y_1,y_2)
 &=-\mathbb{I}(y_1\le q_1,\,y_2\le q_2)\log p-\left\{1-\mathbb{I}(y_1\le q_1,\,y_2\le q_2)\right\}\log(1-p).
 \end{align*}

For each score $\mathrm{m}\in\{\mathrm{QS},\mathrm{ALS},\mathrm{FZ},\mathrm{FZG}\}$,
we report the asset-specific average score
\begin{align*}
 \frac1T\sum_{t=n+1}^{n+T}
 S_{\mathrm{m}}(\widehat Q_{i,t},\widehat e_{i,t};r_{i,t}),\qquad i=1,2,
\end{align*}
with the ES argument omitted for QS. 
For the Bernoulli score, we also report the average score
\begin{align*}
 \frac1T\sum_{t=n+1}^{n+T}
  S_{\mathrm{Bern}}\!\left(
   \widehat p_t,
   \widehat Q_{1,t},
   \widehat Q_{2,t};
   r_{1,t},r_{2,t}
 \right).
\end{align*}
Lower average scores indicate better predictive
performance under the corresponding scoring rule.
These scores evaluate the out-of-sample forecasts
obtained from the sequential Bayesian framework.
More formal comparative backtests can be found in the Supplementary Materials.

\section{Simulation Study}\label{sec:simulation}

This study uses the AL density as a working likelihood rather than as a fully specified return distribution. Consequently, any simulation study requires an externally specified parametric data-generating process (DGP), and the resulting conclusions depend on that particular design. A related strategy is adopted by \citet{Gerlach2011}, who evaluate CAViaR estimation using data generated from a parametric GARCH model.

With this background in mind, we conduct a simulation study to examine whether the proposed copula-based
ES-CAViaR framework can recover one-step-ahead tail-risk measures and the joint shortfall probability when the
true DGP differs from the fitted model. We generate
bivariate returns from the DCC-GARCH-$t$ model. For Asset $i=1,2$, the return
process for DCC-GARCH-$t$ is
\begin{eqnarray}
\nonumber \mathbf{r}_t
&=&
\boldsymbol{\mu}+\mathbf{a}_t,
\quad
\mathbf{a}_t=D_t\boldsymbol{\varepsilon}_t,\\
\label{eq:GARCH:t:innovation}  \boldsymbol\varepsilon_t
 &=&s_\nu\mathbf X_t,
 \quad \mathbf X_t\mid\mathcal F_{t-1}
 \sim t_{2,\nu}(\mathbf 0,\mathbf R_t),
 \quad s_\nu=\sqrt{\frac{\nu-2}{\nu}},
 \quad \nu>2,\\
\nonumber
h_{ii,t}
&=&
\omega_i+\alpha_i a_{i,t-1}^2+\beta_i h_{ii,t-1},
\quad i=1,2,
\nonumber\\
\mathbf{R}_t
&=&
(1-\kappa_1-\kappa_2)\bar{\mathbf{R}}
+\kappa_1{\mathbf{R}}_{t-1,M}
+\kappa_2\mathbf{R}_{t-1},
\nonumber\\
\nonumber 
\boldsymbol{\mu}&=&(\mu_1,\mu_2)', \quad
D_t
=
\operatorname{diag}
\left(\sqrt{h_{11,t}},\sqrt{h_{22,t}}\right),
\quad
\bar{\mathbf{R}}=
\begin{pmatrix}
1 & \rho\\
\rho & 1
\end{pmatrix},
\end{eqnarray}
where $\mathbf{R}_t$ denotes the time-varying conditional correlation matrix, $\bar{\mathbf{R}}$ denotes its unconditional correlation target, and ${\mathbf{R}}_{t-1,M}$ denotes the local correlation matrix computed from the most recent $M=4$ standardized innovations.
In~\eqref{eq:GARCH:t:innovation}, we denote by $t_{2,\nu}(\mathbf 0,\mathbf R_t)$ the bivariate
Student-$t$ distribution with $\nu$ degrees of freedom, location zero,
and scale matrix $\mathbf R_t$. 
The true parameter values used in the simulation are in Table~\ref{table:sim:true:params}.

\begin{table}[t]
\caption{The true parameter values of the DCC-GARCH-$t$ model in the simulation study.}
\vspace{-4mm}
\begin{center}
\small
\setlength{\tabcolsep}{4.2pt}
\renewcommand{\arraystretch}{0.90}
\begin{tabular}{cccccccccc}
\toprule
 $\mu_1$ & $\mu_2$
& $\omega_1$ & $\alpha_1$ & $\beta_1$
& $\omega_2$ & $\alpha_2$ & $\beta_2$
& $(\kappa_1,\kappa_2,\rho)$
& $\nu$ \\
\midrule
 0.08 & 0.02
& 0.25 & 0.07 & 0.80
& 0.15 & 0.02 & 0.70
& $(0.05,0.60,0.65)$
& 7 \\

\bottomrule
\end{tabular}
\end{center}
\label{table:sim:true:params}
\end{table}

Let $T_\nu$ denote the cdf of the univariate Student-$t$ distribution
with $\nu$ degrees of freedom, location zero, and unit scale, and let
$T_{2,\nu}(x_1,x_2;\rho)$ denote the cdf of the corresponding bivariate
Student-$t$ distribution with scale matrix 
\begin{align*}
\begin{pmatrix}
1 & \rho \\
\rho & 1
\end{pmatrix},
\qquad \rho\in(-1,1).
\end{align*}
Then the marginal conditional
VaR for DCC-GARCH-$t$ specification is $Q_{i,t}^{\mathrm{DCC}\text{-}t}
 =\mu_i+\sqrt{h_{ii,t}}\,s_\nu\, T_\nu^{-1}(\alpha)$, $i=1,2$.
Hence, the true joint shortfall probability~\eqref{eq:true:joint:prob} becomes
\begin{align}
 p_t^{\mathrm{DCC}\text{-}t}
 :=
 \mathbb{P}\left(
r_{1,t}\leq Q_{1,t}^{\mathrm{DCC}\text{-}t},
\,r_{2,t}\leq Q_{2,t}^{\mathrm{DCC}\text{-}t}
\mid\mathcal F_{t-1}\right)=
C^{t}(\alpha,\alpha;\nu,\rho_t),
 \label{eq:joint:prob:t}
\end{align}
where 
\begin{align*}
C^{t}(u_1,u_2;\nu,\rho_t)=T_{2,\nu}\!\left(
 T_\nu^{-1}(u_1),T_\nu^{-1}(u_2);\rho_t\right)
 \end{align*}
 is the Student-$t$ copula with correlation parameter $\rho_t\in(-1,1)$ and the degrees of freedom $\nu>0$.

We use 1,000 observations for estimation, which is of a similar magnitude
to the observation rolling window used in the empirical application.
For each simulated sample, we calculate the true one-step-ahead 1\% VaR and ES as well as the joint shortfall probability directly
from the DGP.
We then compare them with the corresponding forecasts from the ES-CAViaR-st with Clayton-copula specification.
We repeat the experiment over $N_s=200$ replications.

Because the structural parameters of the DCC-GARCH-\(t\) model do not have direct counterparts in the copula-based ES-CAViaR-st specification, a direct comparison of parameter estimates is not appropriate. Instead, we evaluate the accuracy of the estimated VaR, ES, and joint VaR violation probability relative to their replication-specific true values.
For replication $k=1,\ldots,N_s$, let $Q_{i,k}^0$ and $e_{i,k}^0$
denote the true one-step-ahead VaR and ES for asset $i=1,2$, and
let $p_k^0$ denote the true joint shortfall probability. 
We suppress the common forecast date $n+1$. The corresponding forecasts are
$\widehat Q_{i,k}$, $\widehat e_{i,k}$, $i=1,2$, and $\widehat p_k$.
For each of the five targets, write $x_k^0$ and $\widehat x_k$ for
its true value and forecast in replication $k$, respectively.
We report the bias, mean absolute error (MAE),
root mean squared error (RMSE), and mean absolute relative error (MARE) defined as follows:
\begin{align*}
\operatorname{Bias}
&=
\frac{1}{N_s}\sum_{k=1}^{N_s}
(\widehat{x}_k-x_k^0),
\quad
&\operatorname{MAE}
&=
\frac{1}{N_s}\sum_{k=1}^{N_s}
|\widehat{x}_k-x_k^0|,\\
\operatorname{RMSE}
&=
\left[
\frac{1}{N_s}\sum_{k=1}^{N_s}
(\widehat{x}_k-x_k^0)^2
\right]^{1/2},
\quad
&\operatorname{MARE}
&=
\frac{1}{N_s}\sum_{k=1}^{N_s}
\left|
\frac{\widehat{x}_k-x_k^0}{x_k^0}
\right| \times 100.
\end{align*}
Note that MARE measures the average absolute forecast error relative to the
replication-specific true quantity, expressed as a percentage.

\begin{table}[t]
\centering
\begin{threeparttable}
\caption{\small Simulation results for the Clayton copula-based
ES-CAViaR-st model under the DCC-GARCH-$t$
data-generating process (200 replications).}
\vspace{-0.3cm}
\label{tab:simulation}
\renewcommand{\arraystretch}{0.70}
\setlength{\tabcolsep}{4pt}
\begin{tabular*}{\linewidth}{@{\extracolsep{\fill}}lrrrr@{}}
\toprule
Quantity & Bias & MAE & RMSE & MARE (\%) \\
\midrule
Asset 1 VaR
& -0.022 & 0.340 & 0.467 & 9.759 \\

Asset 1 ES
& 0.280 & 0.513 & 0.653 & 11.751 \\

Asset 2 VaR
& 0.009 & 0.160 & 0.205 & 8.674 \\

Asset 2 ES
& 0.106 & 0.254 & 0.315 & 10.994 \\

Joint probability
& $0.729\mathrm{E}{-3}$
& $1.125\mathrm{E}{-3}$
& $1.300\mathrm{E}{-3}$
& 35.947 \\
\bottomrule
\end{tabular*}
\begin{tablenotes}
\footnotesize
\item Note:
Bias is defined as the forecast minus the replication-specific true value.
\end{tablenotes}
\end{threeparttable}
\end{table}

The results are summarized in Table~\ref{tab:simulation}, indicating that the fitted
copula-based ES-CAViaR-st model performs reasonably well under the 
DGP. 
Since bias is defined as the fitted forecast minus the
replication-specific true value, the VaR biases of $-0.022$ and $0.009$
indicate slightly more negative forecasts for Asset~1 and slightly more positive forecasts for Asset~2, respectively.
The corresponding VaR MAREs are $9.759\%$ and $8.674\%$.
In contrast, the ES biases are positive for both assets, at $0.280$ and
$0.106$, indicating that the model produces more positive ES forecasts
and therefore underestimates the magnitude of expected tail losses.
The ES MAREs of $11.751\%$ and $10.994\%$ are also slightly larger than
those of the corresponding VaR forecasts.

For the joint shortfall probability, the positive bias of
$0.729\times10^{-3}$ indicates average overestimation relative to the
DGP-implied probabilities.
The larger MARE of $35.947\%$ for the joint probability
forecast suggests that relatively accurate marginal VaR
and ES forecasts do not necessarily ensure equally
accurate joint tail-risk forecasts under model misspecification.

\section{Empirical Study}\label{sec:empirical:study}

\subsection{Data, Models and Bayesian Estimation}

\paragraph{Data description}
To examine joint tail risk in cryptocurrency markets,
we obtain daily Bitcoin and Ethereum price data from
Yahoo Finance using the \textit{quantmod} package
in \textsf{R}. The corresponding tickers are
\texttt{BTC-USD} and \texttt{ETH-USD}.
Both cryptocurrencies trade continuously,
including weekends and holidays.
Yahoo Finance records daily cryptocurrency prices
using Coordinated Universal Time (UTC), with each
daily observation covering the interval from
00:00 to 24:00 UTC. We align the two series by
their daily observation dates and retain observations
available for both assets. The downloaded dataset
contains no missing daily observations, so no
imputation is necessary. We use unadjusted OHLC
prices, including the closing price (\texttt{Close}),
rather than adjusted prices (\texttt{Adjusted}).
Unlike equities, cryptocurrencies do not undergo
conventional corporate actions such as stock splits
or dividends that require price adjustments.

The sample spans January 2018 to July 2025 and
contains 2{,}769 daily observations for the
bivariate Bitcoin--Ethereum series. We divide
the sample into an initial estimation period
from January 2018 to December 2019, containing
730 observations, and an out-of-sample period
from January 2020 to July 2025, containing
2{,}039 observations. We use a rolling estimation
window of 730 observations to generate
one-step-ahead forecasts throughout the
out-of-sample period.

We select this sample period to capture distinct phases of cryptocurrency
market development, including the rapid expansion of trading activity
following 2017 and subsequent episodes of heightened volatility and
market stress. The out-of-sample period includes the COVID-19-induced
market turmoil in March 2020, which triggered sharp price declines
across traditional financial and cryptocurrency markets, followed by
the Terra-Luna collapse in May 2022 and the FTX collapse in November
2022. These episodes provide a relevant setting for assessing the
forecasting performance of competing VaR and ES models under
changing market conditions.
By evaluating forecasts over this period, we assess the ability of the proposed models to capture joint tail risk under prolonged and severe market stress.

\begin{figure}[!th]
	\centering
	\includegraphics[width=1\textwidth]{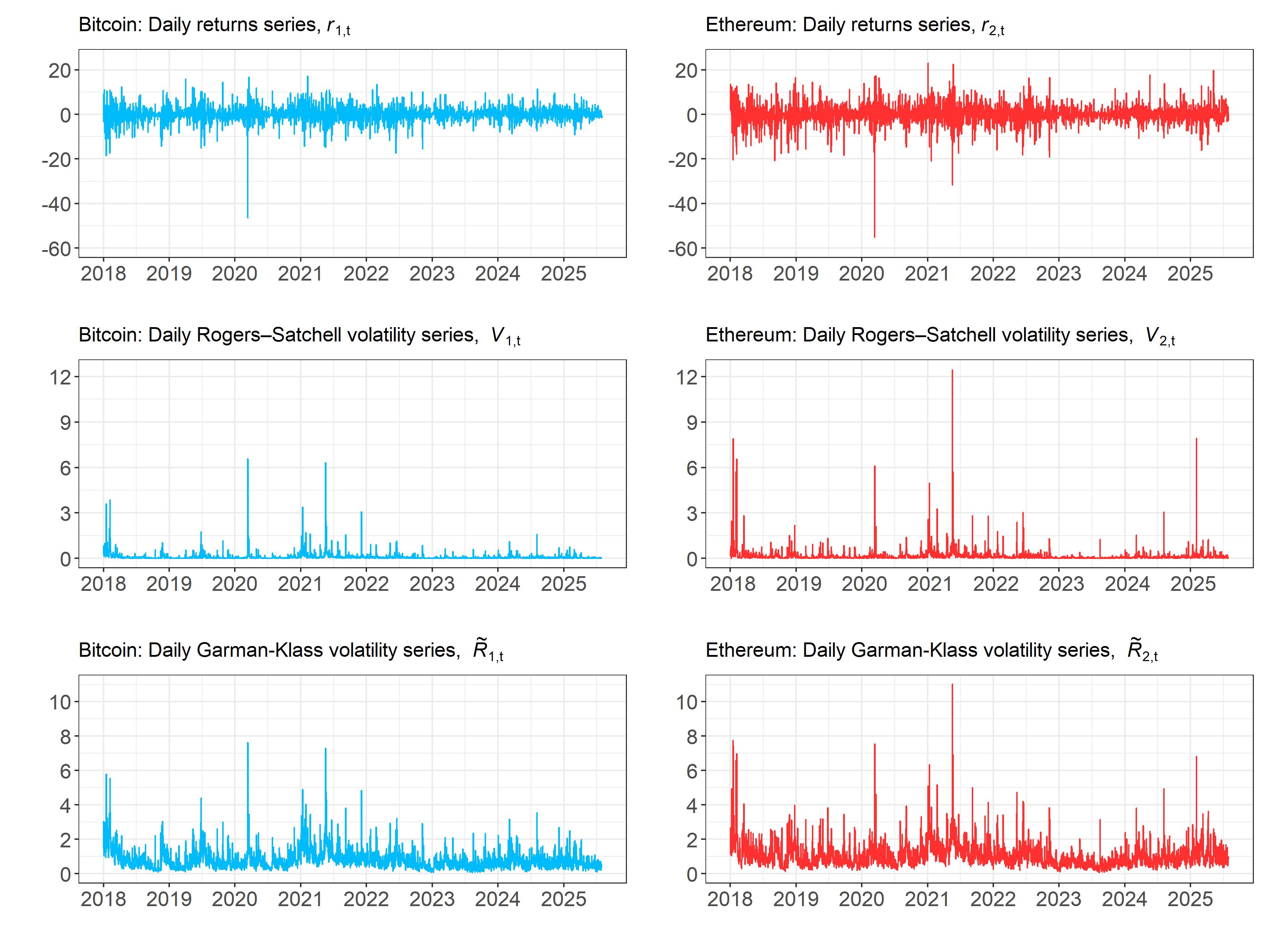} 
	
   \caption{Time-series plots of daily returns, Rogers-Satchell volatility, and Garman-Klass volatility for Bitcoin and Ethereum.}
\label{fig:BTC_ETH_matrix}   
\end{figure}

From Figure~\ref{fig:BTC_ETH_matrix}, which plots daily returns, Rogers-Satchell volatility, and Garman-Klass volatility for Bitcoin and Ethereum, we observe notable volatility clustering and periods of elevated volatility in both cryptocurrencies. Table~\ref{tab:Summary_BTC_ETH} reports the summary statistics for Bitcoin and Ethereum, including the daily return series $r_{i,t}$, the Rogers-Satchell volatility series $V_{i,t}$ defined in equation~\eqref{eq:ES-CAViaR-v}, and the Garman-Klass volatility series $\tilde{R}_{i,t}$ defined in equation~\eqref{eq:ES-CAViaR-R}.
Both Bitcoin and Ethereum returns exhibit negative skewness and leptokurtosis, indicating substantial departures from normality and motivating semiparametric CAViaR-type models. The Rogers-Satchell volatility $V_{i,t}$ and Garman-Klass volatility $\tilde{R}_{i,t}$ also show substantial variation, reflecting the high volatility of cryptocurrency markets.

\begin{figure}[!th]
	\centering
	\includegraphics[width=0.8\textwidth]{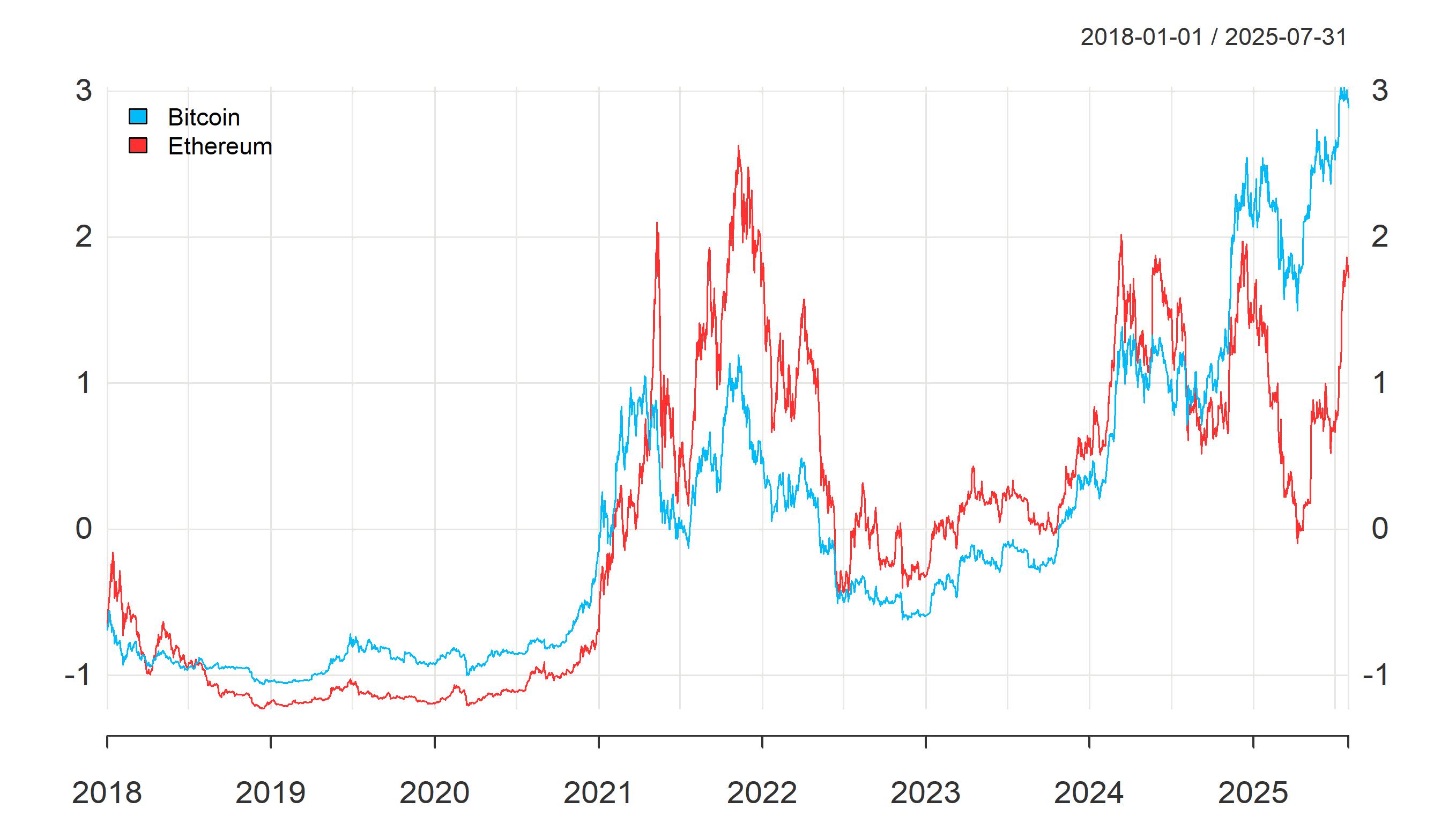} 
    \caption{Standardized price time series of Bitcoin and Ethereum.}\label{fig:BTC_ETH_SP}    
\end{figure}

Figure \ref{fig:BTC_ETH_SP} shows the standardized price time series of Bitcoin and Ethereum from January 2018 to July 2025. The two assets exhibit strong co-movement during major market episodes, while also displaying noticeable differences in the magnitude and timing of extreme price movements, particularly during periods of heightened volatility.
These features suggest that the dependence structure may be time-varying and asymmetric, especially within the tails. Such observations motivate the subsequent use of copula-based models, which can flexibly capture cross-asset dependence beyond linear correlation.

\begin{figure}[!th]
	\centering
\includegraphics[width=0.8\textwidth]{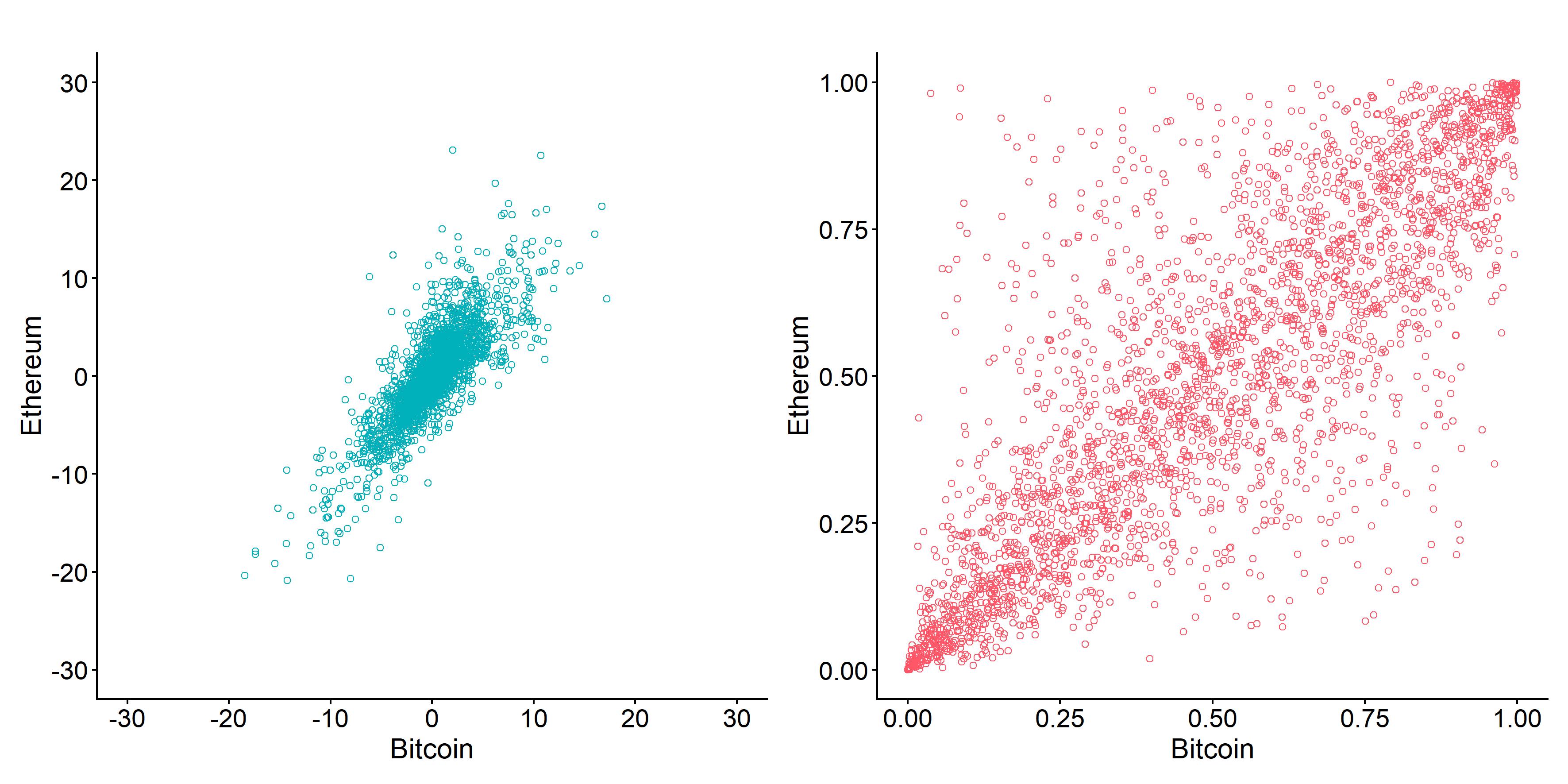} 
\caption{Scatter plots of Bitcoin and Ethereum
daily returns (left) and their empirical
rank transformations on the unit square (right).}
\label{fig:BTC_ETH_RT}
\end{figure}

\begin{table}[!th]
\centering
\caption{Summary statistics for Bitcoin and Ethereum returns and volatility measures.}
\label{tab:Summary_BTC_ETH}
\vspace{-0.4cm}
\renewcommand{\arraystretch}{0.70}
\setlength{\tabcolsep}{4pt}
\makebox[\textwidth][c]{%
\scalebox{0.92}{%
\begin{tabular}{llcrrrrrrr}
\midrule
Asset
& \multicolumn{1}{c}{Period}
& \multicolumn{1}{c}{Series}
& \multicolumn{1}{c}{\shortstack{Sample\\Size}}
& \multicolumn{1}{c}{Min.}
& \multicolumn{1}{c}{Max.}
& \multicolumn{1}{c}{Mean}
& \multicolumn{1}{c}{Std.}
& \multicolumn{1}{c}{Skewness}
& \multicolumn{1}{c}{\shortstack{Excess\\Kurtosis}} \\
\midrule

\multicolumn{10}{l}{Asset 1: Bitcoin} \\

& Full
& $r_{1,t}$ & 2769 & -46.4730 & 17.1821 & 0.0759 & 3.4874 & -0.96 & 14.30 \\

&
& $V_{1,t}$ & 2769 & 0.0000 & 6.5880 & 0.1073 & 0.2861 &  &  \\

&
& $\tilde{R}_{1,t}$ & 2769 & {0.0735} & {7.6206} & 0.8266 & {0.6211} &  &  \\

& In-sample
& $r_{1,t}$ & 730 & -18.4582 & 16.0042 & -0.0927 & 3.9362 & -0.25 & 3.03 \\

&
& $V_{1,t}$ & 730 & 0.0000 & 3.8708 & 0.1219 & 0.2963 &  &  \\

&
& $\tilde{R}_{1,t}$ & 730 & {0.0968} & {5.7895} & {0.8672} & {0.6902} &  &  \\

& Out-of-sample
& $r_{1,t}$ & 2039 & -46.4730 & 17.1821 & 0.1363 & 3.3110 & -1.36 & 21.66 \\

&
& $V_{1,t}$ & 2039 & 0.0000 & 6.5880 & 0.1021 & 0.2823 &  &  \\

&
& $\tilde{R}_{1,t}$ & 2039 & {0.0735} & {7.6206} & {0.8120} & {0.5940} &  &  \\

\midrule

\multicolumn{10}{l}{Asset 2: Ethereum} \\

& Full
& $r_{2,t}$ & 2769 & -55.0732 & 23.0695 & 0.0573 & 4.5387 & -0.89 & 11.11 \\

&
& $V_{2,t}$ & 2769 & 0.0000 & 12.4483 & 0.1812 & 0.5082 &  &  \\

&
& $\tilde{R}_{2,t}$ & 2769 & {0.0774} & {11.0103} & {1.0860} & {0.7882} &  &  \\

& In-sample
& $r_{2,t}$ & 730 & -20.6860 & 16.6150 & -0.2417 & 4.9587 & -0.39 & 2.25 \\

&
& $V_{2,t}$ & 730 & 0.0000 & 7.8950 & 0.2032 & 0.5642 &  &  \\

&
& $\tilde{R}_{2,t}$ & 730 & {0.1249} & {7.7490} & {1.1382} & {0.8405} &  &  \\

& Out-of-sample
& $r_{2,t}$ & 2039 & -55.0732 & 23.0695 & 0.1643 & 4.3749 & -1.12 & 16.14 \\

&
& $V_{2,t}$ & 2039 & 0.0000 & 12.4483 & 0.1733 & 0.4864 &  &  \\

&
& $\tilde{R}_{2,t}$ & 2039 & {0.0774} & {11.0103} & {1.0673} & {0.7679} &  &  \\
\midrule
\end{tabular}
}}
\begin{tablenotes}
\small
\item \textit{Notes:}
\(r_{i,t}
=
[\log(P_{i,t})-\log(P_{i,t-1})]\times100\),
where \(P_{i,t}\) denotes the closing price at time \(t\).
The Rogers-Satchell volatility, \(V_{i,t}\), is transformed and rescaled as
\(\log(V_{i,t}+1)\times100\).
The Garman-Klass volatility \(\tilde{R}_{i,t}\) is transformed as
\(\sqrt{\tilde{R}_{i,t}\times1000}\).
\end{tablenotes}
\end{table}

Figure~\ref{fig:BTC_ETH_RT} presents scatter plots
of Bitcoin and Ethereum daily returns and their
empirical rank transformations on the unit square.
The raw returns exhibit strong positive association,
while the empirical ranks provide a clearer view
of their dependence pattern. The concentration
of observations near the lower-left corner suggests
joint lower-tail association, motivating the
investigation of tail dependence through the
copula-based framework in the empirical analysis.

\paragraph{Competing models}
Within the proposed copula-based ES-CAViaR framework,
we model tail dependence between Bitcoin and Ethereum
using the Clayton and BB1 copulas. Each copula is
combined with four marginal specifications:
(i) the ES-CAViaR model of \citet{Taylor2019JBES};
(ii) the ES-CAViaR-st model, where the exceedance
term is normalized by conditional volatility;
(iii) the ES-CAViaR-v model, which incorporates the
Rogers--Satchell volatility measure; and
(iv) the ES-CAViaR-R model, which incorporates the
Garman--Klass volatility measure.

For comparison, we include two external parametric
multivariate benchmarks, DCC-GARCH-$t$ as in Section~\ref{sec:simulation} and DCC-GARCH-N.
Importantly, we conduct both estimation and forecasting
for all competing specifications within a Bayesian
framework, including the two DCC-GARCH benchmarks.
We obtain posterior parameter estimates using MCMC
methods and construct the corresponding one-step-ahead
VaR, ES, and joint shortfall probability forecasts
under a common rolling-window design and
out-of-sample evaluation period. Thus, the comparison
maintains a consistent Bayesian forecasting framework
across the proposed models and external benchmarks.

The DCC-GARCH-N benchmark retains the conditional-variance
and correlation dynamics of DCC-GARCH-$t$ but replaces
the innovation specification in~\eqref{eq:GARCH:t:innovation}
with
\[
\boldsymbol{\varepsilon}_t\mid\mathcal F_{t-1}
\sim \operatorname{N}_2(\mathbf{0},\mathbf{R}_t).
\]
Under this Gaussian specification, the marginal
conditional VaR is
\[
Q_{i,t}^{\mathrm{DCC-N}}
=
\mu_i+\sqrt{h_{ii,t}}\,\Phi^{-1}(\alpha),
\quad i=1,2,
\]
and the joint shortfall probability
in~\eqref{eq:true:joint:prob} becomes
\begin{equation}
\label{eq:joint:prob:gaussian}
p_t^{\mathrm{DCC-N}}
=
C^{\mathrm{Ga}}(\alpha,\alpha;\rho_t)
=
\Phi_2\left(
\Phi^{-1}(\alpha),
\Phi^{-1}(\alpha);\rho_t
\right),
\end{equation}
where $\rho_t=(\mathbf R_t)_{12}$,
$C^{\mathrm{Ga}}$ denotes the Gaussian copula,
$\Phi$ is the standard normal cdf, and
$\Phi_2(\cdot,\cdot;\rho)$ is the bivariate
standard normal cdf with correlation $\rho$.

For both DCC-GARCH benchmarks, the Bayesian
posterior estimates determine the conditional
variance and correlation forecasts used in the
corresponding model-implied VaR, ES, and joint
exceedance probability expressions.
We construct the joint probability forecasts
using~\eqref{eq:joint:prob:t}
and~\eqref{eq:joint:prob:gaussian},
respectively.

\paragraph{Bayesian estimation}

We generate $20,000$ MCMC iterations, discard the first
$8,000$ as burn-in, and retain every fourth draw
thereafter, yielding 3{,}000 posterior draws.
For the Clayton copula, we employ a Gaussian random-walk MH (RWMH) algorithm on Kendall's tau
$\tau=\theta/(\theta+2)$, with the proposal scale
adaptively tuned during burn-in and fixed thereafter.
Proposed points outside the range $(0,1)$
are rejected.

For the marginal parameter vectors
$\boldsymbol{\beta}_i$ and $\boldsymbol{\gamma}_i$,
$i=1,2$, we employ an adaptive Gaussian RWMH algorithm during burn-in.
After burn-in, we use a mixture of independence-kernel
MH (IKMH) and RWMH
updates with probabilities 0.8 and 0.2,
respectively.
The independent-kernel component employs a Gaussian proposal
whose mean vector and covariance matrix are estimated
from the burn-in draws to account for posterior
dependence among the parameters within each block.

We apply the same mixture sampling scheme to the
BB1 copula, jointly updating the dependence
parameters on the transformed scale
$\boldsymbol{\phi}
=(\log\theta,\log(\delta-1))^{\prime}$.
The independence proposals use the burn-in draws
to estimate their mean vectors and covariance
matrices, while the random-walk proposal covariance
matrices remain fixed after burn-in.
Further implementation details are provided in
the Supplementary Materials.

Convergence and mixing are assessed using trace plots and autocorrelation
functions. The corresponding diagnostic plots, reported in the Supplementary
Materials, show stable chains with satisfactory mixing and no apparent trends.

\begin{table}[t]
\centering
\captionsetup{font={small,stretch=0.9},skip=5pt}
\caption{
Pseudo-posterior summaries of ES-CAViaR-v marginal
and copula parameters under Clayton and BB1 copulas.}
\label{tab:ES-CAViaR-v-pa}
{\renewcommand{\arraystretch}{0.75}
\setlength{\tabcolsep}{7pt}
\scalebox{0.9}{%
\begin{tabular}{llrrrrrr}
\toprule
 & & \multicolumn{3}{c}{Clayton}
     & \multicolumn{3}{c}{BB1} \\
\cmidrule(lr){3-5} \cmidrule(lr){6-8}
Return series & Parameter
& Mean & 2.5\% & 97.5\%
& Mean & 2.5\% & 97.5\% \\
\midrule
Bitcoin
& $\beta_{1,1}$  & -0.3814 & -0.5834 & -0.2338
                  & -0.3850 & -0.5957 & -0.2377 \\
& $\beta_{1,2}$  & 0.9266 & 0.8953 & 0.9491
                  & 0.9259 & 0.8937 & 0.9479 \\
& $\beta_{1,3}$  & -1.5639 & -1.9405 & -1.1490
                  & -1.5756 & -1.9557 & -1.1789 \\
& $\gamma_{1,1}$ & 0.2065 & 0.0147 & 0.4913
                  & 0.2023 & 0.0120 & 0.4847 \\
& $\gamma_{1,2}$ & 0.1664 & 0.0051 & 0.4499
                  & 0.1575 & 0.0047 & 0.4269 \\
& $\gamma_{1,3}$ & 0.1744 & 0.0054 & 0.4923
                  & 0.1671 & 0.0062 & 0.4652 \\
Ethereum
& $\beta_{2,1}$  & -0.0789 & -0.1070 & -0.0560
                  & -0.0794 & -0.1092 & -0.0565 \\
& $\beta_{2,2}$  & 0.9797 & 0.9747 & 0.9839
                  & 0.9797 & 0.9745 & 0.9839 \\
& $\beta_{2,3}$  & -0.8578 & -1.0379 & -0.6813
                  & -0.8575 & -1.0285 & -0.6781 \\
& $\gamma_{2,1}$ & 0.2567 & 0.0203 & 0.5769
                  & 0.2532 & 0.0231 & 0.5610 \\
& $\gamma_{2,2}$ & 0.1594 & 0.0067 & 0.4473
                  & 0.1620 & 0.0080 & 0.4380 \\
& $\gamma_{2,3}$ & 0.1940 & 0.0069 & 0.5156
                  & 0.1947 & 0.0106 & 0.4876 \\
\midrule
Copula
& $\theta$      & 1.5479 & 0.9777 & 2.1456
                 & 1.2353 & 0.7854 & 1.7425 \\
& $\delta$      & -- & -- & --
                 & 1.1427 & 1.0275 & 1.3633 \\
\bottomrule
\end{tabular}}}
\end{table}

\begin{table}[t]
\captionsetup{font={small,stretch=0.9},skip=5pt}
\caption{Pseudo-posterior summaries of ES-CAViaR-R marginal
and copula parameters under Clayton and BB1 copulas.}
\label{tab:ES-CAViaR-R-pa}
\centering
{\renewcommand{\arraystretch}{0.75}
\setlength{\tabcolsep}{7pt}
\scalebox{0.9}{%
\begin{tabular}{llrrrrrr}
\toprule
 & & \multicolumn{3}{c}{Clayton}
     & \multicolumn{3}{c}{BB1} \\
\cmidrule(lr){3-5} \cmidrule(lr){6-8}
Return series & Parameter
& Mean & 2.5\% & 97.5\%
& Mean & 2.5\% & 97.5\% \\
\midrule
Bitcoin
& $\beta_{1,1}$  & -0.6342 & -0.8931 & -0.3930
                  & -0.6399 & -0.8945 & -0.4080 \\
& $\beta_{1,2}$  & 0.8055 & 0.7551 & 0.8569
                  & 0.8044 & 0.7534 & 0.8532 \\
& $\beta_{1,3}$  & -0.8664 & -1.0797 & -0.6805
                  & -0.8678 & -1.0899 & -0.6895 \\
& $\gamma_{1,1}$ & 0.1622 & 0.0084 & 0.4291
                  & 0.1603 & 0.0077 & 0.3982 \\
& $\gamma_{1,2}$ & 0.1773 & 0.0059 & 0.4647
                  & 0.1787 & 0.0089 & 0.4750 \\
& $\gamma_{1,3}$ & 0.1652 & 0.0079 & 0.4419
                  & 0.1700 & 0.0087 & 0.4294 \\
Ethereum
& $\beta_{2,1}$  & -0.0043 & -0.0172 & -0.0001
                  & -0.0039 & -0.0151 & -0.0001 \\
& $\beta_{2,2}$  & 0.9728 & 0.9642 & 0.9794
                  & 0.9729 & 0.9646 & 0.9796 \\
& $\beta_{2,3}$  & -0.2531 & -0.3312 & -0.1886
                  & -0.2523 & -0.3308 & -0.1872 \\
& $\gamma_{2,1}$ & 0.2955 & 0.0274 & 0.6199
                  & 0.2835 & 0.0236 & 0.5876 \\
& $\gamma_{2,2}$ & 0.1766 & 0.0087 & 0.4746
                  & 0.1787 & 0.0092 & 0.4814 \\
& $\gamma_{2,3}$ & 0.2133 & 0.0097 & 0.5264
                  & 0.2096 & 0.0103 & 0.5431 \\
\midrule
Copula
& $\theta$      & 2.1804 & 1.4575 & 2.9516
                 & 1.3865 & 0.8343 & 2.0588 \\
& $\delta$      & -- & -- & --
                 & 1.3455 & 1.1123 & 1.6818 \\
\bottomrule
\end{tabular}
}}
\end{table}

\subsection{Results and Discussion}\label{sec:results:discussion}

\paragraph{Posterior summaries}

The pseudo-posterior parameter summaries for the ES-CAViaR-v
and ES-CAViaR-R models are reported in
Tables~\ref{tab:ES-CAViaR-v-pa} and
\ref{tab:ES-CAViaR-R-pa}, respectively, while those
for the remaining specifications are provided in the
Supplementary Materials. Both models exhibit persistent
VaR dynamics, as reflected by the relatively large
posterior estimates of $\beta_{i,2}$.
Under the Clayton copula, the posterior means of $\theta$
are 1.5479 and 2.1804 for ES-CAViaR-v and ES-CAViaR-R,
respectively, indicating stronger implied lower-tail
dependence for the latter specification.
Under the BB1 copula, the corresponding posterior
means are
$(\theta,\delta)^{\prime}=(1.2353,1.1427)^{\prime}$
and $(1.3865,1.3455)^{\prime}$, indicating variation
in the estimated dependence structure across
marginal specifications.
The posterior intervals of the copula parameters
account for uncertainty in the marginal parameters,
because the ranks of the standardized residuals
$z_{i,t}$ are recomputed at each MCMC draw.
The BB1 copula reduces to the Clayton copula when
$\delta=1$. The posterior mean of \(\delta\) is higher under ES-CAViaR-R than under ES-CAViaR-v, indicating a greater departure from the Clayton copula.

\paragraph{Marginal performance evaluation}

\begin{table}[t]
\centering
{\fontsize{9pt}{10pt}\selectfont
\renewcommand{\arraystretch}{0.9}
\setlength{\tabcolsep}{10pt}
\caption{{Marginal 1\% VaR violation rates for Bitcoin and Ethereum, 2020--2025.}}
\label{tab:violation}
\small
\begin{tabular}{llrrrr}
\toprule
& & \multicolumn{2}{c}{Bitcoin}
    & \multicolumn{2}{c}{Ethereum}\\
\cmidrule(lr){3-4}
\cmidrule(lr){5-6}
Copula & Marginal model
& Viol. & VRate (\%)
& Viol. & VRate (\%)\\
\midrule
Clayton & ES-CAViaR    & 19 & {0.932} & 23 & 1.128\\
        & ES-CAViaR-st & 19 & {0.932} & 23 & 1.128\\
        & ES-CAViaR-v  & 18 & 0.883 & 22 & 1.079\\
        & ES-CAViaR-R  & 18 & 0.883 & 19 & {0.932}\\
\addlinespace
BB1
& ES-CAViaR
& 19 & 0.932 & 23 & 1.128 \\
& ES-CAViaR-st
& 19 & 0.932 & 23 & 1.128 \\
& ES-CAViaR-v
& 18 & 0.883 & 22 & 1.079 \\
& ES-CAViaR-R
& 18 & 0.883 & 19  & 0.932 \\
\midrule
\multicolumn{2}{l}{DCC-GARCH-N} & 34	& 1.667  & {35} & {1.717} \\
\multicolumn{2}{l}{DCC-GARCH-$t$}
& 25 & 1.226 & 28 & 1.373 \\
\bottomrule
\end{tabular}
}
\vspace{0.5ex}
\begin{minipage}{0.95\linewidth}
\footnotesize
\textit{Note:} The out-of-sample period contains $T=2039$ days, so the
expected number of violations at the nominal 1\% level is 20.4.
Viol.\ denotes the number of VaR violations, and VRate denotes the
corresponding violation rate. 
\end{minipage}
\end{table}

Table~\ref{tab:violation} reports the marginal 1\%
VaR violation rates for Bitcoin and Ethereum.
The copula-based ES-CAViaR specifications produce
rates of 0.883\%--0.932\% for Bitcoin and
0.932\% and 1.128\% for Ethereum.
The Bitcoin rates are all below the nominal level,
whereas the Ethereum rates lie on both sides of it.
For Bitcoin, the baseline and ES-CAViaR-st
specifications under both copulas yield the rates
closest to 1\%, at 0.932\%.
For Ethereum, two ES-CAViaR-R specifications are closest to
the nominal level with a violation rate of 0.932\%.
The corresponding rates are 1.667\% and 1.717\% for DCC-GARCH-N and 1.226\% and 1.373\%
for DCC-GARCH-$t$.
All proposed specifications therefore yield
empirical violation rates closer to the nominal
level than either benchmark.

\begin{figure}[t]
    \centering
    \includegraphics[width=1.0\textwidth]
    {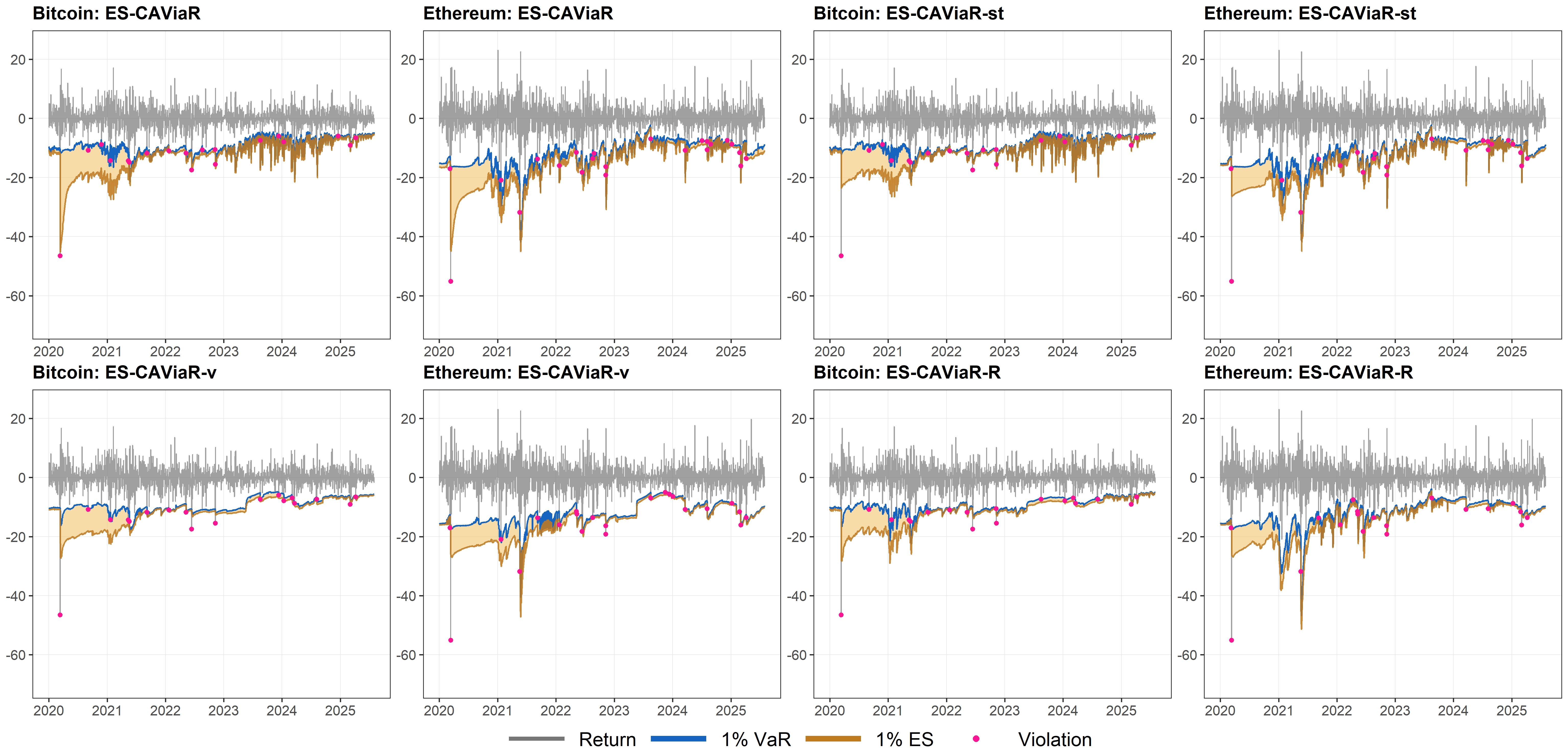}
\vspace{-0.5cm}
    \caption{\small
    Daily returns and one-day-ahead 1\% VaR and ES forecasts
    for Bitcoin and Ethereum under four ES-CAViaR models
    linked by a Clayton copula.
    Shaded areas represent VaR-ES gaps, and dots indicate
    VaR violations.
    }
\label{fig:VaR_ES_Clayton_all}
\end{figure}

\begin{figure}[h!]
    \centering
    \includegraphics[width=1.0\textwidth]
    {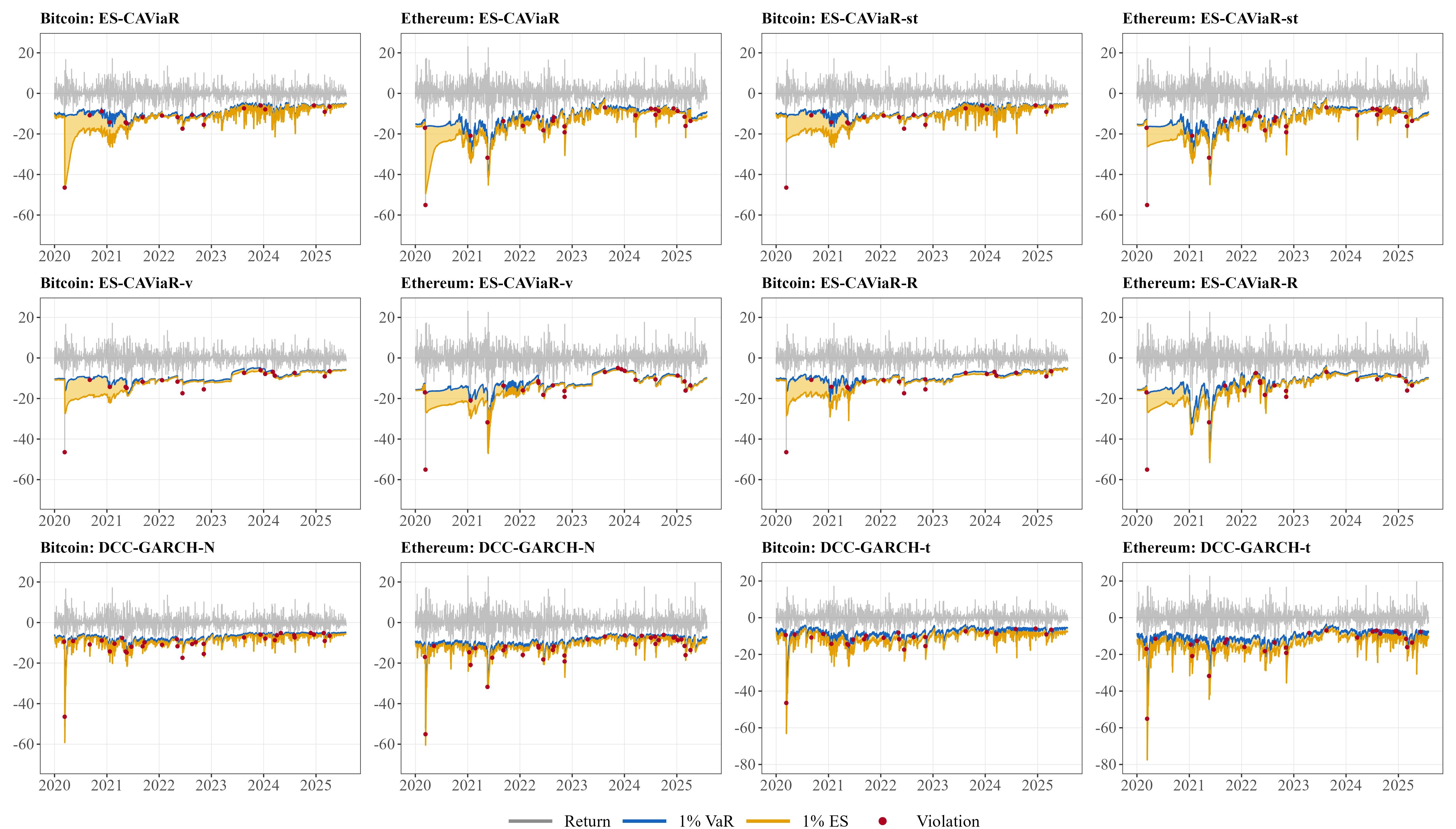}
\vspace{-0.5cm}   
\caption{\small  
One-day-ahead 1\% VaR and ES forecasts for Bitcoin and Ethereum
under four copula-based ES-CAViaR models and two DCC-GARCH benchmarks. Shaded areas represent VaR–ES gaps, and dots indicate VaR violations.
}\label{fig:VaR_ES_BB1_all}
\end{figure}

Figures~\ref{fig:VaR_ES_Clayton_all} and~\ref{fig:VaR_ES_BB1_all} present the one-step-ahead 1\% VaR and ES forecasts for Bitcoin and Ethereum under the Clayton and BB1 copulas, respectively, alongside the DCC-GARCH benchmarks. Both copulas produce similar forecasting patterns, with pronounced increases in downside risk during the COVID-19 market turbulence in early 2020. The VaR-ES gaps widen substantially during this period, particularly for Ethereum, indicating greater expected loss severity beyond VaR. Among the four marginal specifications, ES-CAViaR generally produces the largest gaps, while ES-CAViaR-st, ES-CAViaR-v, and ES-CAViaR-R yield more moderate estimates. VaR violations, indicated by dots, occur mainly during periods of heightened market volatility.

The sharp market decline on March 12, 2020, amid the COVID-19 financial turmoil, provides a notable example of differences in the models' responses to extreme losses. On that day, the S\&P 500 triggered a circuit breaker, while Bitcoin and Ethereum also experienced substantial declines. Following VaR violations for both cryptocurrencies, DCC-GARCH-t produces exceptionally large negative ES forecasts on March 13, 2020, reaching $-63.142$ and $-77.562$ for Bitcoin and Ethereum, respectively. In contrast, the copula-based ES-CAViaR models generate less extreme forecasts, with varying degrees of persistence across marginal specifications. As neither cryptocurrency experiences comparably severe losses in the subsequent days, the DCC-GARCH-t forecasts appear particularly conservative during this episode, highlighting differences in the models' sensitivity to extreme market shocks.

\begin{table}[t]
\centering
\caption{Out-of-sample average scoring-rule losses
for marginal VaR and ES forecasts of Bitcoin and Ethereum,
2020--2025.}
\label{tab:scoring}
\renewcommand{\arraystretch}{0.7}
\setlength{\tabcolsep}{4pt}
\resizebox{\textwidth}{!}{%
\begin{tabular}{@{}llrrrr@{\hspace{12pt}}rrrr@{}}
\toprule
& & \multicolumn{4}{c}{Bitcoin (BTC)}
  & \multicolumn{4}{c}{Ethereum (ETH)} \\
\cmidrule(lr){3-6}\cmidrule(lr){7-10}
& & VaR & \multicolumn{3}{c}{VaR and ES}
  & VaR & \multicolumn{3}{c}{VaR and ES} \\
\cmidrule(lr){3-3}\cmidrule(lr){4-6}
\cmidrule(lr){7-7}\cmidrule(lr){8-10}
Copula & Marginal model
& QS & ALS & FZ & FZG
& QS & ALS & FZ & FZG \\
\midrule

Clayton
& ES-CAViaR
& 13.076 & 357.071 & 255.096 & 82.214
& 17.786 & 391.465 & 289.804 & 87.468 \\

& ES-CAViaR-st
& 13.076 & 357.023 & 254.999 & 82.212
& 17.847 & 392.311 & 290.597 & 87.653 \\

& ES-CAViaR-v
& 12.985 & 358.302 & 256.162 & 82.157
& 17.625 & 385.711 & 283.801 & 86.798 \\

& ES-CAViaR-R
& 12.666
& \underline{354.398}
& \underline{252.308}
& 81.808
& 17.638 & 387.429 & 285.708 & 86.945 \\

\addlinespace[3pt]

BB1
& ES-CAViaR
& 13.085 & 357.399 & 255.407 & 82.224
& 17.833 & 392.235 & 290.571 & 87.578 \\

& ES-CAViaR-st
& 13.077 & 357.061 & 255.039 & 82.213
& 17.846 & 392.271 & 290.573 & 87.643 \\

& ES-CAViaR-v
& 12.973 & 358.209 & 256.069 & 82.145
& \underline{17.618}
& \underline{385.570}
& \underline{283.650}
& \underline{86.791} \\

& ES-CAViaR-R
& \underline{\textbf{12.666}}
& 354.422
& 252.340
& \underline{\textbf{81.807}}
& 17.678 & 387.764 & 286.052 & 86.984 \\

\midrule

\multicolumn{2}{l}{DCC-GARCH-N}
& 13.247 & 363.328 & 260.601 & 82.434
& 17.519 & 390.093 & 287.466 & 86.700 \\

\multicolumn{2}{l}{DCC-GARCH-$t$}
& 13.024
& \textbf{353.624}
& \textbf{251.163}
& 82.232
& \textbf{16.960}
& \textbf{376.947}
& \textbf{274.560}
& \textbf{86.216} \\
\bottomrule
\end{tabular}%
}
\par\vspace{5pt}
\begin{minipage}{0.98\textwidth}
\footnotesize
\textit{Note:}
QS evaluates VaR, whereas ALS, FZ, and FZG jointly
evaluate VaR and ES.
Scores are averaged over time separately for each asset
and multiplied by 100; lower values indicate better
forecasting performance.
For each asset and scoring rule, boldface indicates
the lowest score among all models, whereas underlining
indicates the lowest score among the eight proposed
models across both copula families.
\end{minipage}
\end{table}

Table~\ref{tab:scoring} reports the out-of-sample
average scoring-rule losses for marginal VaR and ES
forecasts of Bitcoin and Ethereum.
For Bitcoin, Clayton ES-CAViaR-R achieves the lowest
QS and FZG losses, whereas DCC-GARCH-$t$ yields the
lowest ALS and FZ losses.
Among the proposed models, Clayton ES-CAViaR-R
performs best under all four scoring rules.
For Ethereum, BB1 ES-CAViaR-v yields the lowest
QS, ALS, FZ and FZG losses among the proposed models.
For a given marginal specification, differences
between the Clayton and BB1 results are generally small.
For Bitcoin, all eight proposed models outperform DCC-GARCH-N across all four scoring rules and DCC-GARCH-$t$ under FZG, whereas DCC-GARCH-$t$
achieves the lowest ALS and FZ losses.
For Ethereum, DCC-GARCH-$t$ yields the lowest
losses under all four scoring rules.

Although QS is a consistent scoring function for quantile forecasts, a lower QS loss does not necessarily imply that the VaR violation rate is closer to the nominal level.
As an asymmetrically weighted absolute error,
QS depends on the magnitudes of forecast errors
on both sides of the predicted quantile rather
than directly measuring deviations from the
nominal violation rate.
For example, DCC-GARCH-N produces VaR violation
rates of 1.667\% and 1.717\% for Bitcoin and
Ethereum, respectively, substantially exceeding
the nominal 1\% level, while maintaining
relatively competitive QS losses.
Similarly, DCC-GARCH-$t$ achieves the lowest
ALS and FZ losses for Bitcoin and all four
scoring-rule losses for Ethereum, despite
violation rates of 1.226\% and 1.373\%,
respectively, exceeding the nominal 1\% level.
These findings highlight the importance of
evaluating scoring-rule performance alongside
empirical violation rates.

\begin{figure}
\centering
\includegraphics[width=0.98\textwidth]
{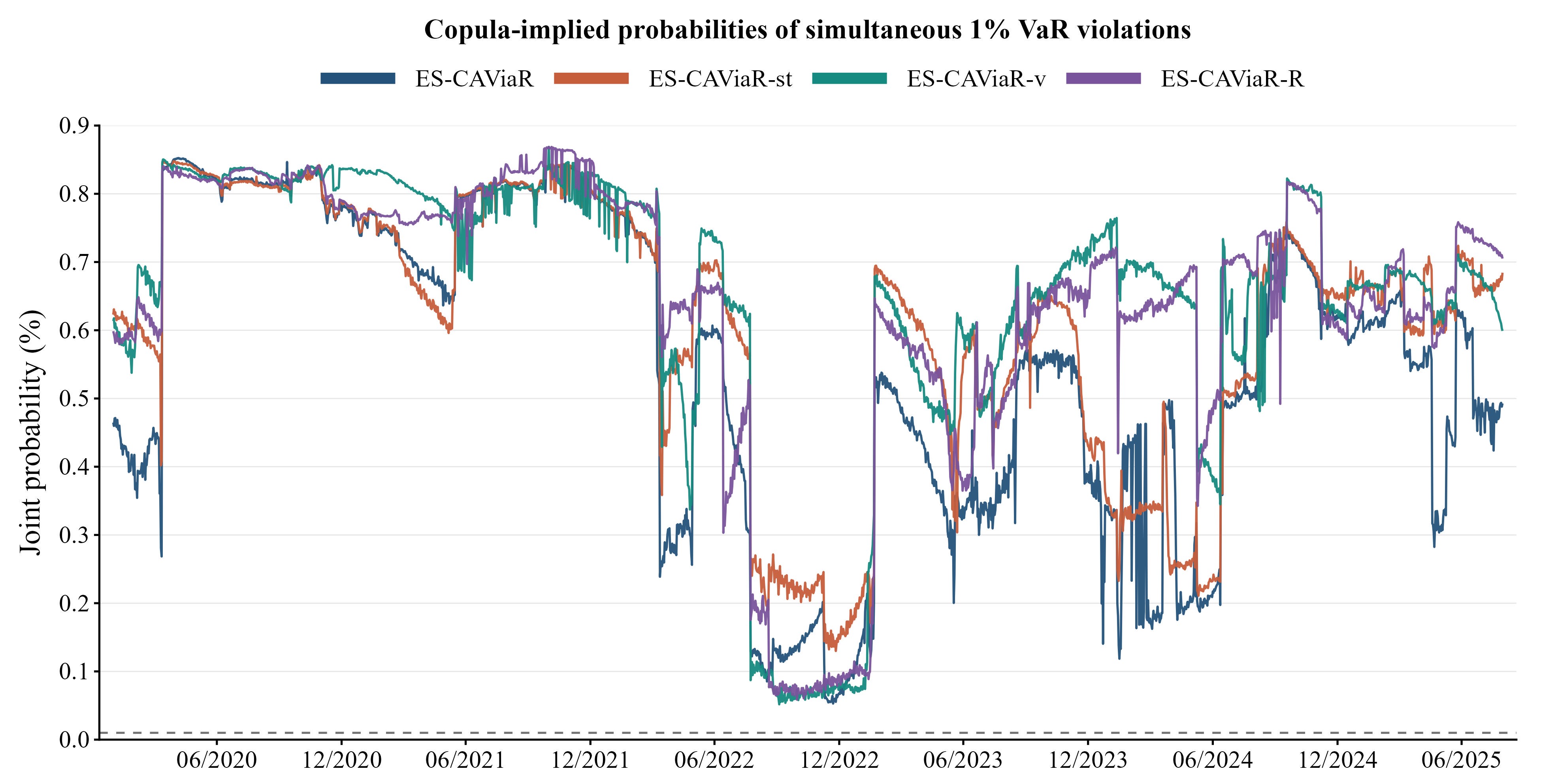}
\vspace{-0.2cm}
\caption{\small
Predicted probabilities of simultaneous 1\% VaR violations
under four Clayton copula-based ES-CAViaR models.
The dashed line indicates the joint shortfall probability under independence ($0.01\%$).
}
\label{fig:joint_prob_clayton}
\end{figure}

\paragraph{Joint performance evaluation}
Figure \ref{fig:joint_prob_clayton} presents the one-step-ahead
probabilities of simultaneous 1\% VaR violations under the four
Clayton copula-based ES-CAViaR models. 
These probabilities are obtained by evaluating
the Clayton copula at $(0.01,0.01)$ using the
corresponding rolling parameter point estimates,
as defined in Equation~\eqref{eq:joint_prob:forecast}.

Despite differences in their marginal specifications,
all four models exhibit similar temporal patterns, with relatively high joint tail-event probabilities during the COVID-19 period in 2020-2021,
followed by a substantial decline in the second half of 2022. This decline may reflect changes
in the lower-tail dependence between Bitcoin and Ethereum, rather than solely a reduction
in their individual market risks.
The predicted probabilities subsequently increase during 2023-2024 and generally remain well above
the independence benchmark of $0.01\%$, suggesting that an independence assumption
would underestimate joint downside risk.
These findings indicate time-varying cross-asset tail dependence and highlight the
importance of modeling joint tail risk beyond marginal VaR and ES forecasts.

\begin{figure}[!ht]
\centering
\includegraphics[width=1.0\textwidth]{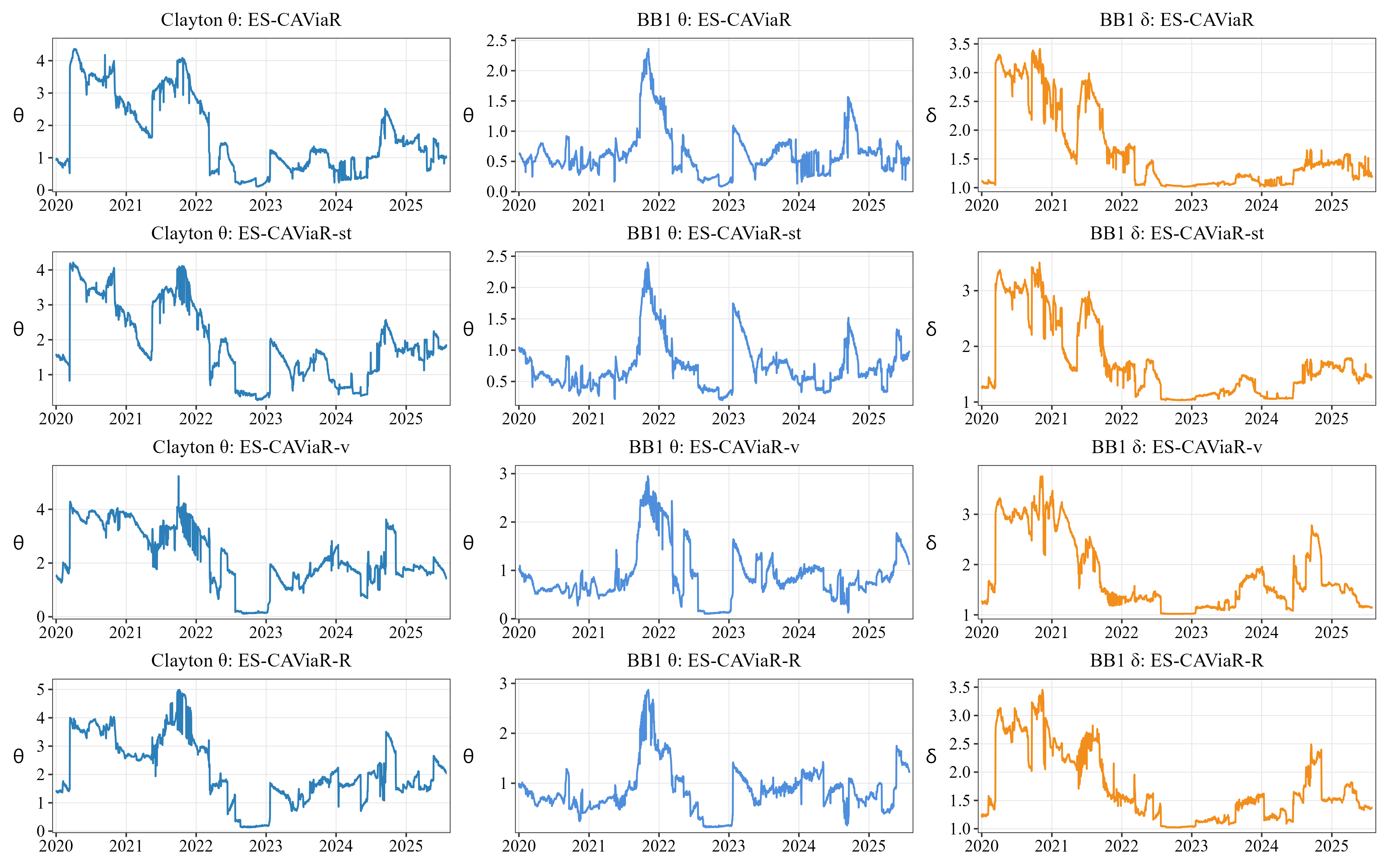}
\caption{\small
Rolling posterior estimates of the Clayton parameter $\theta_t$ and
BB1 parameters $(\theta_t,\delta_t)$ for four ES-CAViaR specifications
over 2,039 one-step-ahead forecast origins, using a rolling window
of 730 observations.}
\label{fig:Copula_parameters}
\end{figure}

\begin{figure}[!ht]
\centering
\includegraphics[width=1.0\textwidth]{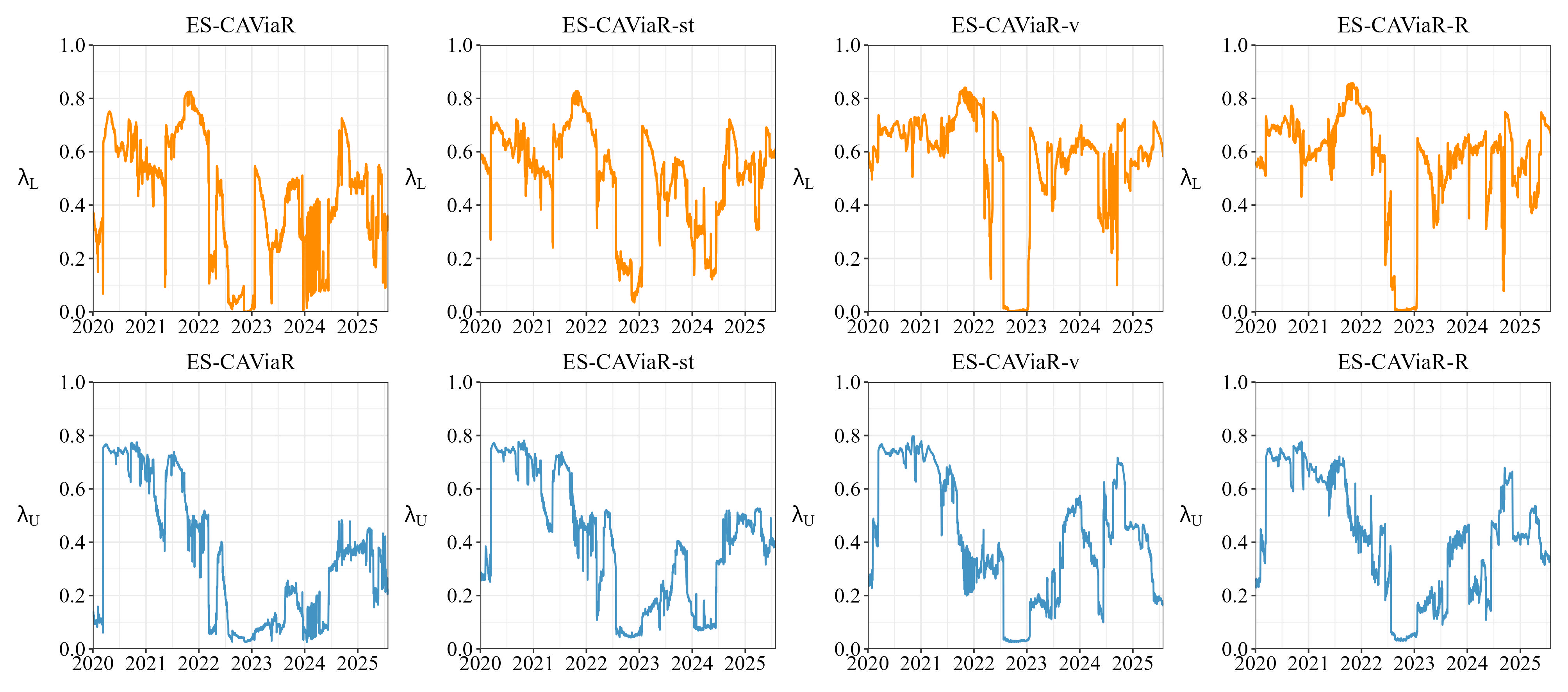}
\caption{\small
Rolling estimates of the lower-tail and upper-tail dependence
coefficients implied by the BB1 copula across the four ES-CAViaR
specifications over 2,039 one-step-ahead forecasts.
At each forecast origin, all model parameters are re-estimated
using a rolling window of 730 observations.}
\label{fig:BB1_tail_dependence}
\end{figure}

Figure~\ref{fig:Copula_parameters} presents the rolling posterior
estimates of the Clayton copula parameter $\theta_t$ and the BB1
copula parameters $(\theta_t,\delta_t)^{\prime}$ across the four ES-CAViaR
specifications.
For the Clayton copula, larger values of $\theta_t$ indicate
stronger lower-tail dependence, with $\theta_t=0.5$, $4$, and $5$
corresponding to lower-TDCs of
$0.250$, $0.841$, and $0.871$, respectively.
The estimates suggest stronger lower-tail dependence during
2020-2021, weaker dependence during 2022-2023, and occasional
increases in 2024-2025.
In contrast, the BB1 copula allows for asymmetric tail dependence,
with $\theta_t$ and $\delta_t$ jointly determining lower-tail
dependence, while $\delta_t$ controls upper-tail dependence.
The time-varying BB1 parameter estimates reflect changes in
both the strength and asymmetry of dependence, allowing greater
flexibility than the Clayton specification.
Across the four ES-CAViaR specifications, the parameter estimates
display broadly similar temporal patterns, although differences
in their magnitudes suggest that the marginal specification
influences the estimated dependence structure.

Figure~\ref{fig:BB1_tail_dependence} presents the implied lower- and
upper-tail TDCs of the BB1 copula. Both
coefficients are high (around 0.7-0.8) in 2020-2021 but decline
markedly during 2022-2023, approaching zero across all four
specifications. This period coincides with major cryptocurrency market
disruptions, including the Terra-Luna collapse in May 2022 and the FTX
collapse in November 2022. From 2023 onward, the two tails evolve
differently: lower-tail dependence generally exceeds upper-tail dependence, indicating stronger co-crash than co-boom behaviour between Bitcoin and Ethereum in the later sample. This highlights the ability of the BB1 copula to capture time-varying asymmetric tail dependence, although this additional flexibility does not necessarily translate
into more accurate joint downside-risk forecasts.


\begin{table}
\caption{Out-of-sample joint tail-event forecast evaluation, 2020--2025.} \label{tab:S_Bern}
\vspace{-0.5cm}
\vspace{7pt}
\centering
{\fontsize{10pt}{11pt}\selectfont
\renewcommand{\arraystretch}{0.85}
\setlength{\tabcolsep}{4pt}
\begin{tabular}{@{}llrrrrr@{}}
\toprule
Copula & Marginal model & Joint viol.
& $V_{12}$ (\%) & Mean $\hat p_t$ (\%)
& Obs. $-$ Pred.
& $\bar S_{\mathrm{Bern}}$ \\
\midrule

Clayton
& ES-CAViaR
& 13 & 0.638 & 0.558
& 0.080 & 4.046 \\

& ES-CAViaR-st
& 13 & 0.638 & 0.620
& \underline{0.018} & \textbf{3.987} \\

& ES-CAViaR-v
& 14 & 0.687 & 0.661
& 0.026 & 4.161 \\

& ES-CAViaR-R
& 13 & 0.638 & 0.653
& \textbf{-0.015} & 4.032 \\

\addlinespace[3pt]

BB1
& ES-CAViaR
& 13 & 0.638 & 0.457
& 0.180 & 4.114 \\

& ES-CAViaR-st
& 13 & 0.638 & 0.530
& 0.107 & \underline{4.018} \\

& ES-CAViaR-v
& 14 & 0.687 & 0.580
& 0.106 & 4.159 \\

& ES-CAViaR-R
& 13 & 0.638 & 0.572
& 0.066 & 4.034 \\

\midrule
\multicolumn{2}{l}{DCC-GARCH-N}
& 24 & 1.177 & 0.446
& 0.731 & 6.977 \\
\multicolumn{2}{l}{DCC-GARCH-$t$}
& 19 & 0.932 & 0.557
& 0.375 & 5.452 \\
\bottomrule
\end{tabular}
\par\vspace{4pt}
\begin{minipage}{0.98\linewidth}
\footnotesize
\textit{Notes:}
The observed joint violation rate $V_{12}$,
and the mean predicted
probability are both multiplied by
100 for display.
Obs. $-$ Pred. reports the signed difference between the observed
and predicted rates, measured in percentage points.
Bern.\,score is $100\bar S_{\mathrm{Bern}}$,
the mean Bernoulli logarithmic score multiplied
by 100.
Across all ten models, boldface and underlining indicate the best and second-best values, respectively, ranked by the absolute difference
for Obs. $-$ Pred. and by the score itself
for Bern.\,score.
Rankings use unrounded values.
\end{minipage}
}
\end{table}

Table~\ref{tab:S_Bern} evaluates forecasts of simultaneous
1\% VaR violations. The copula-based ES-CAViaR models
produce joint violation rates of 0.638\%--0.687\% and
mean predicted probabilities of 0.457\%--0.661\%.
Clayton ES-CAViaR-R exhibits the closest agreement,
with an observed rate of 0.638\%, a mean predicted
probability of 0.653\%, and a difference of
$-0.015$ percentage points.
Clayton ES-CAViaR-st ranks second by absolute
difference and achieves the lowest Bernoulli score of 3.987.
BB1 ES-CAViaR-R ranks second by the unrounded Bernoulli
score, slightly ahead of Clayton ES-CAViaR-R.
Across the eight proposed models, Bernoulli scores
range from 3.987 to 4.161, compared with 6.977 for DCC-GARCH-N and 5.452 for DCC-GARCH-$t$.
All eight models also exhibit smaller absolute differences
between observed joint violation rates and mean predicted
probabilities than either benchmark.
In particular, DCC-GARCH-N predicts an average probability
of 0.446\%, well below its observed joint violation rate
of 1.177\%.
These results provide descriptive evidence on
joint tail-event probability forecasts within the proposed
semiparametric framework.

\begin{table}[t]
\centering
\caption{\small Bernoulli logarithmic scores for joint 1\%
VaR exceedances and differences relative to DCC-GARCH-$t$,
2020--2025.}
\label{tab:bern_diff}
\renewcommand{\arraystretch}{0.7}
\setlength{\tabcolsep}{12pt}
\begin{tabular}{llcrrr}
\toprule
Copula & Marginal
& $\bar S_{\mathrm{Bern}}$
& $\Delta^{A,\mathrm{DCC}\text{-}t}_{\mathrm{Bern}}$
& \multicolumn{2}{c}{95\% BB CI} \\
\cmidrule(lr){5-6}
& & & & Lower & Upper \\
\midrule
Clayton & ES-CAViaR
& 4.046 & \textbf{$-$1.406}
& $-$2.906 & $-$0.053 \\
 & ES-CAViaR-st
& 3.987 & \textbf{$-$1.465}
& $-$2.953 & $-$0.136 \\
 & ES-CAViaR-v
& 4.161 & $-$1.291
& $-$2.813 & 0.147 \\
 & ES-CAViaR-R
& 4.032 & \textbf{$-$1.420}
& $-$2.977 & $-$0.009 \\
\addlinespace
BB1 & ES-CAViaR
& 4.114 & $-$1.338
& $-$2.848 & 0.038 \\
 & ES-CAViaR-st
& 4.018 & \textbf{$-$1.434}
& $-$2.920 & $-$0.091 \\
 & ES-CAViaR-v
& 4.159 & $-$1.293
& $-$2.814 & 0.160 \\
 & ES-CAViaR-R
& 4.034 & $-$1.418
& $-$2.981 & 0.013 \\
\addlinespace
\multicolumn{2}{l}{DCC-GARCH-N}
& 6.977 & \textbf{1.525}
& 0.426 & 2.817 \\
\addlinespace
\multicolumn{2}{l}{DCC-GARCH-$t$}
& 5.452 & --
& --& --\\
\bottomrule
\end{tabular}
\vspace{0.5ex}
\begin{minipage}{0.95\linewidth}
\footnotesize
\textit{Notes:}
All scores, differences, and intervals are multiplied by 100.
Lower Bernoulli scores indicate better predictive performance.
We denote by $\Delta^{A,\mathrm{DCC}\text{-}t}_{\mathrm{Bern}}$
the mean loss difference relative to DCC-GARCH-$t$;
positive values favor DCC-GARCH-$t$.
The 95\% confidence intervals use a paired stationary
block bootstrap with mean block length 12 and 20{,}000 resamples.
Bold indicates differences whose intervals exclude zero.
\end{minipage}
\end{table}

To assess uncertainty in joint-event forecast comparisons,
Table~\ref{tab:bern_diff} reports Bernoulli score differences relative to
DCC-GARCH-$t$, together with 95\% stationary block-bootstrap confidence
intervals~\citep{PolitisRomano1994}.
All eight Bayesian copula specifications achieve lower mean Bernoulli
losses than DCC-GARCH-$t$.
The 95\% confidence intervals for the
score differences exclude zero for three Clayton specifications and BB1-ES-CAViaR-st.
These results favor the proposed copula models, but do not
establish a uniquely superior specification.
Additional uncertainty assessments can be found in the
Supplementary Materials.

\section{Conclusion and future research}
\label{sec:conclusion}

This study develops a Bayesian copula-based bivariate
semiparametric framework that integrates marginal VaR
and ES forecasting with cross-asset tail dependence.
Unlike conventional parametric multivariate models,
the proposed approach does not require a fully
specified joint return distribution. By combining
ES-CAViaR marginal dynamics with Clayton and BB1
copulas through rank-based standardized residuals,
the framework provides one-step-ahead forecasts
of marginal tail-risk measures and the conditional
probability of simultaneous VaR violations.
The Bayesian estimation procedure accommodates
posterior uncertainty in model parameters.

An important contribution is the extension of
the ES-CAViaR framework through a
volatility-standardized exceedance term and
alternative range-based volatility measures.
These modifications allow the marginal VaR and ES
dynamics to respond differently to recent market
volatility and extreme losses. 
The proposed framework also distinguishes
the evaluation of marginal VaR and ES forecasts
from that of joint tail-event probabilities,
providing complementary information on marginal
and joint tail-risk forecasts.

The empirical analysis of Bitcoin and Ethereum
from 2020 to 2025 reveals differences in forecasting
performance across assets and evaluation criteria.
For Bitcoin, Clayton ES-CAViaR-R and DCC-GARCH-$t$ achieve the lowest
marginal losses depending on the choice of scoring rules, whereas DCC-GARCH-$t$ achieves
the lowest losses under all four scoring rules for Ethereum.
For simultaneous VaR violation probability forecasts,
Clayton ES-CAViaR-st yields the lowest average Bernoulli score.
BB1 ES-CAViaR-st ranks second using unrounded scores.
Clayton ES-CAViaR-R retains the smallest absolute difference
between the observed joint violation rate and the mean
predicted probability.
All eight proposed models yield lower Bernoulli
scores than both DCC-GARCH benchmarks in the
reported evaluation.
Because the issued VaR thresholds may differ
across models, these Bernoulli score comparisons
are interpreted descriptively alongside the
marginal forecast evaluation.
The simulation study under the DCC-GARCH-$t$ DGP
further indicates that estimating joint tail-event
probabilities accurately is more challenging than
estimating marginal VaR and ES.

The rolling copula estimates reveal substantial
changes in dependence across market conditions,
including the COVID-19 period and the
cryptocurrency market disruptions of 2022--2023.
The BB1 copula accommodates asymmetric
lower- and upper-tail dependence, but its
additional flexibility does not yield a uniform
reduction in average Bernoulli scores relative
to the corresponding Clayton specifications.
These findings suggest that the choice of
marginal dynamics and dependence specification
should reflect the forecasting objective rather
than model complexity alone.

The analysis focuses on Bitcoin and Ethereum,
with copula parameters held constant within
each rolling estimation window. Future research
may extend the framework to additional assets
and portfolios, incorporate explicitly
time-varying copula parameters, and investigate
alternative dependence structures. Further
extensions may consider systemic risk measures and
posterior predictive assessments of joint
tail-risk uncertainty.

\nonblind{
\section*{Acknowledgements}
The authors are sincerely grateful to the Editor-in-Chief,
the Associate Editor, and the referees for their careful
evaluation, insightful comments, and constructive suggestions,
which have substantially improved the quality and clarity
of the manuscript. Cathy W.~S. Chen's research is supported by the National Science and Technology Council, Taiwan (NSTC 115-2123-M-035-001).
Takaaki Koike is supported by the Japan Society for the Promotion of Science
(JSPS) KAKENHI grant numbers JP24K00273 and JP26K21178.
}

\bibliography{bibliography_arxiv}
\bibliographystyle{agsm}

\newpage

\hypersetup{pageanchor=false}   
\setcounter{section}{0}
\setcounter{equation}{0}
\setcounter{table}{0}
\setcounter{figure}{0}
\renewcommand{\thesection}{S.\arabic{section}}
\renewcommand{\theHsection}{S.\arabic{section}}
\renewcommand{\theequation}{S.\arabic{equation}}
\renewcommand{\theHequation}{S.\arabic{equation}}
\renewcommand{\thetable}{S.\arabic{table}}
\renewcommand{\theHtable}{S.\arabic{table}}
\renewcommand{\thefigure}{S.\arabic{figure}}
\renewcommand{\theHfigure}{S.\arabic{figure}}
\makeatletter
\def\hyper@natlinkstart#1{\Hy@backout{#1}\hyper@linkstart{cite}{citesupp.#1}\def\hyper@nat@current{#1}}
\def\hyper@natlinkbreak#1#2{\hyper@linkend#1\hyper@linkstart{cite}{citesupp.#2}}
\def\hyper@natanchorstart#1{\Hy@raisedlink{\hyper@anchorstart{citesupp.#1}}}
\makeatother

\begin{center}
{\LARGE Supplementary Material for\\[6pt]
``Modeling Joint Tail Risk and Dependence for Cryptocurrencies via a Copula-Based Semiparametric Framework''\par}
\end{center}
\vspace{1em}

This document consists of two sections.
Section~\ref{sec:Bayesian:estimation}  provides additional details on the Bayesian estimation procedure and MCMC implementation for the proposed copula-based ES-CAViaR models.
Section~\ref{sec:additional:backtesting} then presents additional backtesting results including  formal comparative and calibration backtests for the forecasts of VaR, ES and the joint shortfall probability.

\section{Bayesian Estimation}
\label{sec:Bayesian:estimation}

We employ blockwise Metropolis-Hastings (MH) algorithms
for posterior simulation. For the marginal dynamics,
we update the parameter vectors $\boldsymbol{\beta}_i$
and $\boldsymbol{\gamma}_i$, $i=1,2$, separately.
For the BB1 copula, we jointly update the transformed
parameter vector
\[
\boldsymbol{\phi}
=
\left(\log\theta,\log(\delta-1)\right)^{\prime}.
\]
These parameter blocks share the same adaptive MCMC
procedure. 
In the description below, $\boldsymbol{\xi}$ denotes the block being updated, that is, $\boldsymbol{\beta}_i$ or $\boldsymbol{\gamma}_i$ for marginal updates and $\boldsymbol{\phi}$ for the BB1 copula.
Let $d$ denote the dimension of $\boldsymbol{\xi}$.

At iteration $j$, let $\boldsymbol{\xi}^{[j-1]}$ denote the
current value of the block and $\boldsymbol{\xi}_j^*$ its
proposed value. During burn-in, we use an adaptive random-walk
MH (RWMH) update with proposal
\begin{align*}
\boldsymbol{\xi}_j^*
&\sim \operatorname{N}_d\bigl(\boldsymbol{\xi}^{[j-1]},\boldsymbol{\Sigma}_j\bigr),
\end{align*}
where $\boldsymbol{\Sigma}_j$ denotes the adaptive proposal covariance matrix in effect at the start of iteration $j$. 
We adjust the proposal
step sizes and covariance structure during burn-in
and fix the proposal covariance matrix thereafter.

After burn-in, we use a mixture of independence-kernel MH (IKMH) and RWMH
updates, selected with probabilities 0.8 and 0.2,
respectively.
The proposals are
\begin{align*}
\text{IKMH:}\qquad
\boldsymbol{\xi}_j^*
&\sim \operatorname{N}_d\bigl(\widehat{\boldsymbol{\mu}},
               \widehat{\boldsymbol{\Sigma}}\bigr),\\
\text{RWMH:}\qquad
\boldsymbol{\xi}_j^*
&\sim \operatorname{N}_d\bigl(\boldsymbol{\xi}^{[j-1]},
               \boldsymbol{\Sigma}_{\mathrm{RW}}\bigr).
\end{align*}
Here $\widehat{\boldsymbol{\mu}}$ and $\widehat{\boldsymbol{\Sigma}}$
are the mean vector and covariance matrix estimated from the burn-in draws for this block. Both are held fixed after burn-in.
The matrix $\boldsymbol{\Sigma}_{\mathrm{RW}}$
denotes the proposal covariance matrix fixed
at the end of burn-in.

Let $q(\cdot\mid\widehat{\boldsymbol{\mu}},
\widehat{\boldsymbol{\Sigma}})$ denote the density of the Gaussian
independence proposal. Let $\pi_j$ denote the target density of $\boldsymbol{\xi}$
at iteration $j$.
The acceptance probabilities for the two updates are
\begin{align*}
P_{I,j}
&=
\min\left\{1,\,
\frac{\pi_j(\boldsymbol{\xi}_j^*)}
     {\pi_j(\boldsymbol{\xi}^{[j-1]})}
\frac{q(\boldsymbol{\xi}^{[j-1]}
          \mid\widehat{\boldsymbol{\mu}},\widehat{\boldsymbol{\Sigma}})}
     {q(\boldsymbol{\xi}_j^*
          \mid\widehat{\boldsymbol{\mu}},\widehat{\boldsymbol{\Sigma}})}
\right\},\\
P_{R,j}
&=
\min\left\{1,\,
\frac{\pi_j(\boldsymbol{\xi}_j^*)}
     {\pi_j(\boldsymbol{\xi}^{[j-1]})}
\right\}.
\end{align*}
We apply the acceptance probability corresponding
to the selected proposal kernel.
If the proposal is accepted, we set
$\boldsymbol{\xi}^{[j]}=\boldsymbol{\xi}_j^*$.
Otherwise, we retain
$\boldsymbol{\xi}^{[j]}=\boldsymbol{\xi}^{[j-1]}$.

\paragraph{Clayton copula update}

For the Clayton copula, we employ a Gaussian
RWMH algorithm on Kendall's tau
$\tau=\theta/(\theta+2)$ throughout posterior
simulation. At iteration $j$, we propose
\[
\tau_j^{*}
\sim \operatorname{N}(\tau^{[j-1]},s_j^2),
\]
where $s_j$ denotes the proposal step size.
We adapt $s_j$ during burn-in and keep it fixed
thereafter. 
We reject proposals outside $(0,1)$
without resampling.

\subsection{Posterior summaries and MCMC diagnostics}
\label{sec:posterior:summaries:diagnostics}

\paragraph{Posterior summaries}

The posterior summaries for ES-CAViaR and ES-CAViaR-st are  given in Tables~\ref{tab:ES-CAViaR} and~\ref{tab:ES-CAViaR-st}, respectively.

\paragraph{MCMC convergence diagnostics}

\begin{figure}[t]
\centering
\includegraphics[width=0.48\textwidth]{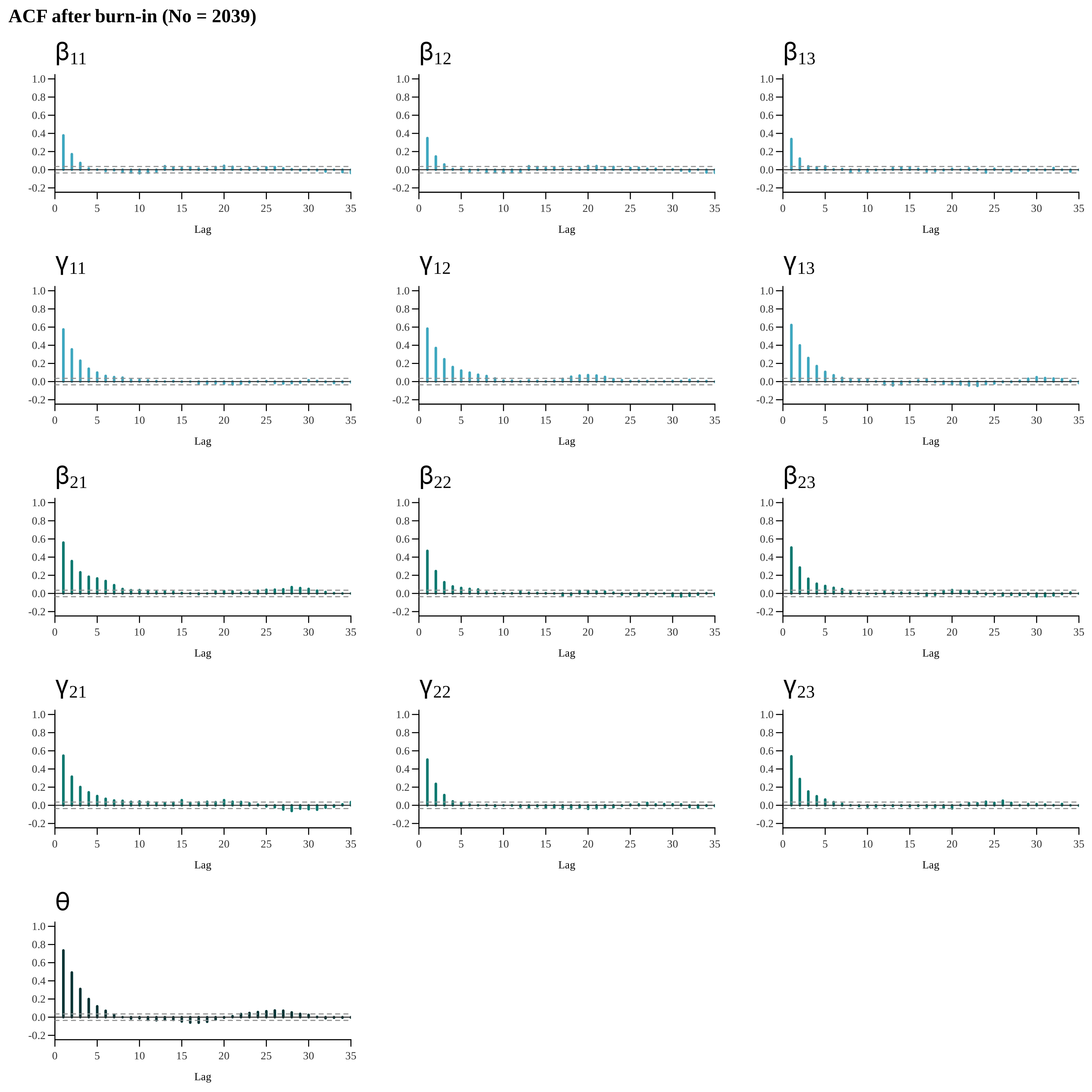} 
\includegraphics[width=0.48\textwidth]{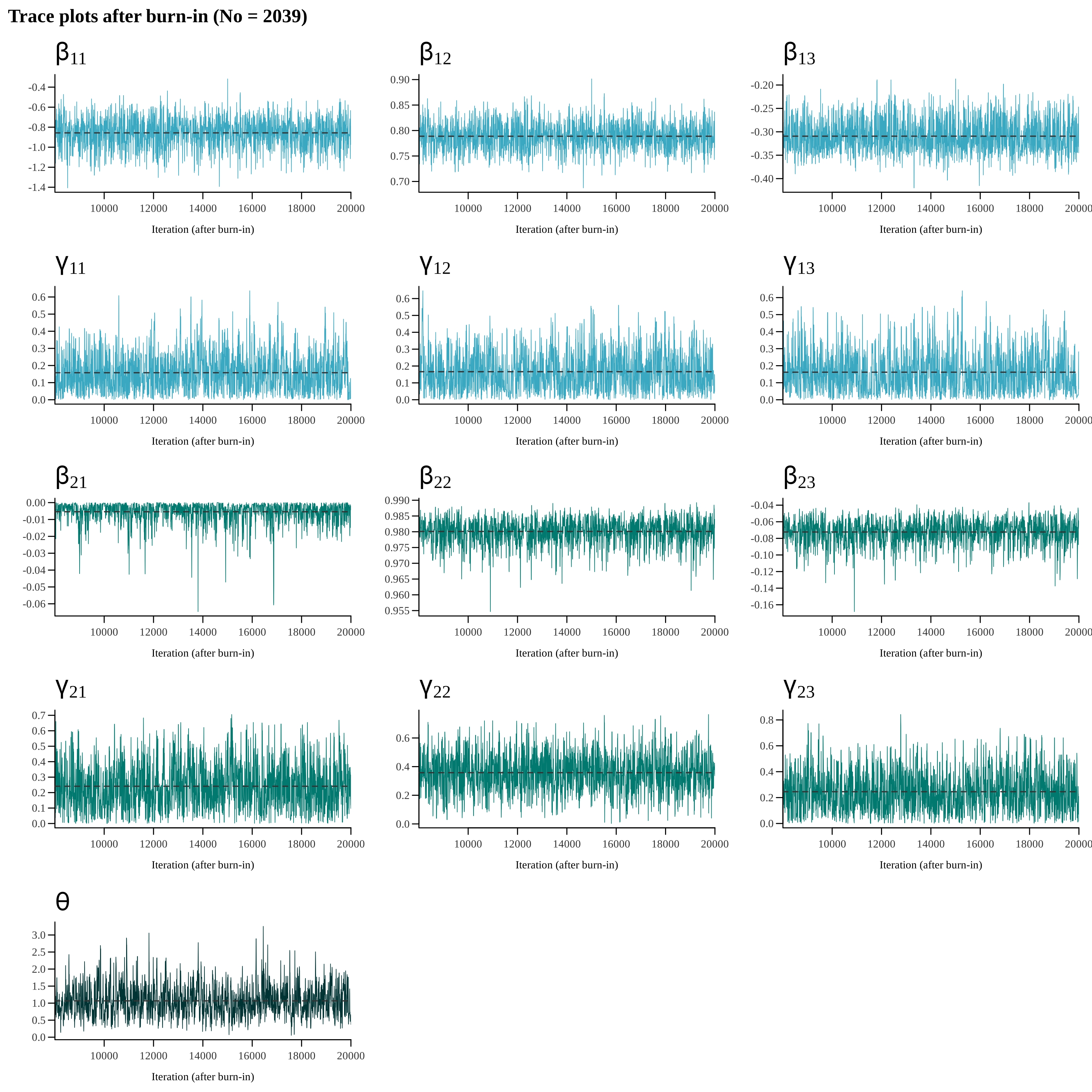} 
\caption{ACF and trace plots of the MCMC draws under the ES-CAViaR model with a Clayton copula.}
\label{fig:ES-CAViaR-ACF}
\end{figure}

\begin{figure}[t]
\centering   
\includegraphics[width=0.48\textwidth]{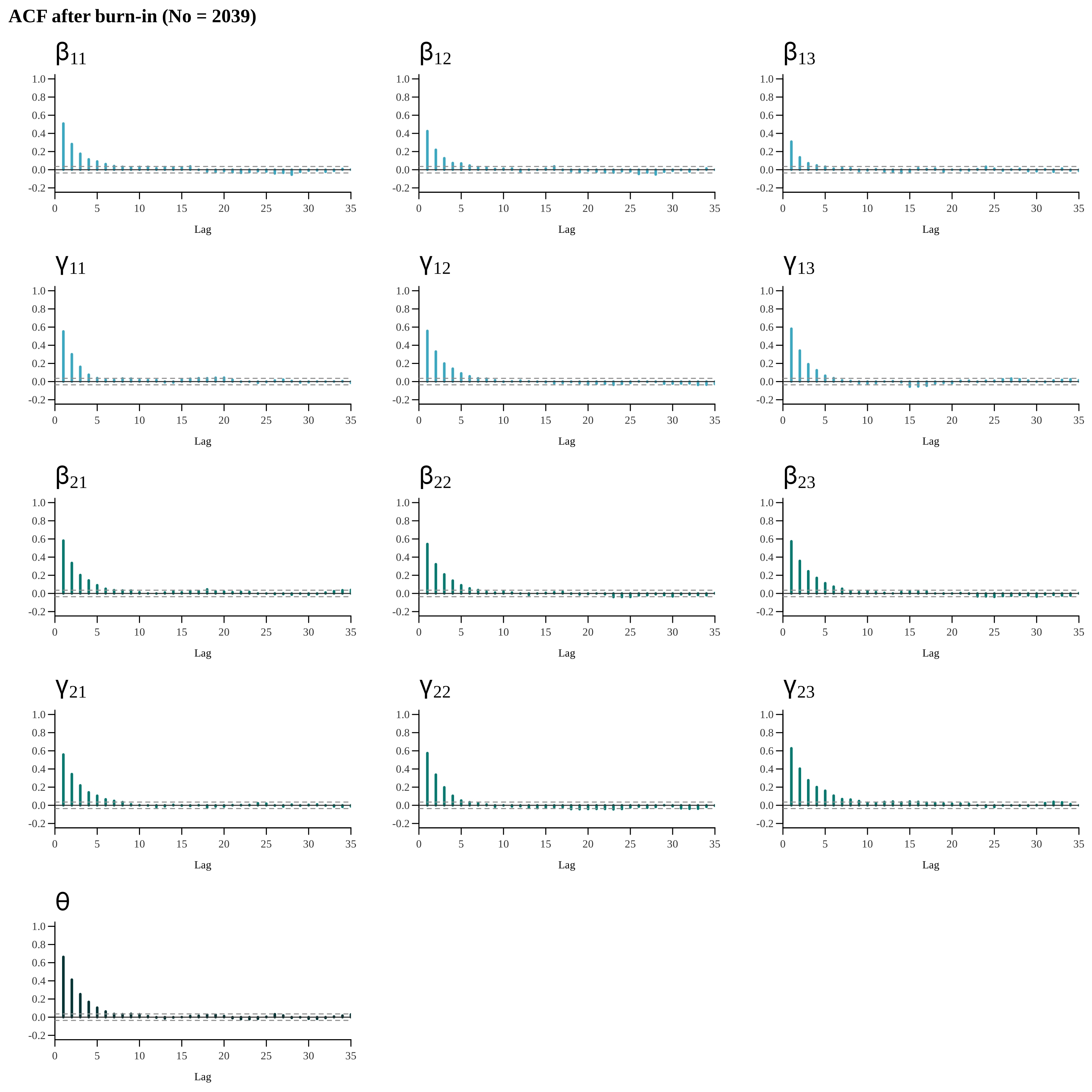} 
\includegraphics[width=0.48\textwidth]{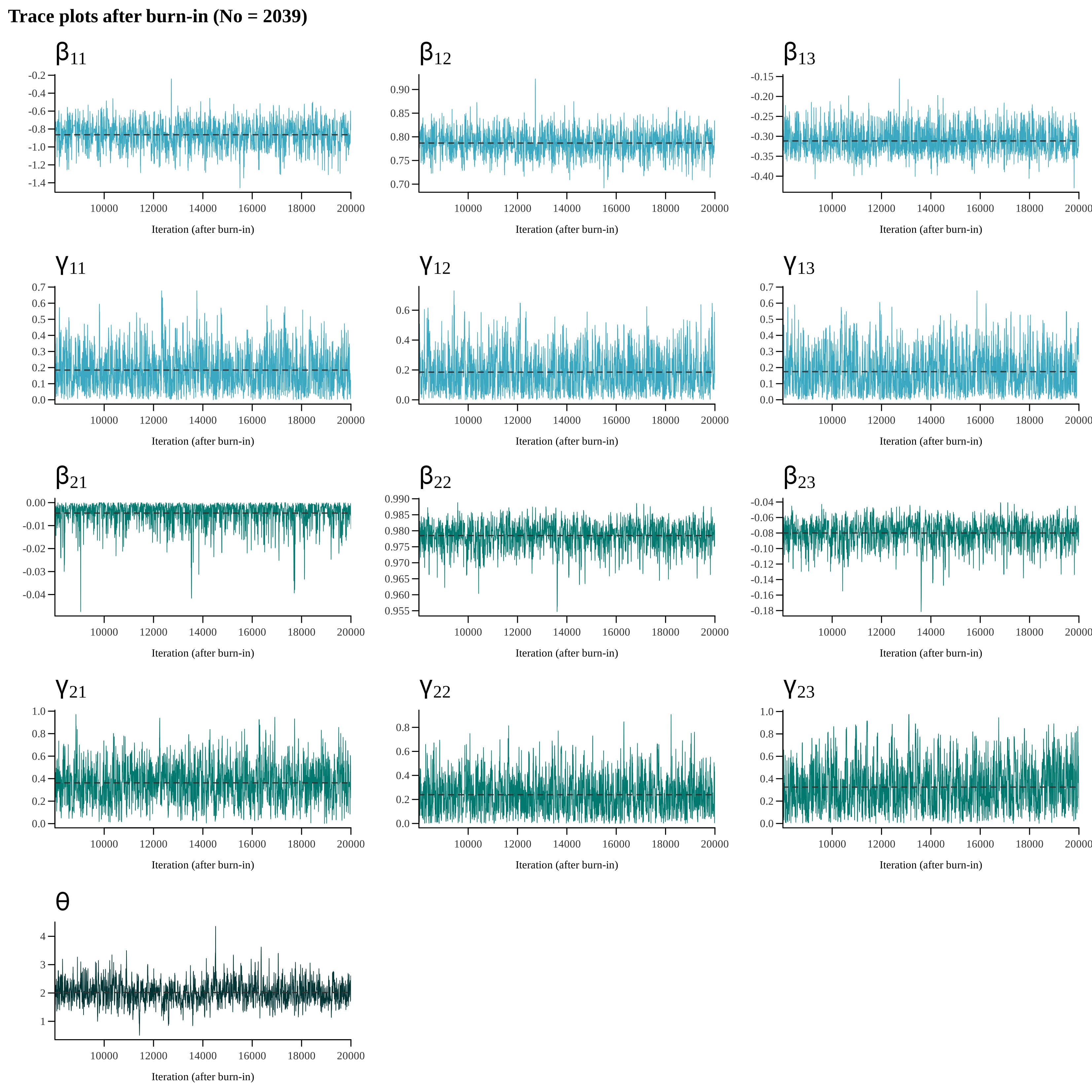} 
\caption{ACF and trace plots of the MCMC draws under the ES-CAViaR-st model with a Clayton copula.}
\label{fig:ES-CAViaR-st-ACF-TRACE}
\end{figure}

\begin{figure}[t]
\centering   
\includegraphics[width=0.48\textwidth]{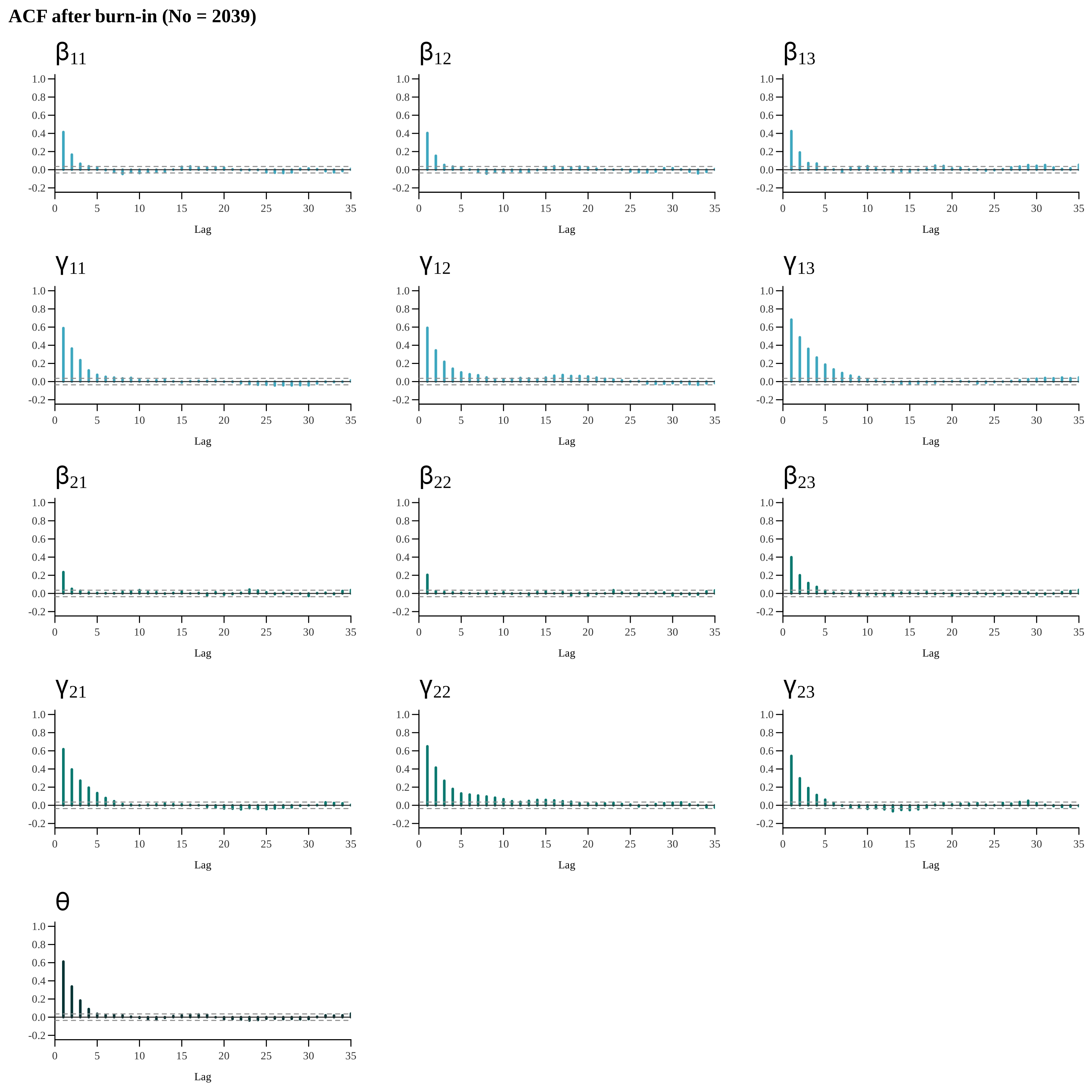} 
\includegraphics[width=0.48\textwidth]{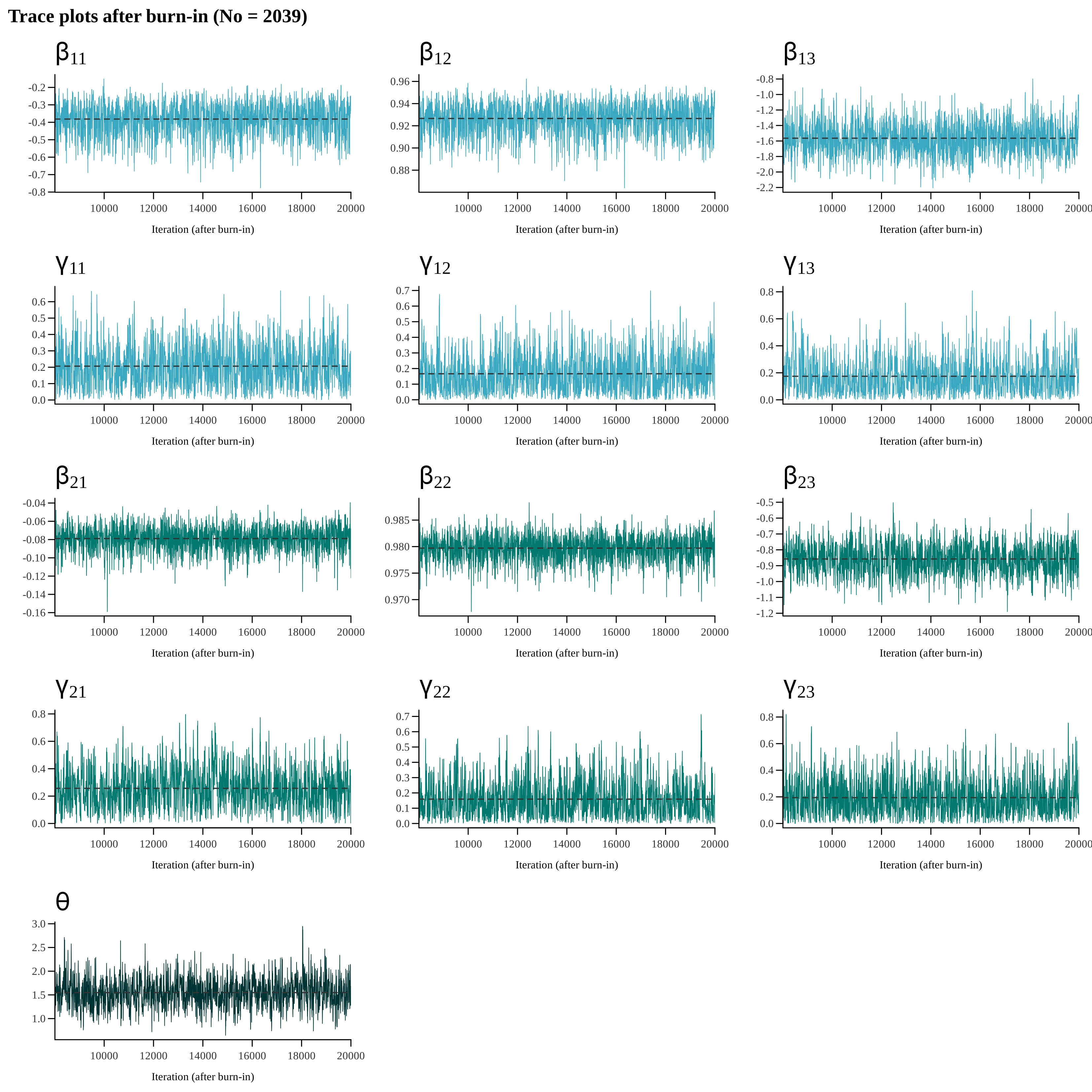} 
\caption{ACF and trace plots of the MCMC draws under the ES-CAViaR-v model with a Clayton copula.}
\label{fig:ES-CAViaR-v-ACF-TRACE}
\end{figure}

\begin{figure}[!ht]
\centering   
\includegraphics[width=0.48\textwidth]{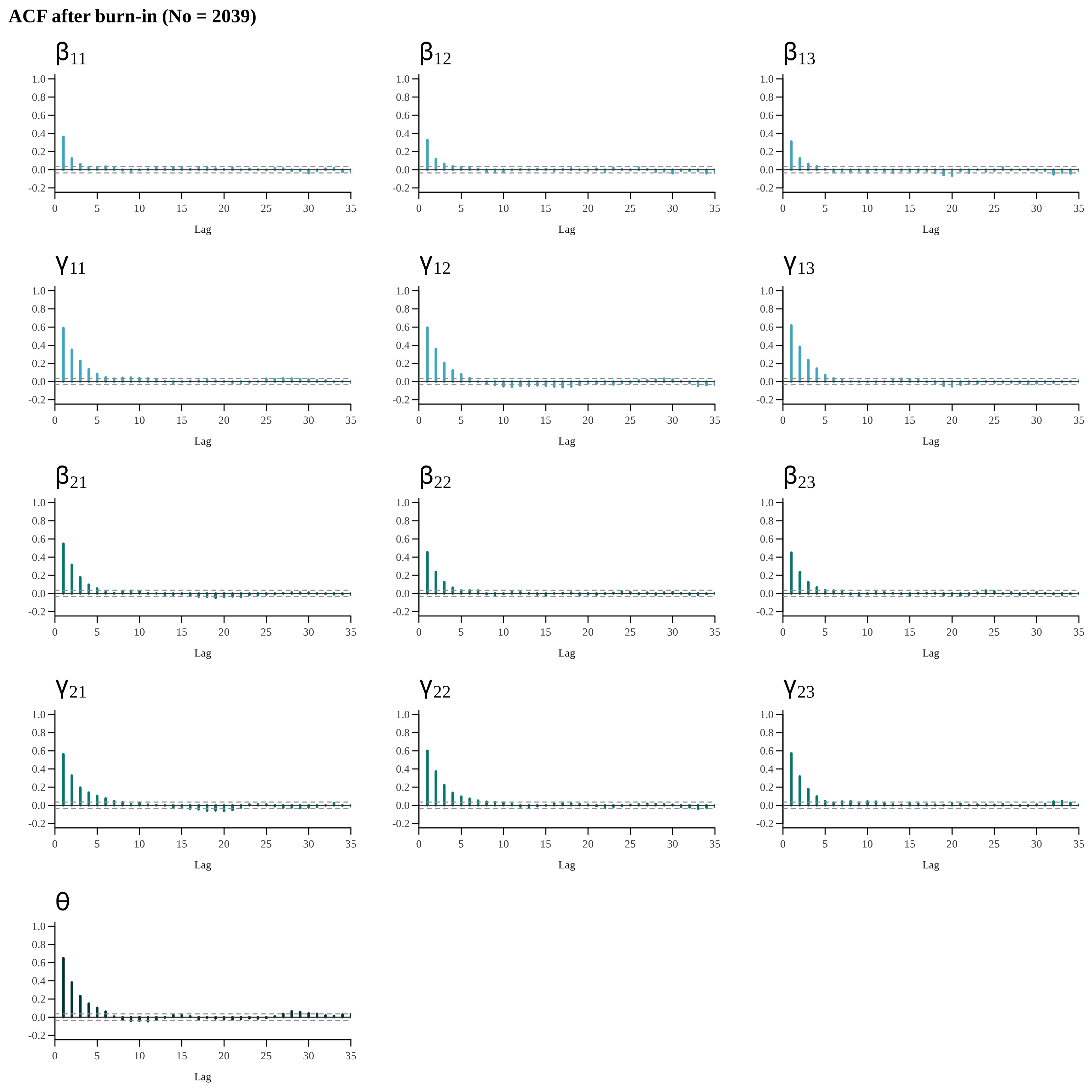} 
\includegraphics[width=0.48\textwidth]{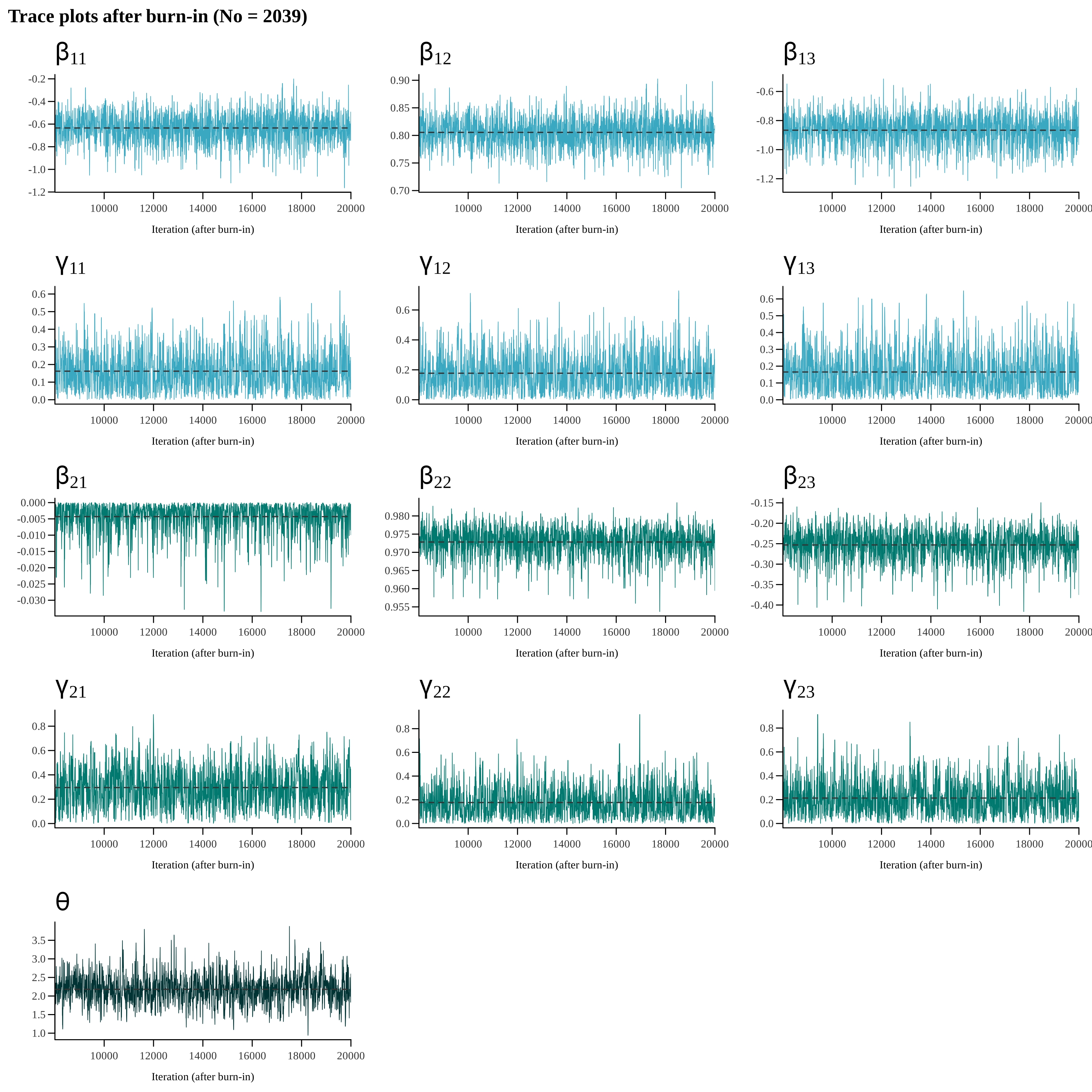} 
\caption{ACF and trace plots of the MCMC draws under the ES-CAViaR-R model with a Clayton copula.}
\label{fig:ES-CAViaR-R-ACF-TRACE}
\end{figure}

\begin{figure}[t]
	\centering   
	\includegraphics[width=0.48\textwidth]{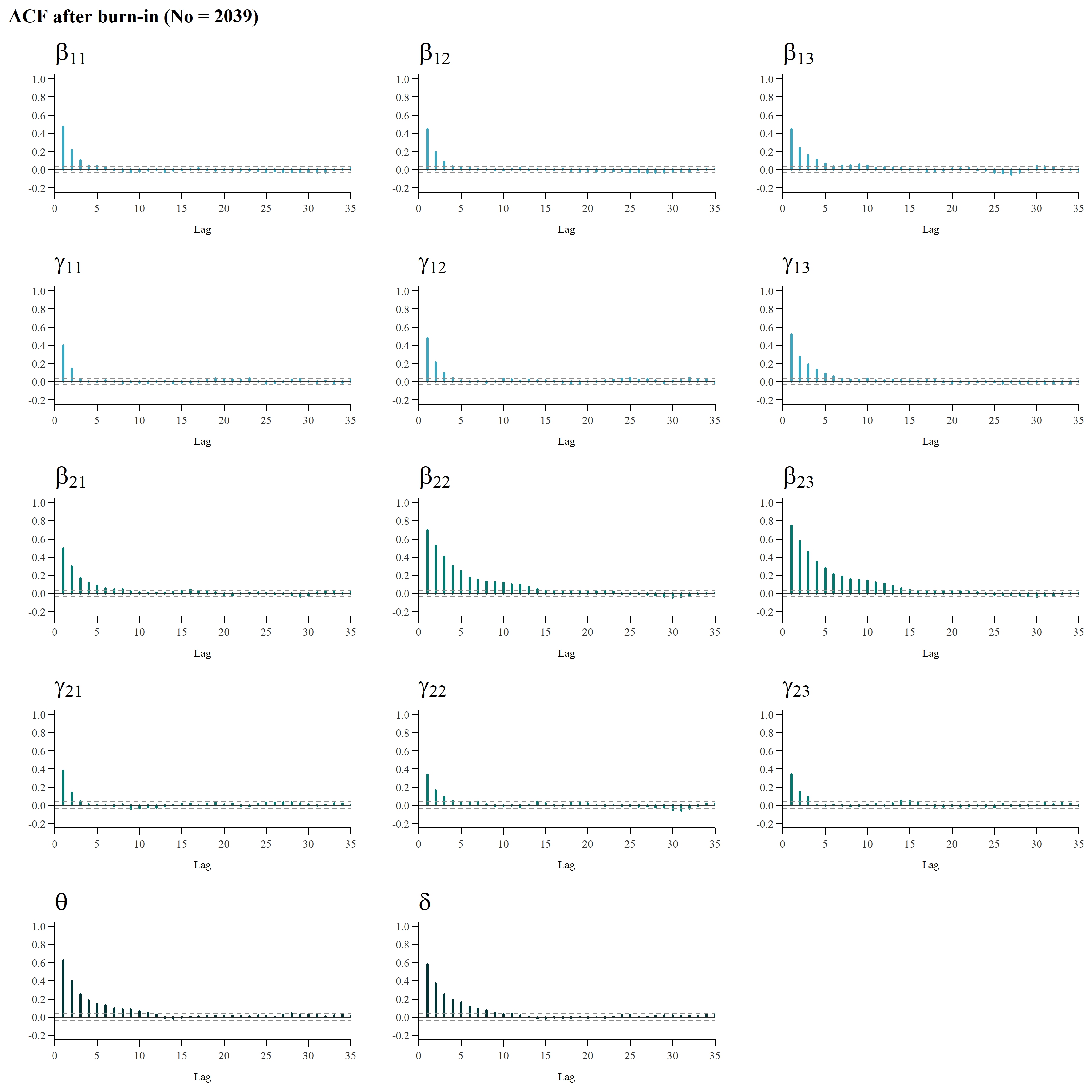} 
	\includegraphics[width=0.48\textwidth]{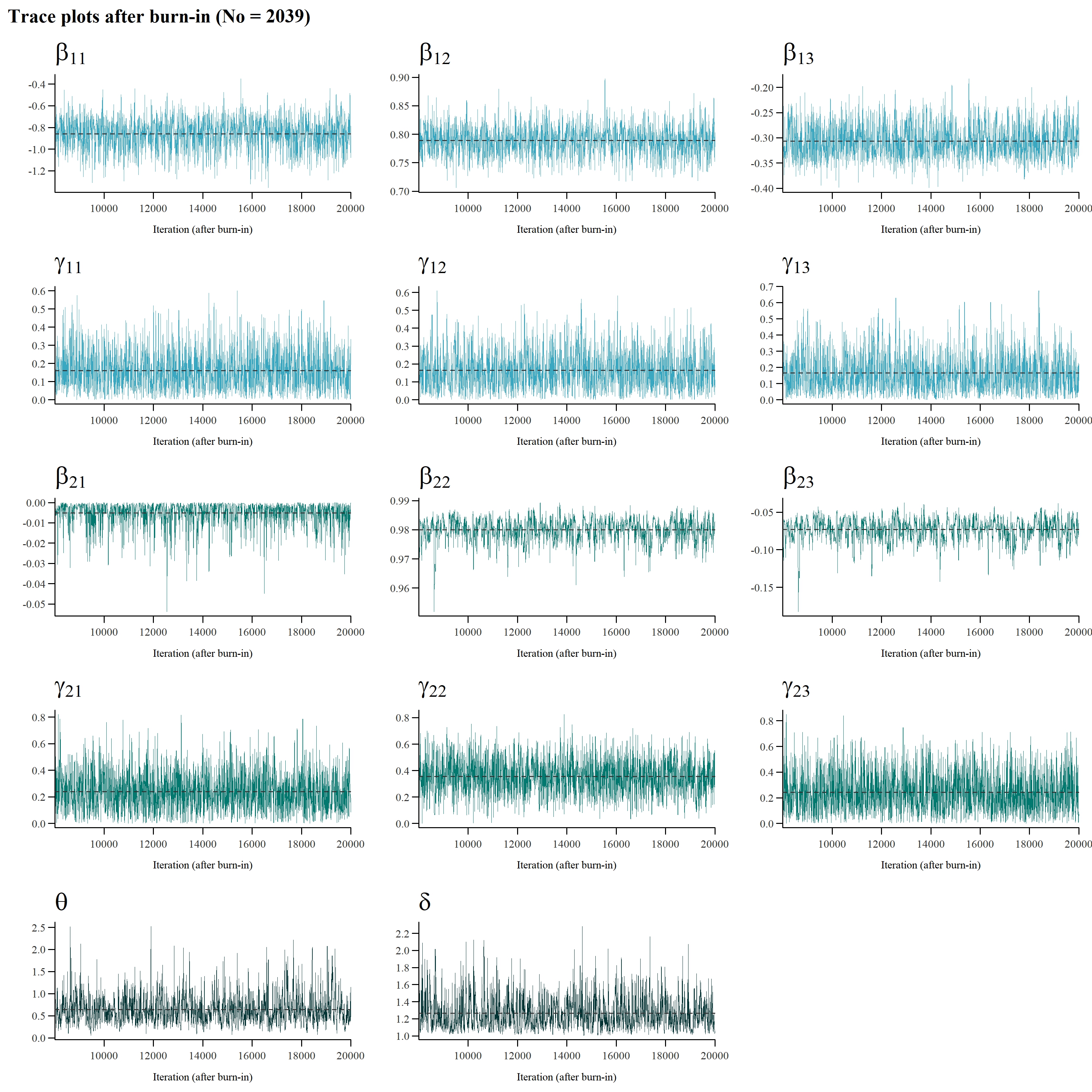}    
    \caption{ACF and trace plots of the MCMC draws under the ES-CAViaR model with a BB1 copula.}
    \label{fig:ES_BB_acf_all} 
\end{figure}

\begin{figure}[t]
	\centering   
	\includegraphics[width=0.48\textwidth]{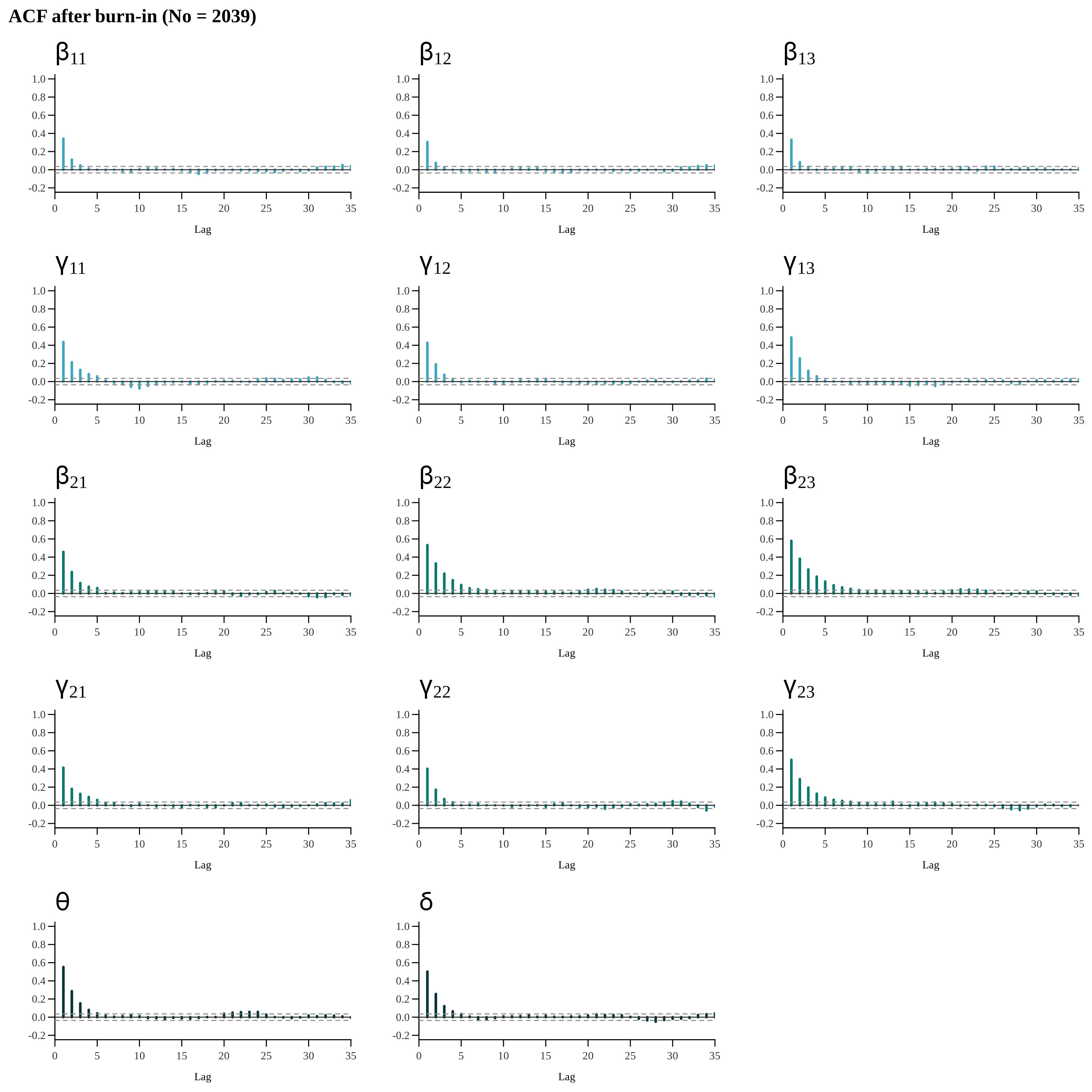}
    \includegraphics[width=0.48\textwidth]{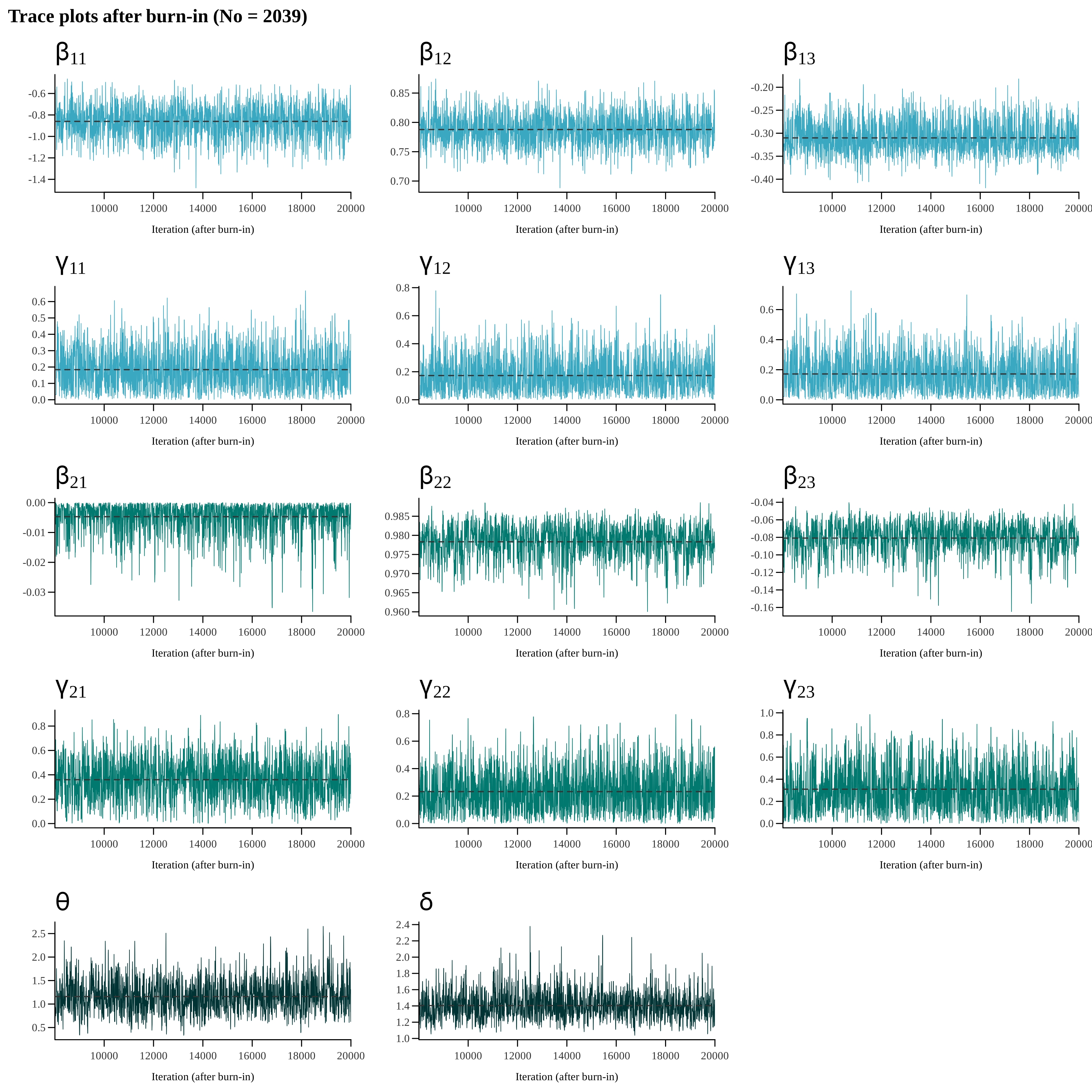} 
    \caption{ACF and trace plots of the MCMC draws under the ES-CAViaR-st model with a BB1 copula.}\label{fig:BB1-ES-CAViaR-st-ACF} 
\end{figure}

\begin{figure}[t]
	\centering   
	\includegraphics[width=0.48\textwidth]{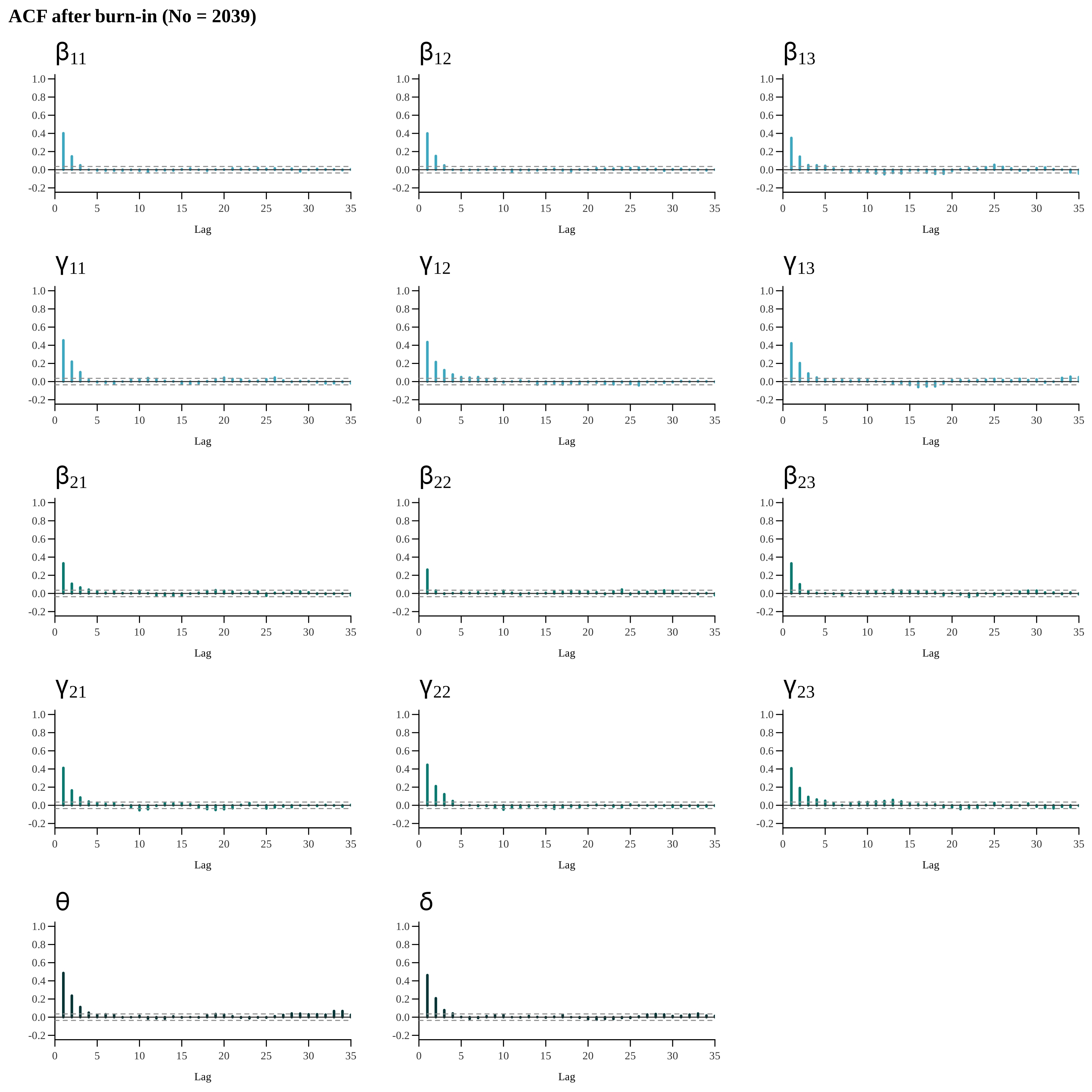}
    \includegraphics[width=0.48\textwidth]{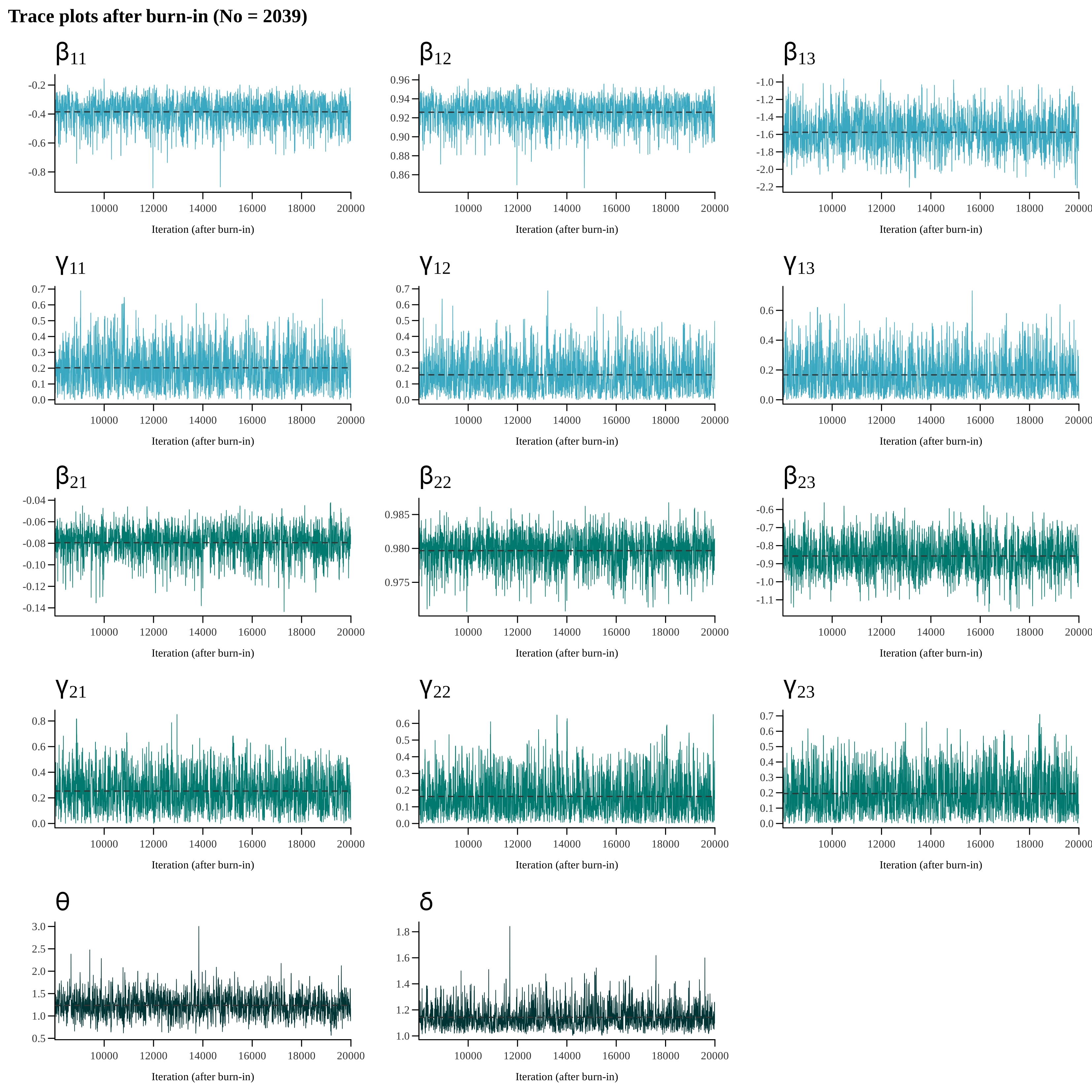}     
    \caption{ACF and trace plots of the MCMC draws under the ES-CAViaR-v model with a BB1 copula.}\label{fig:BB1-ES-CAViaR-v-ACF} 
\end{figure}

\begin{figure}[!h]
	\centering   
	\includegraphics[width=0.48\textwidth]{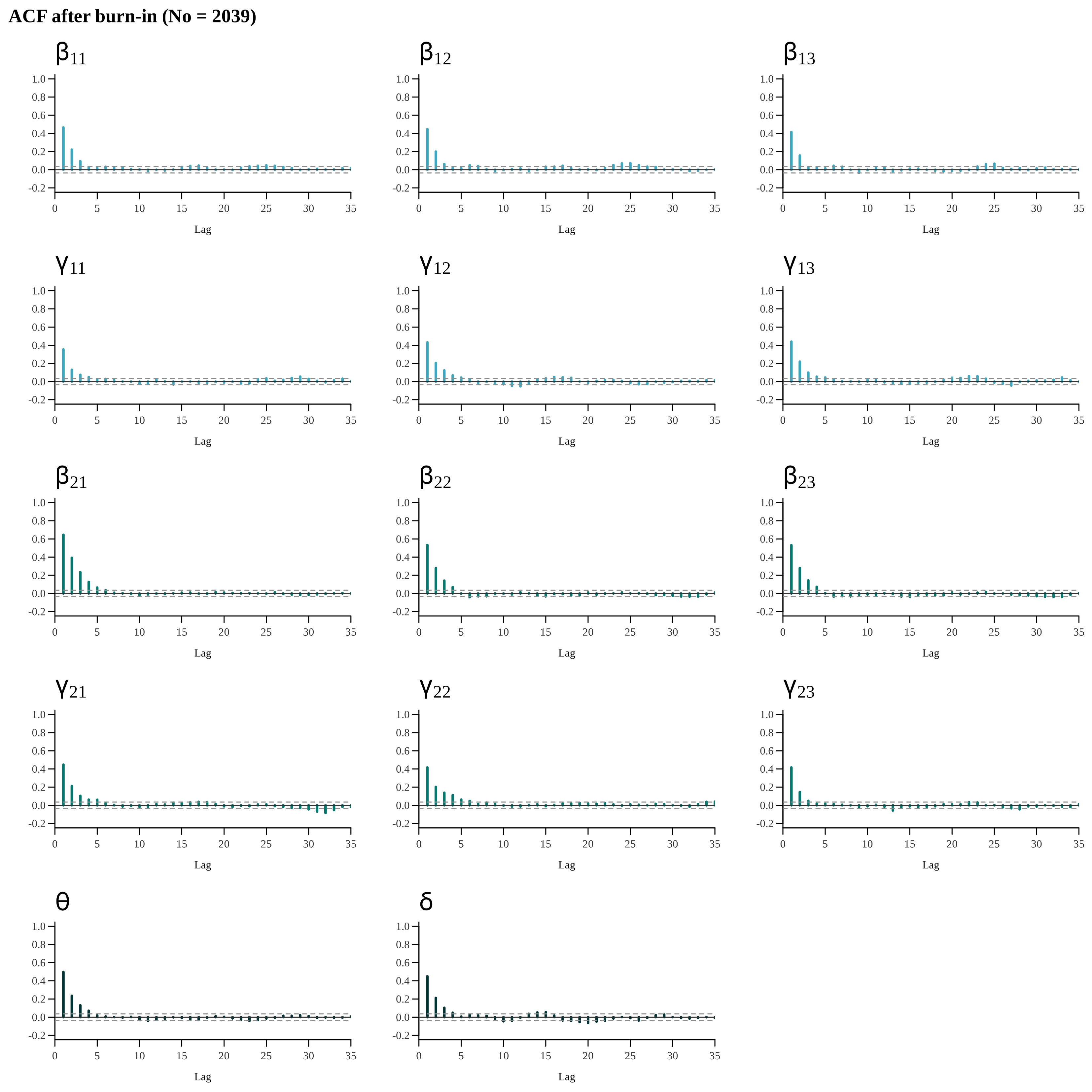}
    \includegraphics[width=0.48\textwidth]{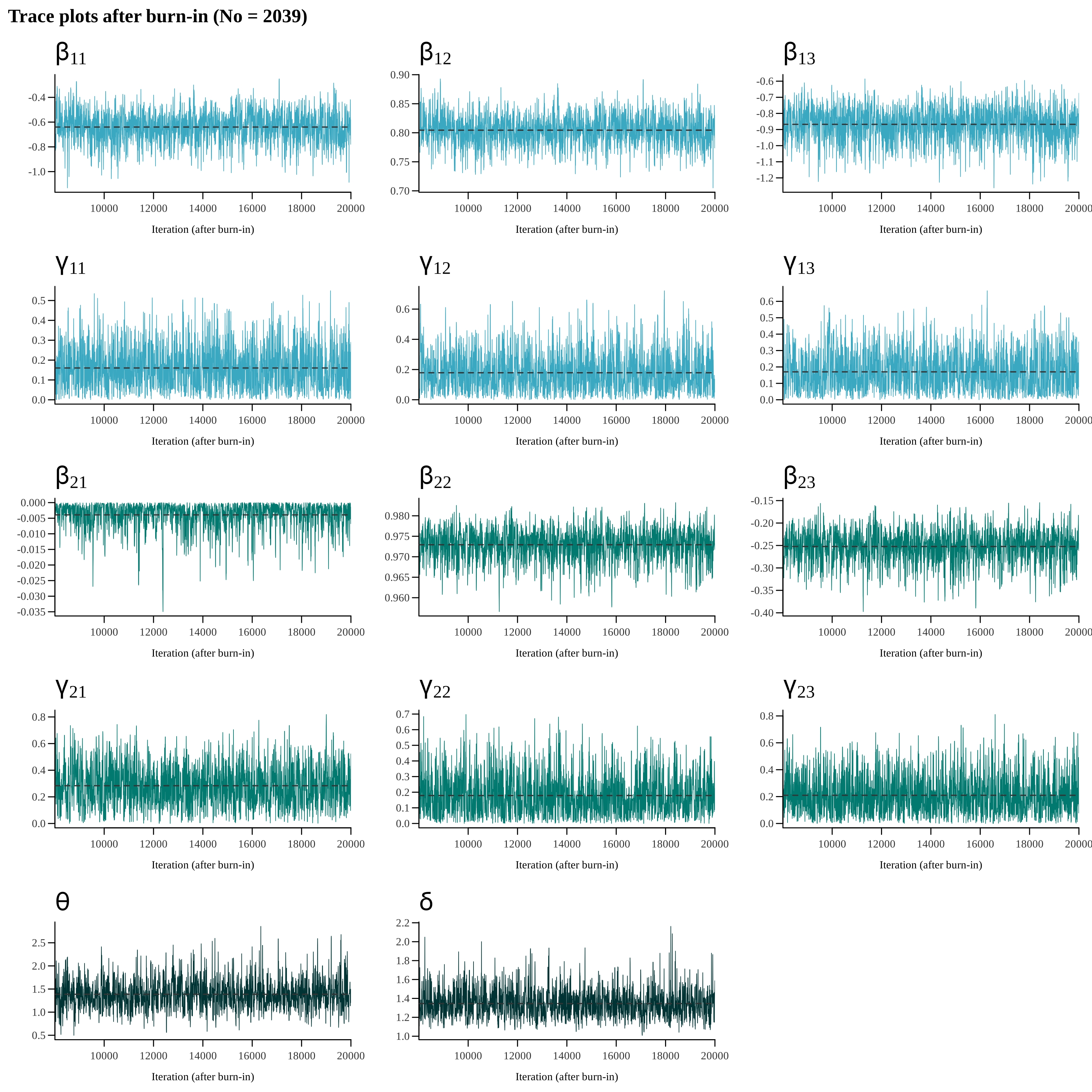} 
    \caption{ACF and trace plots of the MCMC draws under the ES-CAViaR-R model with a BB1 copula.}\label{fig:BB1-ES-CAViaR-R-ACF} 
\end{figure}

We assess MCMC convergence and mixing using trace plots and autocorrelation functions (ACFs) of the retained draws, reported in Figures~\ref{fig:ES-CAViaR-ACF}--\ref{fig:BB1-ES-CAViaR-R-ACF}. The plots show no apparent trends or signs of poor mixing, suggesting satisfactory convergence and exploration of the posterior distributions.

\clearpage
\begin{table}[t]
\centering
\caption{Pseudo-posterior summaries of the marginal
and copula parameters for the ES-CAViaR specification
under Clayton and BB1 copulas.}
\label{tab:ES-CAViaR}
{\renewcommand{\arraystretch}{0.85}
\setlength{\tabcolsep}{6pt}
\scalebox{0.85}{
\begin{tabular}{llrrrrrr}
\toprule
 &  & \multicolumn{3}{c}{Clayton} & \multicolumn{3}{c}{BB1} \\
\cmidrule(lr){3-5} \cmidrule(lr){6-8}
Return series & Parameter & Mean & 2.5\% & 97.5\% & Mean & 2.5\% & 97.5\% \\
\midrule
Bitcoin & $\beta_{1,1}$ & -0.8569 & -1.1539 & -0.6042 & -0.8578 & -1.1567 & -0.6058 \\
 & $\beta_{1,2}$ & 0.7888 & 0.7394 & 0.8385 & 0.7891 & 0.7402 & 0.8404 \\
 & $\beta_{1,3}$ & -0.3093 & -0.3647 & -0.2386 & -0.3066 & -0.3595 & -0.2384 \\
 & $\gamma_{1,1}$ & 0.1579 & 0.0069 & 0.4167 & 0.1609 & 0.0082 & 0.4195 \\
 & $\gamma_{1,2}$ & 0.1664 & 0.0072 & 0.4266 & 0.1638 & 0.0108 & 0.4111 \\
 & $\gamma_{1,3}$ & 0.1619 & 0.0080 & 0.4469 & 0.1647 & 0.0082 & 0.4574 \\
Ethereum & $\beta_{2,1}$ & -0.0054 & -0.0205 & -0.0001 & -0.0052 & -0.0213 & -0.0001 \\
 & $\beta_{2,2}$ & 0.9801 & 0.9711 & 0.9863 & 0.9799 & 0.9712 & 0.9860 \\
 & $\beta_{2,3}$ & -0.0723 & -0.1059 & -0.0493 & -0.0731 & -0.1077 & -0.0500 \\
 & $\gamma_{2,1}$ & 0.2411 & 0.0136 & 0.5655 & 0.2403 & 0.0140 & 0.5608 \\
 & $\gamma_{2,2}$ & 0.3576 & 0.1025 & 0.6241 & 0.3564 & 0.0988 & 0.6179 \\
 & $\gamma_{2,3}$ & 0.2454 & 0.0129 & 0.5779 & 0.2438 & 0.0138 & 0.5823 \\
 \midrule
Copula & $\theta$ & 1.0664 & 0.3551 & 2.0714 & 0.6397 & 0.1901 & 1.4760 \\
   & $\delta$ & -- & -- & -- & 1.2631 & 1.0306 & 1.7120 \\
\bottomrule
\end{tabular}}}
\end{table}

\begin{table}[h!]
\centering
\caption{Pseudo-posterior summaries of the marginal
and copula parameters for the ES-CAViaR-st specification under Clayton and BB1 copulas.}
\label{tab:ES-CAViaR-st}
{\renewcommand{\arraystretch}{0.85}
\setlength{\tabcolsep}{6pt}
\scalebox{0.85}{
\begin{tabular}{llrrrrrr}
\toprule
 &  & \multicolumn{3}{c}{Clayton} & \multicolumn{3}{c}{BB1} \\
\cmidrule(lr){3-5} \cmidrule(lr){6-8}
Return series & Parameter & Mean & 2.5\% & 97.5\% & Mean & 2.5\% & 97.5\% \\
\midrule
Bitcoin & $\beta_{1,1}$ & -0.8643 & -1.1535 & -0.6142 & -0.8600 & -1.1512 & -0.6040 \\
 & $\beta_{1,2}$ & 0.7868 & 0.7398 & 0.8350 & 0.7878 & 0.7390 & 0.8378 \\
 & $\beta_{1,3}$ & -0.3116 & -0.3624 & -0.2421 & -0.3102 & -0.3647 & -0.2395 \\
 & $\gamma_{1,1}$ & 0.1845 & 0.0087 & 0.4473 & 0.1840 & 0.0094 & 0.4451 \\
 & $\gamma_{1,2}$ & 0.1855 & 0.0094 & 0.4770 & 0.1728 & 0.0078 & 0.4620 \\
 & $\gamma_{1,3}$ & 0.1744 & 0.0051 & 0.4548 & 0.1718 & 0.0068 & 0.4626 \\
Ethereum & $\beta_{2,1}$ & -0.0046 & -0.0173 & -0.0001 & -0.0047 & -0.0175 & -0.0001 \\
 & $\beta_{2,2}$ & 0.9785 & 0.9697 & 0.9851 & 0.9783 & 0.9693 & 0.9852 \\
 & $\beta_{2,3}$ & -0.0800 & -0.1158 & -0.0535 & -0.0808 & -0.1198 & -0.0540 \\
 & $\gamma_{2,1}$ & 0.3625 & 0.0650 & 0.6996 & 0.3598 & 0.0538 & 0.6705 \\
 & $\gamma_{2,2}$ & 0.2393 & 0.0126 & 0.5985 & 0.2325 & 0.0166 & 0.5655 \\
 & $\gamma_{2,3}$ & 0.3250 & 0.0210 & 0.7746 & 0.3098 & 0.0206 & 0.7553 \\
 \midrule
Copula & $\theta$ & 2.0147 & 1.3192 & 2.7913 & 1.1590 & 0.5872 & 1.8762 \\
 & $\delta$ & -- & -- & -- & 1.4040 & 1.1309 & 1.7782 \\
\bottomrule
\end{tabular}}}
\end{table}

\clearpage 

\section{Additional backtesting}\label{sec:additional:backtesting}

We complement the forecast summaries in the main text with predictive calibration assessment and Frequentist backtests including calibration and comparative backtests.

\subsection{Predictive calibration assessment}

Let $I_{i,t}=\mathbb{I}(r_{i,t}<\hat Q_{i,t})$, $i=1,2$,
and $J_{t}=I_{1,t}I_{2,t}$.
We assess joint calibration using
\[
V_{12}=\frac{1}{T}\sum_{t=n+1}^{n+T}
(J_{t}-\hat p_{t}), \qquad
V_{i|j}=\frac{1}{T}\sum_{t=n+1}^{n+T}
I_{j,t}\left(I_{i,t}-\frac{\hat p_{t}}{\alpha}\right),
\]
where $i,j=1,2$ and $\hat p_{t}$ denotes the forecast of joint
shortfall probability.
Negative values indicate over-prediction,
while positive values indicate under-prediction.

Under joint calibration, the forecast probabilities
of the four exceedance outcomes $(1,1)$, $(1,0)$,
$(0,1)$, and $(0,0)$ are
$\hat p_{t}$, $\alpha-\hat p_{t}$,
$\alpha-\hat p_{t}$, and
$1-2\alpha+\hat p_{t}$, respectively, where $1$ in the first argument indicates the VaR exceedance of Bitcoin and the second means the exceedance of Ethereum.
We generate 20,000 simulated sequences from these
probabilities to construct 95\% predictive intervals
for $V_{12}$, $V_{1|2}$, and $V_{2|1}$.
These intervals quantify sampling uncertainty
conditional on the estimated forecast probabilities.

Table~\ref{tab:joint_moments} reports joint and conditional exceedance calibration moments, together with 95\% predictive intervals and stationary block-bootstrap confidence intervals~\citep{politis1994stationary}.
All eight copula-based specifications have joint
calibration moments $V_{12}$ within their predictive
intervals, indicating no clear evidence against
joint calibration. Their conditional calibration
moments also generally lie within the corresponding
predictive intervals, with only BB1 combined with ES-CAViaR showing some under-prediction of Ethereum exceedances conditional on Bitcoin distress.

In contrast, both DCC-GARCH benchmarks exhibit
joint and conditional calibration moments outside
their predictive intervals, indicating systematic
under-prediction of joint tail events.
The bootstrap confidence intervals for DCC-GARCH-N also exclude zero for $V_{12}$. 
Overall, these findings provide favorable evidence
for the joint tail-risk calibration of the proposed
copula-based framework relative to the DCC-GARCH
benchmarks, although some conditional calibration
discrepancies are observable.

\begin{table}[H]
\centering
\caption{Joint VaR exceedance calibration moments at the 1\% level,
with 95\% intervals, 2020--2025.}
\label{tab:joint_moments}
\renewcommand{\arraystretch}{0.75}
\setlength{\tabcolsep}{4pt}
\resizebox{\textwidth}{!}{%
\setlength{\tabcolsep}{3pt}
\begin{tabular}{ll c c c c c c c c c c}
\toprule
 & & & & \multicolumn{3}{c}{$V_{12}$ (\%)}
 & \multicolumn{3}{c}{$V_{1|2}$ (\%)}
 & \multicolumn{2}{c}{$V_{2|1}$ (\%)}\\
\cmidrule(lr){5-7}\cmidrule(lr){8-10}\cmidrule(lr){11-12}
Copula & Marginal & $N_{12}$ & $\sum_t\hat p_{t}$
& Value & 95\% PPI & 95\% BB CI
& Value & 95\% PPI & 95\% BB CI
& Value & 95\% PPI \\
\midrule

Clayton & ES-CAViaR & 13 & 11.4
& 0.080 & [$-$0.312, 0.325] & [$-$0.246, 0.458]
& 0.063 & [$-$0.193, 0.190] & [$-$0.203, 0.340]
& 0.131 & [$-$0.194, 0.192] \\

 & ES-CAViaR-st & 13 & 12.6
& 0.018 & [$-$0.325, 0.361] & [$-$0.306, 0.395]
& $-$0.002 & [$-$0.199, 0.191] & [$-$0.264, 0.264]
& 0.096 & [$-$0.198, 0.192] \\

 & ES-CAViaR-v & 14 & 13.5
& 0.026 & [$-$0.317, 0.369] & [$-$0.292, 0.376]
& 0.007 & [$-$0.191, 0.178] & [$-$0.180, 0.188]
& 0.074 & [$-$0.192, 0.181] \\

 & ES-CAViaR-R & 13 & 13.3
& $-$0.015 & [$-$0.310, 0.377] & [$-$0.345, 0.364]
& 0.083 & [$-$0.191, 0.179] & [$-$0.129, 0.329]
& 0.071 & [$-$0.195, 0.177] \\

\addlinespace

BB1 & ES-CAViaR & 13 & 9.3
& 0.180 & [$-$0.261, 0.327] & [$-$0.144, 0.556]
& 0.175 & [$-$0.196, 0.202] & [$-$0.084, 0.466]
& \textbf{0.230} & [$-$0.196, 0.202] \\

 & ES-CAViaR-st & 13 & 10.8
& 0.107 & [$-$0.285, 0.352] & [$-$0.225, 0.479]
& 0.102 & [$-$0.209, 0.205] & [$-$0.151, 0.372]
& 0.187 & [$-$0.202, 0.201] \\

 & ES-CAViaR-v & 14 & 11.8
& 0.106 & [$-$0.286, 0.351] & [$-$0.205, 0.461]
& 0.078 & [$-$0.203, 0.193] & [$-$0.102, 0.257]
& 0.141 & [$-$0.201, 0.192] \\

 & ES-CAViaR-R & 13 & 11.7
& 0.066 & [$-$0.278, 0.360] & [$-$0.264, 0.441]
& 0.148 & [$-$0.205, 0.193] & [$-$0.060, 0.398]
& 0.136 & [$-$0.205, 0.196] \\

\addlinespace

\multicolumn{2}{l}{DCC-GARCH-N} & 24 & 9.1
& \textbf{0.731} & [$-$0.250, 0.290] & [0.285, 1.226]
& \textbf{0.491} & [$-$0.209, 0.214] & [0.187, 0.818]
& \textbf{0.514} & [$-$0.208, 0.214] \\

\addlinespace

\multicolumn{2}{l}{DCC-GARCH-t} & 19 & 11.4
& \textbf{0.375} & [$-$0.312, 0.326] & [$-$0.018, 0.816]
& \textbf{0.230} & [$-$0.221, 0.211] & [$-$0.001, 0.471]
& \textbf{0.279} & [$-$0.212, 0.211] \\

\bottomrule
\end{tabular}}
\vspace{0.5ex}
\begin{minipage}{0.98\linewidth}
\footnotesize
\textit{Notes:}
All calibration moments are multiplied by 100.
We denote by $N_{12}$ the observed number of joint exceedances,
and $\sum_t\hat p_{t}$ the expected number of joint violations.
PPI denotes the 95\% predictive interval based on
20{,}000 simulations.
BB CI denotes the 95\% stationary block-bootstrap
percentile interval of~\citet{politis1994stationary},
using 20{,}000 resamples and a mean block length
of 12 selected following~\citet{politis2004automatic}.
Bold values lie outside the PPI.
\end{minipage}
\end{table}

\subsection{Frequentist backtests}
\label{sec:freq:backtests}

Although our proposed methodology adopts a Bayesian
framework, we additionally provide Frequentist
tests to assess forecast reliability.

\paragraph{Calibration backtests}

We first apply heteroskedasticity and
autocorrelation consistent (HAC) Wald tests to
examine whether calibration moments equal zero.
The HAC covariance estimator accounts for
heteroskedasticity and serial dependence
in these moments; see~\citet{NeweyWest1987,Andrews1991}.
The hypotheses we test are as follows:
\begin{align*}
 H_{0,\mathcal J}:
 \mathbb E[\boldsymbol g_{\mathcal J,t}]=\boldsymbol 0,
 \qquad
 H_{1,\mathcal J}:
 \mathbb E[\boldsymbol g_{\mathcal J,t}]\ne\boldsymbol 0,
\end{align*}
where $\mathcal J\in\{\{1\},\{1,2\},\{3\},\{3,4\},\{1,2,3,4,5\}\}$ is a set of indices and $\boldsymbol g_{\mathcal J,t}$ is the $\mathcal J$-subvector of
\begin{align*}
 \boldsymbol g_t
 &=
 \begin{pmatrix}
 V_{\mathrm{VaR}}(\widehat Q_{1,t};r_{1,t})\\
 V_{\mathrm{ES}}(\widehat Q_{1,t},\widehat e_{1,t};r_{1,t})\\
 V_{\mathrm{VaR}}(\widehat Q_{2,t};r_{2,t})\\
 V_{\mathrm{ES}}(\widehat Q_{2,t},\widehat e_{2,t};r_{2,t})\\
 V_C(\widehat p_t,\widehat Q_{1,t},\widehat Q_{2,t};r_{1,t},r_{2,t})
 \end{pmatrix},
 \qquad t=n+1,\ldots,n+T,
\end{align*}
for the forecast combination $(\widehat Q_{1,t},\widehat e_{1,t},\widehat Q_{2,t},\widehat e_{2,t},\widehat p_{t})^{\prime}$. 
Here, for a realization $y$ and  a forecast pair $(q,e)$ of VaR--ES,
\begin{align*}
 V_{\mathrm{VaR}}(q;y)
 &=\mathbb I(y\leq q)-\alpha,
 \nonumber\\
 V_{\mathrm{ES}}(q,e;y)
 &=e-q+\frac{1}{\alpha}(q-y)\mathbb I(y\leq q),
\end{align*}
are  the VaR and ES identification functions of \citet{FZ2016,NZ2017}; in addition, for a forecast probability $p$ of a simultaneous violation
of the marginal thresholds $q_1$ and $q_2$, define
\begin{align*}
 V_C(p,q_1,q_2;y_1,y_2)
 &=\mathbb I(y_1\leq q_1,\,y_2\leq q_2)-p.
\end{align*}
Consequently, each $\mathcal J$ corresponds to Bitcoin VaR, Bitcoin VaR--ES,
Ethereum VaR, Ethereum VaR--ES, and the combined five
conditions, respectively. 
The tests are based on Wald test statistics, which asymptotically follow a chi-squared distribution with degrees of freedom $|\mathcal J|$; see~\citet{NZ2017} for details.

Table~\ref{tab:calibration:wald} reports HAC-based Wald tests p-values for marginal
VaR and VaR-ES calibration.
None of the individual asset-specific tests rejects
the proposed specifications at the 5\% level.
For the combined five-moment test, only 
ES-CAViaR-R specifications show evidence against
calibration.
DCC-GARCH-N fails four asset-specific tests,
whereas DCC-GARCH-$t$ shows no 
calibration violations.

\paragraph{Comparative backtests}

We also conduct comparative backtests based on Diebold-Mariano test statistics following~\citet{NZ2017}.
Table~\ref{tab:comparative:all} reports HAC-standardized
score differences relative to the DCC-GARCH-$t$ benchmark, where positive values favor the benchmark.
Bold face indicates a two-sided p-value below 0.05, showing evidence against the null hypothesis that the score difference is zero on average.
For Bitcoin, none of the proposed specifications
shows statistically significant differences
under the four scoring rules.
For Ethereum, several ALS and FZ comparisons
favor DCC-GARCH-$t$.
The results suggest that the proposed models remain
competitive in marginal forecasting, although they
do not consistently outperform DCC-GARCH-$t$.

\begin{table}[t]
\centering
\begin{threeparttable}
\caption{HAC Wald-test p-values
for VaR and ES calibration.}
\label{tab:calibration:wald}
\footnotesize
\setlength{\tabcolsep}{6pt}
\renewcommand{\arraystretch}{1.15}

\begin{tabular}{llccccc}
\toprule
 & & \multicolumn{2}{c}{Bitcoin}
 & \multicolumn{2}{c}{Ethereum}
 & Combined \\
\cmidrule(lr){3-4}
\cmidrule(lr){5-6}
Copula & Marginal model
& VaR & (VaR, ES)
& VaR & (VaR, ES)
& Five moments \\
\midrule
Clayton & ES-CAViaR & 0.754 & 0.892 & 0.601 & 0.564 & 0.164 \\
 & ES-CAViaR-st & 0.754 & 0.739 & 0.601 & 0.354 & 0.082 \\
 & ES-CAViaR-v & 0.572 & 0.525 & 0.734 & 0.505 & 0.422 \\
 & ES-CAViaR-R & 0.582 & 0.685 & 0.754 & 0.214 & \textbf{0.012} \\
\addlinespace[4pt]
BB1 & ES-CAViaR & 0.754 & 0.869 & 0.601 & 0.534 & 0.085 \\
 & ES-CAViaR-st & 0.754 & 0.740 & 0.601 & 0.353 & 0.088 \\
 & ES-CAViaR-v & 0.572 & 0.529 & 0.734 & 0.512 & 0.203 \\
 & ES-CAViaR-R & 0.582 & 0.682 & 0.754 & 0.204 & \textbf{0.010} \\
\addlinespace[4pt]
\midrule
\multicolumn{2}{l}{DCC-GARCH-N} & \textbf{0.019} & \textbf{0.033} & \textbf{0.013} & \textbf{0.017} & 0.066 \\
\multicolumn{2}{l}{DCC-GARCH-$t$} & 0.360 & 0.592 & 0.148 & 0.337 & 0.466 \\
\bottomrule
\end{tabular}
\begin{tablenotes}[flushleft]
\footnotesize
\item[] \textit{Note:}
Bold face indicates $p<0.05$.
\end{tablenotes}
\end{threeparttable}
\end{table}

\begin{table}[h!]
\centering
\begin{threeparttable}
\caption{Signed HAC-standardized score differences
relative to the DCC-GARCH-$t$ benchmark.
}
\label{tab:comparative:all}
\renewcommand{\baselinestretch}{1}
\fontsize{8}{10}\selectfont
\setlength{\tabcolsep}{2.5pt}
\renewcommand{\arraystretch}{1.35}

\begin{tabular*}{\textwidth}
{@{\extracolsep{\fill}}llcccc@{\hspace{7pt}}cccc@{}}
\toprule
 & & \multicolumn{4}{c}{Bitcoin (BTC)}
   & \multicolumn{4}{c}{Ethereum (ETH)} \\
\cmidrule(lr){3-6}
\cmidrule(lr){7-10}
Copula & Marginal model
& QS & ALS & FZ & FZG
& QS & ALS & FZ & FZG \\
\midrule
Clayton & ES-CAViaR & 0.130 & 0.777 & 0.879 & $-$0.044 & 1.334 & \textbf{2.466} & \textbf{2.556} & 1.610 \\
 & ES-CAViaR-st & 0.127 & 0.730 & 0.818 & $-$0.046 & 1.441 & \textbf{2.485} & \textbf{2.560} & 1.682 \\
 & ES-CAViaR-v & $-$0.087 & 0.963 & 1.018 & $-$0.164 & 1.086 & 1.534 & 1.598 & 0.920 \\
 & ES-CAViaR-R & $-$0.696 & 0.143 & 0.210 & $-$0.783 & 0.937 & 1.702 & 1.787 & 1.003 \\
\addlinespace[4pt]
\midrule
BB1 & ES-CAViaR & 0.153 & 0.855 & 0.953 & $-$0.021 & 1.426 & \textbf{2.556} & \textbf{2.642} & 1.678 \\
 & ES-CAViaR-st & 0.129 & 0.739 & 0.827 & $-$0.045 & 1.448 & \textbf{2.487} & \textbf{2.564} & 1.688 \\
 & ES-CAViaR-v & $-$0.115 & 0.940 & 0.996 & $-$0.188 & 1.078 & 1.510 & 1.573 & 0.910 \\
 & ES-CAViaR-R & $-$0.699 & 0.148 & 0.216 & $-$0.786 & 1.007 & 1.759 & 1.844 & 1.070 \\
\addlinespace[4pt]
\midrule
\multicolumn{2}{l}{DCC-GARCH-N} & 1.107 & 1.578 & 1.514 & 0.813 & 1.655 & 1.900 & 1.841 & 1.196 \\
\multicolumn{2}{l}{DCC-GARCH-$t$} & -- & -- & -- & -- & -- & -- & -- & -- \\
\bottomrule
\end{tabular*}
\begin{tablenotes}[flushleft]
\fontsize{8}{9.5}\selectfont
\item[] \textit{Notes:}
Negative values favor the row model.
Bold face indicates a two-sided p-value below 0.05.
\end{tablenotes}
\end{threeparttable}
\end{table}

\end{document}